\documentclass[letterpaper,12pt]{article}

\usepackage{mathptmx}

\usepackage{graphics}

\usepackage[numbers,sort&compress]{natbib}

\usepackage{amssymb}

\begin{document}

\begin{titlepage}
\begin{center}
\huge{\bfseries Decomposition of Total Effect with the Notion of Natural Counterfactual Interaction Effect}\\
[1.5cm]
\Large Xin Gao${^{1,2}}$, Li Li${^{1}}$, Li Luo${^{2,3\ast}}$\\
[8.0cm]
\normalsize $^1$ Department of Mathematics and Statistics, University of New Mexico, \\
Albuquerque, New Mexico, USA\\
[0.5cm]
$^2$ Department of Internal Medicine, University of New Mexico,\\
 Albuquerque, New Mexico, USA\\
[0.5cm]
$^3$ Comprehensive Cancer Center,  University of New Mexico, \\
Albuquerque, New Mexico, USA\\
[0.5cm]
$^\ast$ Correspondence Author  ({\ttfamily LLuo@salud.unm.edu})
\end{center}
\end{titlepage}

\thispagestyle{empty}
\clearpage
\section*{Abstract}
Mediation analysis serves as a crucial tool to obtain causal inference based on directed acyclic graphs, which has been widely employed in the areas of biomedical science, social science, epidemiology and psychology. Decomposition of total effect provides a deep insight to fully understand the casual contribution from each path and interaction term. Since the four-way decomposition method was proposed to identify the mediated interaction effect in counterfactual framework, the idea had been extended to a more sophisticated scenario with non-sequential multiple mediators. However, the method exhibits limitations as the causal structure contains direct causal edges between mediators, such as inappropriate modeling of dependence and non-identifiability. We develop the notion of natural counterfactual interaction effect and find that the decomposition of total effect can be consistently realized with our proposed notion. Furthermore, natural counterfactual interaction effect overcomes the drawbacks and possesses a clear and significant interpretation, which may largely improve the capacity of researchers to analyze highly complex causal structures. \\

\noindent
\textbf{Keywords}: causal inference, dependence among mediators, interaction, mediation analysis

\clearpage
\section{Introduction}
\pagenumbering{arabic}
Decomposition of total effect helps researchers to deeply understand the effects through different mechanisms and has gained much attention in literature and application in the last decade \cite{vmul,vpre,d,s,m,v3,v4,vbook,b}. However, the vast majority of research papers investigated on the decomposition into different natural path effects \cite{vmul,vpre,d,s,m}. For example, Steen et al. discussed a flexible approach in a general framework with causally ordered mediators; however they did not evaluate the separate contributions from interaction terms \cite{s}. VanderWeele proposed a four-way decomposition of total effect, which quantifies the interaction effects in counterfactual framework \cite{v4}. He presented methods to decompose total effect into controlled direct effect, reference interaction effect, mediated interaction effect and pure indirect effect \cite{v4}. Bellavia and Valeri extended the idea to a scenario with multiple mediators but they assumed these mediators have no sequential order \cite{b}.  

Since then very limited literature studied the decomposition of total effect including counterfactual interaction effects in a more complex causal structure. A more complex causal structure may refer to the situation when direct causal links among mediators exist, which result in the dependence of one mediator on the other and a sequential order of these mediators. We find that the difficulty comes from two limitations entailed by mediated interaction effect, inappropriate modeling of dependence and non-identifiability. In order to realize the decomposition of total effect in a directed acyclic graph that contains multiple causally ordered mediators, we therefore develop the notion of natural counterfactual interaction effect, and subsequently develop decomposition methods to overcome the two limitations. Furthermore, critical comparisons between the proposed notion and mediated interaction effect are made to demonstrate the advantage of nature of counterfactual interaction effect. 

In the following sections, we first give a brief review on counterfactual definitions, notations and natural path effects. The following section presents the concept of natural counterfactual interaction effect and show that the decomposition of total effect with this notion is mathematically equivalent to the finding from previous report for a single-mediator scenario \cite{v4}. Third, we demonstrate the key differences between natural counterfactual interaction effect and mediated interaction effect when the causal structure includes multiple non-sequential mediators \cite{b}. Finally, we illustrate that the decomposition of total effect with the notion of natural counterfactual interaction effect overcomes the inappropriate modeling of dependence and non-identifiability from mediated interaction effect when direct causal links exist among mediators. The corresponding identification assumptions and real data analysis are also presented at the end. 

\section{Counterfactual definitions, notations and natural path effects}

\subsection{Counterfactual definitions and notations}
We introduce the basic definitions and notations with a single-mediator scenario as shown in Figure 1. The definition of a counterfactual formula is the potential value of outcome $Y$ or mediator $M$ in the causal structure that would have been observed if the exposure $A$ or mediator $M$ were fixed at some certain level that possibly is contrary to the fact \cite{vbook,riden,p01}. Let $Y(a)$ denote the potential value of $Y$ that would have been observed if the exposure $A$ were fixed at a constant level $a$ \cite{vbook}. Similarly, $M(a)$ denotes the potential value of $M$ that would have been observed if $A$ were fixed at $a$ and $Y(a,m)$ denotes the potential value of $Y$ that would have been observed if $A$ and $M$ were fixed at $a$ and $m$, respectively \cite{vbook}. We apply a nested counterfactual formula, e.g. $Y(a, M(a^{\ast}))$, to denote the potential value of $Y$ that would have been observed if the exposure were fixed at $a$ and the mediator were set to what would have been observed when the exposure were fixed at $a^{\ast}$ (Figure 2) \cite{vbook}.\\

\subsection{Natural path effects}
The total effect ($TE$) for an individual in counterfactual framework is defined by the difference between $Y(a)$ and $Y(a^{\ast})$ \cite{vbook}, where $a$ is the treatment level and $a^{\ast}$ is the reference level of the exposure $A$, respectively. Total effect could be decomposed into two parts, natural direct effect ($NDE$) and natural indirect effect ($NIE$) \cite{vbook,p01,rsem}. $NDE$ represents the causal effect along the direct path from $A$ to $Y$ and $NIE$ represents the causal effect along the indirect path from $A$ through $M$ to $Y$. The formulas are given as follows: 
\begin{eqnarray*}
  TE & = & Y(a)-Y(a^\ast)\\
  & = & Y(a,M(a)) - Y(a^{\ast},M(a^\ast))\\
  & = & Y(a,M(a)) - Y(a^\ast,M(a)) + Y(a^\ast,M(a)) - Y(a^{\ast},M(a^\ast))\\
  \\
  NDE & = & Y(a,M(a)) - Y(a^\ast,M(a))\\
  \\
  NIE & = & Y(a^\ast,M(a)) - Y(a^{\ast},M(a^\ast)),
\end{eqnarray*}
where the second equality follows by composition axiom \cite{vbook,a} and the third equality follows by subtracting and adding the same counterfactual formula.  

In this case, $NIE$ is the path-specific effect \cite{p01} along the indirect path or pure indirect effect $PIE$ \cite{riden}, and $NDE$ is total direct effect \cite{riden}. We can also decompose the total effect in a slightly different way:
\begin{eqnarray*}
  TE & = & Y(a)-Y(a^\ast)\\
  & = & Y(a,M(a)) - Y(a^{\ast},M(a^\ast))\\
  & = & Y(a,M(a)) - Y(a,M(a^\ast)) + Y(a,M(a^\ast)) - Y(a^{\ast},M(a^\ast))\\
  \\
  NDE & = & Y(a,M(a^\ast)) - Y(a^\ast,M(a^\ast))\\
  \\
  NIE & = & Y(a,M(a)) - Y(a,M(a^\ast)).
\end{eqnarray*}

Here $NDE$ is the path-specific effect along the direct path or pure direct effect ($PDE$) \cite{riden,p01} and $NIE$ is the total indirect effect \cite{riden}.

\section{Decomposition of total effect in a single-mediator scenario}
VanderWeele proposed a four-way decomposition in a single-mediator scenario (Figure 1) to account for counterfactual interaction effects \cite{v4}, where the total effect can be decomposed into controlled direct effect ($CDE(m^\ast)$), reference interaction effect ($INT_{ref}(m^\ast)$), mediated interaction effect ($INT_{med}$) and pure indirect effect ($PIE$), and $m^\ast$ is the fixed reference level of mediator $M$. The formulas of the four components are \cite{v4}:
\begin{eqnarray*}
  CDE(m^\ast) & := & Y(a,m^\ast)-Y(a^\ast,m^\ast)\\
  \\
  INT_{ref}(m^\ast) & := & \sum_m \left[Y(a,m)-Y(a^\ast,m)-Y(a,m^\ast)+Y(a^\ast,m^\ast)\right]\\
  & &\times I\left(M(a^\ast)=m\right)\\
  \\
  INT_{med} & := & \sum_m \left[Y(a,m)-Y(a^\ast,m)-Y(a,m^\ast)+Y(a^\ast,m^\ast)\right]\\   
  & &\times \left[I\left(M(a)=m\right)- I\left(M(a^\ast)=m\right) \right]\\ 
  \\
  PIE & := & \sum_m \left[Y(a^\ast,m)-Y(a^\ast,m^\ast)\right]   
  \times \left[I\left(M(a)=m\right)- I\left(M(a^\ast)=m\right) \right],\\ 
\end{eqnarray*}
where $TE = CDE(m^\ast)+INT_{ref}(m^\ast)+INT_{med}+PIE$.

We would like to focus on the mediated interaction effect, which has a clear counterfactual interpretation in the sense that it is the portion of total effect due to interaction and mediation \cite{v4,b}. It can be rewritten as:
\begin{eqnarray}
  INT_{med} & := & \sum_m \left[Y(a,m)-Y(a^\ast,m)-Y(a,m^\ast)+Y(a^\ast,m^\ast)\right] \nonumber\\   
  & &\times \left[I\left(M(a)=m\right)- I\left(M(a^\ast)=m\right) \right]\nonumber\\ 
  \nonumber\\
  & = & \sum_m \left[Y(a,m)-Y(a^\ast,m)\right]   
  \times \left[I\left(M(a)=m\right)- I\left(M(a^\ast)=m\right) \right]\nonumber\\
  \nonumber\\
  & = & \sum_m Y(a,m)I\left(M(a)=m\right) - \sum_m Y(a^\ast,m)I\left(M(a)=m\right) \nonumber\\
  & & - \sum_m Y(a,m)I\left(M(a^\ast)=m\right) + \sum_m Y(a^\ast,m)I\left(M(a^\ast)=m\right)\nonumber\\
  \nonumber\\
  & = & Y(a,M(a)) - Y(a^\ast,M(a)) - Y(a,M(a^\ast)) + Y(a^{\ast},M(a^\ast)),
\end{eqnarray}
where the second equality follows by the fact that $Y(a,m^\ast)$ and $Y(a^\ast,m^\ast)$ are constants and can be canceled out through the summation.

To interpret the mediated interaction effect in a different point of view, we first consider a linear model of $Y$ with interaction effect between $A$ and $M$ assuming that both $A$ and $M$ are binary for simplicity, and $Y$ is continuous:
\begin{eqnarray*}
  E(Y|A,M)=\theta_0 + \theta_1 I(A=1)+\theta_2 I(M=1)+\theta_3 I(A=1)I(M=1).
\end{eqnarray*}

We consider the classical definition and notation of additive interaction effect, which measures the magnitude of the joint effects of two factors exceeding the individual effect of each factor, and can be expressed as follows \cite{vbook,rbook}: 
\begin{eqnarray}
  & & E(Y|1,1)-E(Y|0,1)-E(Y|1,0)+E(Y|0,0) \nonumber \\
  \nonumber \\
  & = & p_{11} - p_{01} - p_{10} + p_{00} \\
  \nonumber \\
  & = & (\theta_0 + \theta_1 + \theta_2 + \theta_3)-(\theta_0+\theta_2)-(\theta_0+\theta_1)+(\theta_0) \nonumber \\
  \nonumber \\
  & = & \theta_3, \nonumber
\end{eqnarray}
where $p_{am}=E[Y|A=a,M=m]$ and $\theta_3$ is the interaction effect between $A$ and $M$.

Comparing Eq.(1) and Eq.(2), it can be seen that the counterfactual formulas $M(a^\ast)$ and $M(a)$ in Eq.(1) play the roles of reference level and treatment level for mediator $M$ in counterfactual framework, respectively. Therefore, we propose the following definition of natural counterfactual interaction effect in a single-mediator scenario (Figure 1).\\
\\
\textbf{Definition 1}. 
We follow the classical definition of additive interaction effect to define the natural counterfactual interaction effect ($NatINT$). The natural counterfactual interaction effect in a single-mediator scenario is defined as follows:
\begin{eqnarray*}
NatINT_{AM} := Y(a,M(a)) - Y(a^\ast,M(a)) - Y(a,M(a^\ast)) + Y(a^{\ast},M(a^\ast)),
\end{eqnarray*}
where $M(a^\ast)$ and $M(a)$, the values of $M$ that would have occurred if $A$ were fixed at $a^\ast$ and $a$, are called the natural reference level and natural treatment level of $M$, respectively. \\

Natural counterfactual interaction effect works as the analogue of interaction effect in a linear model. For a single-mediator scenario, natural counterfactual interaction effect equals mediated interaction effect from the four-way decomposition of total effect. Figure 3 graphically illustrates the nature of natural counterfactual interaction effect or mediated interaction effect in the form of classical definition of additive interaction effects \cite{vbook,rbook} with counterfactual formulas of $Y$ with fixed reference level of $A$ at $a^\ast$ and fixed treatment level of $A$ at $a$ as well as with natural reference level of $M$ at $M(a^\ast)$ and natural treatment level of $M$ at $M(a)$. 

Despite the mathematical equivalence, natural counterfactual interaction effect exhibits a dissimilar elucidation in the sense that it measures the simultaneous effect between exposure $A$ with fixed values and mediator $M$ with potential values on the total effect. Another way to put it is that exposure $A$ is set to fixed reference or treatment levels by external intervention \cite{p95} as well as mediator $M$ retains its status as a counterfactual formula and makes advantage of the natural reactions to exposure $A$ for its reference and treatment levels. 

Furthermore, when two or more mediators are present with direct causal edges, natural counterfactual interaction effect takes the dependence among multiple mediators into account and avoid inappropriate modeling and non-identifiability. We use Definition 1 as the foundation to gradually build up the concept for more sophisticated causal structures, which will be studied in subsequent discussions. 

\section{Decomposition of total effect in a non-sequential multiple-mediator scenario}
A non-sequential multiple-mediator scenario investigates the causal structure that has two or more mediators with no direct causal edge between any two of them (Figure 4). Classical definition of n-way additive interaction is defined as the joint effect of all n factors together compared to the combined effects of all n-1 factors separately \cite{vtu}. We define high order natural counterfactual interaction effects in a similar way. \\
\\
\textbf{Definition 2}. 
We denote the reference level and treatment level of exposure $A$ by $a^\ast$ and $a$, respectively. Let the natural reference level of mediator $M_i$ be $M_i(a^\ast)$, which is the value of $M_i$ that would have occurred if $A$ were fixed at $a^\ast$. Let the natural treatment level of $M_i$ be $M_i(a)$, which is the value of $M_i$ that would have occurred if $A$ were fixed at $a$. We follow the definition of classical additive interaction effects to define the natural counterfactual interaction effects with corresponding natural reference levels and natural treatment levels in a non-sequential multiple-mediator scenario. \\

For illustration purpose, we consider a two-mediator causal structure with no sequential order as shown in Figure 5. We have the following formulas from Definition 2: 
\begin{eqnarray*}
  NatINT_{AM_1} & := & Y\left(a,M_1(a),M_2(a^\ast)\right)-Y\left(a^\ast,M_1(a),M_2(a^\ast)\right)\\
  & & - Y\left(a,M_1(a^\ast),M_2(a^\ast)\right)+Y\left(a^\ast,M_1(a^\ast),M_2(a^\ast)\right)\\
  \\
  NatINT_{AM_2} & := & Y\left(a,M_1(a^\ast),M_2(a)\right)-Y\left(a,M_1(a^\ast),M_2(a^\ast)\right)\\
  & & - Y\left(a^\ast,M_1(a^\ast),M_2(a)\right)+Y\left(a^\ast,M_1(a^\ast),M_2(a^\ast)\right)\\
  \\
  NatINT_{AM_1M_2} & := & Y\left(a,M_1(a),M_2(a)\right)-Y\left(a,M_1(a),M_2(a^\ast)\right)\\
  & & - Y\left(a,M_1(a^\ast),M_2(a)\right)-Y\left(a^\ast,M_1(a),M_2(a)\right)\\
  & & +Y\left(a^\ast,M_1(a^\ast),M_2(a)\right)+Y\left(a^\ast,M_1(a),M_2(a^\ast)\right)\\
  & & +Y\left(a,M_1(a^\ast),M_2(a^\ast)\right)-Y\left(a^\ast,M_1(a^\ast),M_2(a^\ast)\right)\\
  \\
  NatINT_{M_1M_2} & := & Y\left(a^\ast,M_1(a),M_2(a)\right)-Y\left(a^\ast,M_1(a),M_2(a^\ast)\right)\\
  & & -Y\left(a^\ast,M_1(a^\ast),M_2(a)\right)+Y\left(a^\ast,M_1(a^\ast),M_2(a^\ast)\right), 
\end{eqnarray*}
where the subscript of $NatINT$ indicates the factors involved in the natural counterfactual interaction effect.

Table 1 presents the 2-way and 3-way interaction effects in a linear model regressed on binary $A$, $M_1$ and $M_2$, and their parallels in counterfactual framework, i.e., natural counterfactual interaction effects.

We show in Appendix A that total effect can be consistently decomposed into 10 components at individual level including the natural counterfactual interaction effects: 
\begin{eqnarray*}
  TE & = & CDE(m_1^\ast,m_2^\ast)+INT_{ref\mbox{-}AM_1}(m_1^\ast,m_2^\ast)+INT_{ref\mbox{-}AM_2}(m_1^\ast,m_2^\ast)\\
  & & +INT_{ref\mbox{-}AM_1M_2}(m_1^\ast,m_2^\ast)+ NatINT_{AM_1} + NatINT_{AM_2}+ NatINT_{AM_1M_2}\\
  & & + NatINT_{M_1M_2} + PIE_{M_1} + PIE_{M_2},
\end{eqnarray*}
where $m_1^\ast$ and $m_2^\ast$ are fixed reference levels for $M_1$ and $M_2$, respectively, $CDE$ denotes controlled direct effect, $INT_{ref}$ denotes reference interaction effect, $NatINT$ denotes natural counterfactual interaction effect and $PIE$ denotes pure indirect effect. 

Bellavia and Valeri \cite{b} proposed the extension of mediated interaction effect in this scenario. The key difference needs to be pointed out between their approach and the notion we developed. The mediated interaction effect between $A$ and $M_1$, for example, is obtained by assigning $M_2$ a fixed reference level at $m_2^\ast$ and allowing $M_1$ to naturally react to exposure $A$ (Appendix B). On the other hand, the natural counterfactual interaction effect between $A$ and $M_1$ allows both mediators to naturally react to exposure $A$. Figure 6 presents a graphical comparison on this key difference. The method from Bellavia and Valeri supports a meaningful interpretation in a non-sequential multiple-mediator scenario by controlling a certain mediator at a fixed level. However, the limitations of mediated interaction effect start to obstruct a valid decomposition of total effect as the direct causal links appear among mediators, which will be discussed in next section. 

\section{Decomposition of total effect in a one-path multiple-mediator scenario}
A one-path multiple-mediator scenario means that the causal structure has two or more mediators, and there only exists direct causal links pointing from mediator $M_i$ to $M_{i+1}$, where $1 \leq i \leq n-1$ if the diagram contains a total of n mediators (Figure 7). This type of causal structure is a special case of the situation with multiple mediators in a sequential order. 
We can again propose the definition of natural counterfactual interaction effect for this setting. \\
\\
\textbf{Definition 3}. 
We denote the reference level and treatment level of exposure $A$ by $a^\ast$ and $a$, respectively. Let the natural reference level of mediator $M_i$ be:
\begin{enumerate}
\item 	$M_i(a^\ast)$ if $i=1$, which is the value of $M_i$ that would have occurred if $A$ were fixed at $a^\ast$,\\
\item   $M_i(a^\ast,M_{i-1}(\cdots))$ if $i\neq1$, which is the value of $M_i$ that would have occurred if $A$ were fixed at $a^\ast$ and with the corresponding potential value of $M_{i-1}$;
\end{enumerate}
Let the natural treatment level of mediator $M_i$ be:
\begin{enumerate}
\item 	$M_i(a)$ if $i=1$, which is the value of $M_i$ that would have occurred if $A$ were fixed at $a$,\\
\item   $M_i(a,M_{i-1}(\cdots))$ if $i\neq1$, which is the value of $M_i$ that would have occurred if $A$ were fixed at $a$ and with the corresponding potential value of $M_{i-1}$;
\end{enumerate}
We follow the definition of classical additive interaction effects and avoid non-identifiability to define the natural counterfactual interaction effects with corresponding natural reference levels and natural treatment levels for a one-path multiple-mediator scenario. \\

We consider a structure with two sequential mediators (Figure 8) for simplicity. According to Definition 3, $M_1(a^\ast)$ and $M_1(a)$ are natural reference level and natural treatment level of $M_1$, respectively. For the second mediator $M_2$, a more complicated situation needs to be tackled since there exists a direct causal link pointing from $M_1$ to $M_2$. It can be seen that both $M_2(a^\ast,M_1(a))$ and $M_2(a^\ast,M_1(a^\ast))$ may be used as the natural reference level of $M_2$ as well as both $M_2(a,M_1(a))$ and $M_2(a,M_1(a^\ast))$ may be used as the natural treatment level of $M_2$. Before making a choice, we inevitably have to discuss the non-identifiability for the counterfactual formula of outcome $Y$. To be concise, $Y\left(a,M_1(a),M_2(a,M_1(a^\ast))\right)$ is not identifiable where the two counterfactual formulas of $M_1$ have different values of exposure $A$. The reason is that the path $A\rightarrow M_1\rightarrow Y$ and the path $A\rightarrow M_1\rightarrow M_2 \rightarrow Y$ in Figure 8 form up a kite graph in which $M_1$ cannot be activated by two different values of $A$ in the mean time \cite{a}. Otherwise, the counterfactual formula of outcome $Y$ is referred to as a problematic formula and implies non-identifiability \cite{a}. Namely the counterfactual formula of $M_1$ in the counterfactual formula of $M_2$ has to be the same as the one in the second input argument of the counterfactual formula of $Y$. Figure 9 presents a graphical illustration for the non-identifiable $Y\left(a,M_1(a),M_2(a,M_1(a^\ast))\right)$. Accordingly, the following formulas can be obtained from Definition 3:
\begin{eqnarray*}
  NatINT_{AM_1} & := & Y\left(a,M_1(a),M_2(a^\ast,M_1(a))\right)-Y\left(a^\ast,M_1(a),M_2(a^\ast,M_1(a))\right)\\
  & & - Y\left(a,M_1(a^\ast),M_2(a^\ast,M_1(a^\ast))\right)+Y\left(a^\ast,M_1(a^\ast),M_2(a^\ast,M_1(a^\ast))\right)\\
  \\
  NatINT_{AM_2} & := & Y\left(a,M_1(a^\ast),M_2(a,M_1(a^\ast))\right)-Y\left(a,M_1(a^\ast),M_2(a^\ast,M_1(a^\ast))\right)\\
  & & - Y\left(a^\ast,M_1(a^\ast),M_2(a,M_1(a^\ast))\right)+Y\left(a^\ast,M_1(a^\ast),M_2(a^\ast,M_1(a^\ast))\right)\\
  \\
  NatINT_{AM_1M_2} & := & Y\left(a,M_1(a),M_2(a,M_1(a))\right)-Y\left(a,M_1(a),M_2(a^\ast,M_1(a))\right)\\
  & & - Y\left(a,M_1(a^\ast),M_2(a,M_1(a^\ast))\right)-Y\left(a^\ast,M_1(a),M_2(a,M_1(a))\right)\\
  & & +Y\left(a^\ast,M_1(a^\ast),M_2(a,M_1(a^\ast))\right)+Y\left(a^\ast,M_1(a),M_2(a^\ast,M_1(a))\right)\\
  & & +Y\left(a,M_1(a^\ast),M_2(a^\ast,M_1(a^\ast))\right)-Y\left(a^\ast,M_1(a^\ast),M_2(a^\ast,M_1(a^\ast))\right)\\
  \\
  NatINT_{M_1M_2} & := & Y\left(a^\ast,M_1(a),M_2(a,M_1(a))\right)-Y\left(a^\ast,M_1(a),M_2(a^\ast,M_1(a))\right)\\
  & & -Y\left(a^\ast,M_1(a^\ast),M_2(a,M_1(a^\ast))\right)+Y\left(a^\ast,M_1(a^\ast),M_2(a^\ast,M_1(a^\ast))\right). 
\end{eqnarray*}

We show in Appendix C that the total effect can be consistently decomposed into 9 components at individual level including the natural counterfactual interaction effects: 
\begin{eqnarray*}
  TE & = & CDE(m_1^\ast,m_2^\ast)+INT_{ref\mbox{-}AM_1}(m_1^\ast,m_2^\ast)+INT_{ref\mbox{-}AM_2+AM_1M_2}(m_2^\ast)\\
  & & + NatINT_{AM_1} + NatINT_{AM_2}+ NatINT_{AM_1M_2}+ NatINT_{M_1M_2}\\
  & & + PIE_{M_1} + PIE_{M_2}.
\end{eqnarray*}

As a side note, the reference interaction effects $INT_{ref\mbox{-}AM_2}(m_1^\ast,m_2^\ast)$ and $INT_{ref\mbox{-}AM_1M_2}(m_1^\ast,m_2^\ast)$ are not separately identifiable under this circumstance because both of them contain a non-identifiable counterfactual formula of outcome $Y$ but the sum of these two components is identifiable and becomes a function of $m_2^\ast$ (Appendix D). 

Considering a general setting that includes a total number of n mediators in a one-path pattern, an example of counterfactual formula of outcome $Y$ can be written as:
\begin{eqnarray*}
  Y\left(a,M_1(a^\ast),\cdots,M_{i-1}(a^\ast,M_{i-2}(\cdots)),M_i(a,M_{i-1}(\cdots)),\cdots,M_n(a,M_{n-1}(\cdots))\right),
\end{eqnarray*}
where the potential value of $M_i(a,M_{i-1}(\cdots))$ depends on the fixed exposure value $a$ and the potential value of its immediate preceding mediator $M_{i-1}$. In order to avoid non-identifiability, all counterfactual formulas of any certain mediator have to be identical in the counterfactual formula of $Y$. 

If the natural reference level of $M_n$ is required in the counterfactual formula of $Y$, then assign $a^\ast$ to the first input argument of $M_n$ and write $M_n(a^\ast,\ M_{n-1}(\cdots))$. if a natural treatment level of $M_{n-1}$, for example, is needed in the same counterfactual formula of $Y$, then assign $a$ to the first input argument of $M_{n-1}$ and write $M_{n-1}(a,\ M_{n-2}(\cdots))$. The process can be repeated until reaching $M_1$. Once $M_1$ is considered, choose $M_1(a^\ast)$ for the natural reference level or $M_1(a)$ for the natural treatment level. This is the spirit of natural counterfactual interaction effect. The mediators are not fixed at certain levels and instead are naturally determined based on the causal structure. By following the classical definition of additive interaction effect, the desired natural counterfactual interaction effect can be obtained with corresponding counterfactual formulas of outcome $Y$ that are derived conforming to the above steps. 

The notion of natural counterfactual interaction effect surmounts the inappropriate modeling of dependence from employing mediated interaction effect. For example, the mediated interaction effect between $A$ and $M_1$, $INT_{med\mbox{-}AM_1}(m_2^*)$, requires a fixed reference level of $M_2$ at $m_2^*$ by external intervention and does not take into account the direct causal link pointing from $M_1$ to $M_2$ which conveys the important feature of two sequential mediators. We show in Appendix E that the nature of $INT_{med\mbox{-}AM_1}$ is identical to the modeling of two non-sequential mediators illustrated in Figure 6A where the mechanism does not evaluate the dependence of $M_2$ on $M_1$. In contrast, natural counterfactual interaction effect allows appropriate modeling of the dependence among mediators and provides a valid interpretation on how interaction terms in the structural models contribute to the total effect. Figure 10 illustrates the crucial difference between the two concepts.

The mediated interaction effect may impede identifiability. In Appendix E, we show that both $INT_{med\mbox{-}AM_2}(m_1^*)$ and $INT_{med\mbox{-}AM_1M_2}(m_1^*,m_2^*)$ are non-identifiable. The notion of natural counterfactual interaction effect overcomes such limitation by preventing problematic counterfactual formulas from the kite graph. 

\section{Identification Assumptions}
We first consider a single-mediator scenario as shown in Figure 1. Four identification assumptions are required \cite{vcon}, which are listed below as ($A^\prime 1$) -- ($A^\prime 4$): 
\begin{eqnarray*}
  & & Y(a,m) \perp A|C     \hspace{2.4cm}    (A^\prime 1)\\  
  \\
  & & Y(a,m) \perp M|\{A,C\}     \hspace{1.5cm}    (A^\prime 2)\\ 
  \\
  & & M(a) \perp A|C     \hspace{2.8cm}    (A^\prime 3)\\
  \\
  & & Y(a,m) \perp M(a^\ast)|C.     \hspace{1.5cm}    (A^\prime 4)\\
\end{eqnarray*}

The assumptions above state that, given a covariate set $C$ or $\{A,C\}$, there exists no unmeasured variables confounding the association between exposure $A$ and outcome $Y$ ($A^\prime 1$), there exists no unmeasured variables confounding the association between mediator $M$ and outcome $Y$ ($A^\prime 2$) and there exists no unmeasured variables confounding the association between exposure $A$ and mediator $M$ ($A^\prime 3$) \cite{vbook}. ($A^\prime 4$) is a strong assumption and a few researchers published their works on this topic \cite{v4,s,ralt}. It could be interpreted as there exists no variables that are causal descendants of exposure $A$, and in the meantime, are confounding the association between mediator $M$ and outcome $Y$ \cite{s,p01}. 

The analogues of ($A^\prime 1$) -- ($A^\prime 4$) for a directed acyclic graph with two mediators in a sequential order (Figure 8) can be found by considering $M_1$ and $M_2$ as a set \cite{s}. Namely, we have the corresponding identification assumptions ($A1$) -- ($A4$):
\begin{eqnarray*}
  & & Y(a,m_1,m_2) \perp A|C     \hspace{5.1cm}    (A1)\\  
  \\
  & & Y(a,m_1,m_2) \perp \{M_1,M_2\}|\{A,C\}     \hspace{2.9cm}    (A2)\\ 
  \\
  & & \{M_1(a),M_2(a,m_1)\} \perp A|C     \hspace{3.85cm}    (A3)\\
  \\
  & & Y(a,m_1,m_2) \perp \{M_1(a^\ast),M_2(a^\ast,m_1)\}|C.     \hspace{1.55cm}    (A4)\\
\end{eqnarray*}

Similarly, the assumptions above state that, given a covariate set $C$ or $\{A,C\}$, there exists no unmeasured variables confounding the association between exposure $A$ and outcome $Y$ ($A1$), there exists no unmeasured variables confounding the association between mediator set $\{M_1,M_2\}$ and outcome $Y$ ($A2$), there exists no unmeasured variables confounding the association between exposure $A$ and mediator set $\{M_1,M_2\}$ ($A3$) and there exists no unmeasured variables that are causal children of exposure $A$, and in the meantime, are confounding the association between mediator $M$ and outcome $Y$ \cite{s, vcon}. 

In order to account for the confounding between $M_1$ and $M_2$, two more assumptions are required other than ($A1$) -- ($A4$): 
\begin{eqnarray*}
  & & M_2(a,m_1) \perp M_1|\{A,C\}     \hspace{2.2cm}    (A5)\\  
  \\
  & & M_2(a,m_1) \perp M_1(a^\ast)|C,     \hspace{2.2cm}    (A6)
\end{eqnarray*}
where ($A5$) and ($A6$) state, respectively, that there exists no unmeasured variables confounding the association between $M_1$ and $M_2$ given $\{A,C\}$, and there exists no unmeasured variables that are causal descendants of exposure $A$, and in the meantime, are confounding the association between $M_1$ and $M_2$ \cite{s}. 

Steen et al \cite{s} presented very comprehensive identification conditions for the causal structures with multiple mediators in a sequential order including the special one-path situation. Steen et al \cite{s} also pointed out that weaker identification assumptions than ($A1$) -- ($A6$) can be considered under certain decompositions, which does not violate the findings of VanderWeele and Vansteelandt \cite{vmul}. We do not offer a further discussion here on this topic since it is not the focus of this article. 

\section{Empirical formulas}
At individual level, each component of total effect generally cannot be estimated; however, if a particular population is considered, we can obtain the expected value of each component as long as certain identification assumptions about confounding are satisfied \cite{v3}. Here we present the empirical formulas without covariates for the decomposition of total effect in a structure with two sequential mediators as shown in Figure 8, where $M_1$ and $M_2$ are categorical random variables:
\begin{eqnarray*}
  E\left[CDE(m_1^\ast,m_2^\ast)\right]& = & p_{am_1^\ast m_2^\ast}-p_{a^\ast m_1^\ast m_2^\ast}\\
  \\
  E[INT_{ref\mbox{-}AM_1}(m_1^\ast,m_2^\ast)] & = & \sum_{m_1}(p_{am_1 m_2^\ast}-p_{a m_1^\ast m_2^\ast}-p_{a^\ast m_1 m_2^\ast}+p_{a^\ast m_1^\ast m_2^\ast})\\
  & & \times Pr(M_1=m_1|A=a^\ast)\\
  \\
  E[INT_{ref\mbox{-}AM_2+AM_1M_2}(m_2^\ast)] & = & \sum_{m_2}\sum_{m_1}(p_{am_1 m_2}-p_{a m_1 m_2^\ast}-p_{a^\ast m_1 m_2}+p_{a^\ast m_1 m_2^\ast})\\
  & & \times Pr(M_1=m_1|A=a^\ast)\\
  & & \times Pr(M_2=m_2|A=a^\ast,M_1=m_1)\\
  \\
  E[NatINT_{AM_1}] & = & \sum_{m_2}\sum_{m_1}(p_{am_1 m_2}-p_{a^\ast m_1 m_2})\\
  & & \times Pr(M_2=m_2|A=a^\ast,M_1=m_1)\\
  & & \times [Pr(M_1=m_1|A=a)-Pr(M_1=m_1|A=a^\ast)]\\
  \\
  E[NatINT_{AM_2}] & = & \sum_{m_2}\sum_{m_1}(p_{am_1 m_2}-p_{a^\ast m_1 m_2})\\
  & & \times Pr(M_1=m_1|A=a^\ast)\\
  & & \times [Pr(M_2=m_2|A=a,M_1=m_1)-Pr(M_2=m_2|A=a^\ast,M_1=m_1)]\\
  \\
  E[NatINT_{AM_1M_2}] & = & \sum_{m_2}\sum_{m_1}(p_{am_1 m_2}-p_{a^\ast m_1 m_2})\\
  & & \times [Pr(M_1=m_1|A=a)-Pr(M_1=m_1|A=a^\ast)]\\
  & & \times [Pr(M_2=m_2|A=a,M_1=m_1)-Pr(M_2=m_2|A=a^\ast,M_1=m_1)]\\
  \\
  E[NatINT_{M_1M_2}] & = & \sum_{m_2}\sum_{m_1}p_{a^\ast m_1 m_2}\\
  & & \times [Pr(M_1=m_1|A=a)-Pr(M_1=m_1|A=a^\ast)]\\
  & & \times [Pr(M_2=m_2|A=a,M_1=m_1)-Pr(M_2=m_2|A=a^\ast,M_1=m_1)]\\
  \\
  E[PIE_{M_1}] & = & \sum_{m_2}\sum_{m_1}p_{a^\ast m_1 m_2}\\
  & & \times Pr(M_2=m_2|A=a^\ast,M_1=m_1)\\
  & & \times [Pr(M_1=m_1|A=a)-Pr(M_1=m_1|A=a^\ast)]\\
  \\
  E[PIE_{M_2}] & = & \sum_{m_2}\sum_{m_1}p_{a^\ast m_1 m_2}\\
  & & \times Pr(M_1=m_1|A=a^\ast)\\
  & & \times [Pr(M_2=m_2|A=a,M_1=m_1)-Pr(M_2=m_2|A=a^\ast,M_1=m_1)],
\end{eqnarray*}
where $p_{am_1m_2}=E[Y|A=a,M_1=m_1,M_2=m_2]$.

Researchers can obtain the estimated average for each component by using the formulas above and the observed data. Nonetheless, causal interpretations cannot be drawn without corresponding identification assumptions on confounding \cite{v4}.

\section{Illustration with real data}
In order to illustrate the notion of natural counterfactual interaction effect, we used the data from a population based study, which focused on the hazard of drinking alcohol as a contribution to the abnormal pattern in mortality \cite{l}, where exposure $A$ is alcohol drinking, mediator $M_1$ is Body Mass Index (BMI), mediator $M_2$ is the log-transformed Gamma Glutamyl Transferase (GGT), outcome $Y$ is Systolic Blood Pressure (SBP), and two confounders are Sex and Age, respectively. The corresponding causal diagram is shown in Figure 11. We use the 2015-2016 data from the National Health and Nutrition Examination Survey  downloaded at {\ttfamily http://www.cdc.gov/nhanes} to illustrate the proposed approach. 

As the exposure is a binary variable, the total effect was obtained by using the contrast $Y(1)-Y(0)$. Log transformation was performed for $M_2$ due to the skewness of the data. The fixed reference levels of $M_1$ and $\log(M_2)$ were chosen at the mean levels. Namely, $m_1^\ast=29.5$ and ${\log{(m_2)}}^\ast=3.05$. The results are conditional on either male or female, and the mean level of Age at $48.3$. Three linear models were fit for $Y$, $\log(M_2)$ and $M_1$. The 95\% confidence intervals were obtained by using bootstrapping \cite{v}. The formula derivations are shown in Appendix F.

Table 2 presents the decomposition of total effect conditional on male and mean level of Age. The controlled direct effect is $0.238$ $(-0.969,1.429)$; the reference interaction effect between $A$ and $M_1$ is $-0.059$ $(-0.203,0.039)$; the sum of two reference interaction effect is $-0.115$ $(-0.516,0.219)$; the natural counterfactual interaction effect between $A$ and $M_1$ is $-0.018$ $(-0.125,0.056)$; the natural counterfactual interaction effect between $A$ and $\log(M_2)$ is $-0.026$ $(-0.194,0.095)$; the natural counterfactual interaction effect among $A$, $M_1$ and $\log(M_2)$ is $0.000386$ $(-0.0059,0.0082)$; the natural counterfactual interaction effect between $M_1$ and $\log(M_2)$ is $0.000873$ $(-0.0094,0.0123)$; the pure direct effect is $0.0636$ $(-1.226,1.317)$; the pure indirect effect through $M_1$ is $-0.0409$ $(-0.206,0.109)$; the pure indirect effect through $\log(M_2)$ is $0.143$ $(0.00803,0.363)$; the total effect is $0.123$ $(-1.178,1.396)$. The results of decomposition of total effect conditional on female and mean level of Age are shown in Table 3. It can be seen that the pure indirect effect through $\log(M_2)$ is the only considerable effect contributing to the outcome for both females and males. The main purpose of the application is to illustrate the method with real data. We do not excessively interpret or concern the significant findings. 

\section{Conclusion}
We developed the concept of natural counterfactual interaction effect, which allows mediators to naturally vary in compliance with exposure. By using this concept, we further presented methods for decomposition of total effect in different directed acyclic graphs.

In a single-mediator situation, mediated interaction effect is a special case of natural counterfactual interaction effect where the two effects are mathematically equivalent. Both effects have sound interpretations but with different perspectives. The divergence starts to appear in the non-sequential multiple-mediator scenario. Mediated interaction effect requires a fixed reference level for each mediator while natural counterfactual interaction effect does not have such a requirement and allows all mediators to naturally vary along with different values of exposure. This crucial difference between the previously developed mediated interaction effect and our proposed natural counterfactual interaction effect lies in that the former is partially controlled and the latter is completely natural. The property of partially controlled effect renders mediated interaction effect inappropriate and non-identifiable in a general directed acyclic graph involving sequential mediators. The above conclusion can be made because the two limitations arise in a causal diagram with two sequential mediators, the simplest causal structure of this type. The inappropriate modeling of dependence will still remain in a more sophisticated situation that incorporates the simplest causal structure as a subgraph. Moreover, the non-identifiability problem also will occur since any additional edges would not improve identifiability \cite{a}. 

Another interesting finding reveals that natural counterfactual interaction effect could be recognized as the parallel notion to natural path effect in the sense that the counterfactual formulas used to establish the two types of effects do not control any mediator at a certain level. The extension may offer researchers new insights into theory developments and applications on causal inference. 

\clearpage
\begin{table}[!h]
\centering
\caption{A comparison between the classical interaction effects in a linear model and natural counterfactual interaction effects in a non-sequential two-mediator scenario}
\begin{tabular}{lll}
\hline
Factors        & Interaction Effects    & Natural Counterfactual \\
               & in Linear Model        & Interaction Effects    \\
\hline
$AM_1$      & $p_{110}-p_{010}-p_{100}+p_{000}$      & $Y\left(a,M_1(a),M_2(a^\ast)\right)-Y\left(a^\ast,M_1(a),M_2(a^\ast)\right)$\\
            &                                        & $- Y\left(a,M_1(a^\ast),M_2(a^\ast)\right)+Y\left(a^\ast,M_1(a^\ast),M_2(a^\ast)\right)$\\
            &                                        &\\
$AM_2$       & $p_{101}-p_{100}-p_{001}+p_{000}$       & $Y\left(a,M_1(a^\ast),M_2(a)\right)-Y\left(a,M_1(a^\ast),M_2(a^\ast)\right)$\\
             &                                         & $- Y\left(a^\ast,M_1(a^\ast),M_2(a)\right)+Y\left(a^\ast,M_1(a^\ast),M_2(a^\ast)\right)$\\
             &                                         &\\
$M_1M_2$      & $p_{011}-p_{010}-p_{001}+p_{000}$      & $Y\left(a^\ast,M_1(a),M_2(a)\right)-Y\left(a^\ast,M_1(a),M_2(a^\ast)\right)$\\
              &                                        &$-Y\left(a^\ast,M_1(a^\ast),M_2(a)\right)+Y\left(a^\ast,M_1(a^\ast),M_2(a^\ast)\right)$\\
              &                                        &\\
$AM_1M_2$      & $p_{111}-p_{110}-p_{101}-p_{011}$      & $Y\left(a,M_1(a),M_2(a)\right)-Y\left(a,M_1(a),M_2(a^\ast)\right)$\\
               &$+p_{001}+p_{010}+p_{100}-p_{000}$      & $- Y\left(a,M_1(a^\ast),M_2(a)\right)-Y\left(a^\ast,M_1(a),M_2(a)\right)$\\
               &                                        & $+Y\left(a^\ast,M_1(a^\ast),M_2(a)\right)+Y\left(a^\ast,M_1(a),M_2(a^\ast)\right)$\\
               &                                        &$+Y\left(a,M_1(a^\ast),M_2(a^\ast)\right)-Y\left(a^\ast,M_1(a^\ast),M_2(a^\ast)\right)$\\
\hline
\end{tabular}
\end{table}

\clearpage
\begin{table}[!h]
\centering
\caption{Illustration with Real Data: Decomposition of Total Effect Conditional on Male and Mean Age}
\begin{tabular}{lll}
\hline
Component        & Estimate    & 95\% C.I. \\
\hline
$ CDE(m_1^\ast,\log(m_2)^\ast)$      & $0.238$      & $-0.969, 1.429$\\
$INT_{ref\mbox{-}AM_1}(m_1^\ast,\log(m_2)^\ast)$       & $-0.059$    &$ -0.203, 0.039$\\
$INT_{ref\mbox{-}A\log(M_2)+AM_1\log(M_2)}(\log(m_2)^\ast)$  & $-0.115$ & $-0.516,0.219$ \\
$NatINT_{AM_1}$      & $-0.018$      & $-0.125,0.056$\\
$NatINT_{A\log(M_2)}$      & $-0.026$      & $-0.194,0.095$\\
$NatINT_{AM_1\log(M_2)}$      & $0.000386$      & $-0.0059,0.0082$\\
$NatINT_{M_1\log(M_2)}$      & $0.000873$      & $-0.0094,0.0123$\\
$PDE$      & $0.0636$      & $-1.226,1.317$\\
$PIE_{M_1}$      & $-0.0409$      & $-0.206,0.109$\\
$PIE_{\log(M_2)}$      & $0.143$      & $0.00803,0.363$\\
$TE$      & $0.123$      & $-1.178,1.396$\\
\hline
\end{tabular}
\end{table}

\clearpage
\begin{table}[!h]
\centering
\caption{Illustration with Real Data: Decomposition of Total Effect Conditional on Female and Mean Age}
\begin{tabular}{lll}
\hline
Component        & Estimate    & 95\% C.I. \\
\hline
$ CDE(m_1^\ast,\log(m_2)^\ast)$      & $0.238$      & $-0.969, 1.429$\\
$INT_{ref\mbox{-}AM_1}(m_1^\ast,\log(m_2)^\ast)$       & $0.087$    &$ -0.0359, 0.263$\\
$INT_{ref\mbox{-}A\log(M_2)+AM_1\log(M_2)}(\log(m_2)^\ast)$  & $0.0658$ & $-0.395,0.533$ \\
$NatINT_{AM_1}$      & $-0.0207$      & $-0.135,0.060$\\
$NatINT_{A\log(M_2)}$      & $-0.0286$      & $-0.206,0.0896$\\
$NatINT_{AM_1\log(M_2)}$      & $0.000377$      & $-0.00586,0.00863$\\
$NatINT_{M_1\log(M_2)}$      & $0.000860$      & $-0.00936,0.0117$\\
$PDE$      & $0.391$      & $-0.828,1.581$\\
$PIE_{M_1}$      & $-0.0448$      & $-0.219,0.114$\\
$PIE_{\log(M_2)}$      & $0.137$      & $0.00752,0.353$\\
$TE$      & $0.435$      & $-0.788,1.629$\\
\hline
\end{tabular}
\end{table}

\clearpage
\begin{figure}[!h]
\centering
\scalebox{0.5}[0.5]{\includegraphics{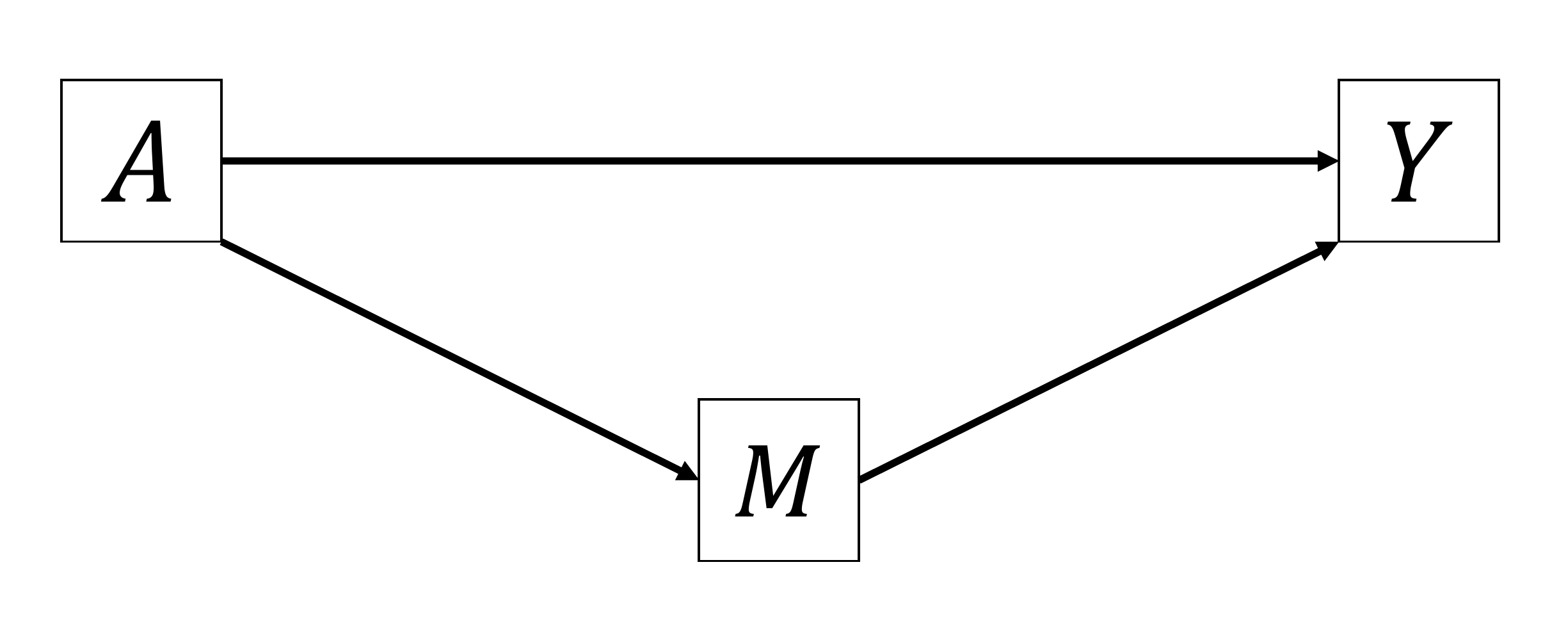}}
\caption{A directed acyclic graph of a single-mediator scenario.}
\label{fig1}
\end{figure}

\clearpage
\begin{figure}[!h]
\centering
\scalebox{0.5}[0.5]{\includegraphics{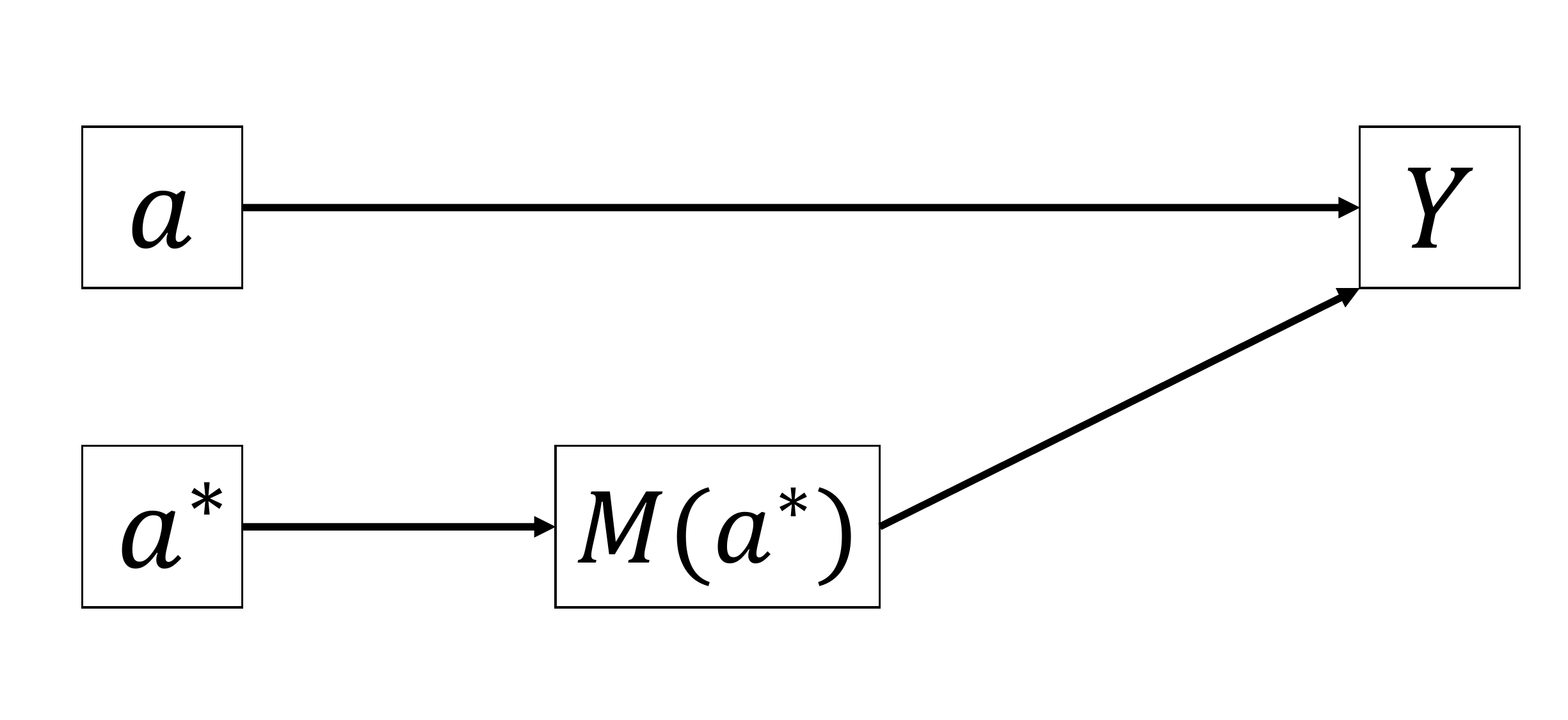}}
\caption{The diagram illustrates the nested counterfactual formula $Y(a, M_1(a^\ast))$.}
\label{fig2}
\end{figure}

\clearpage
\begin{figure}[!h]
\centering
\scalebox{0.44}[0.5]{\includegraphics{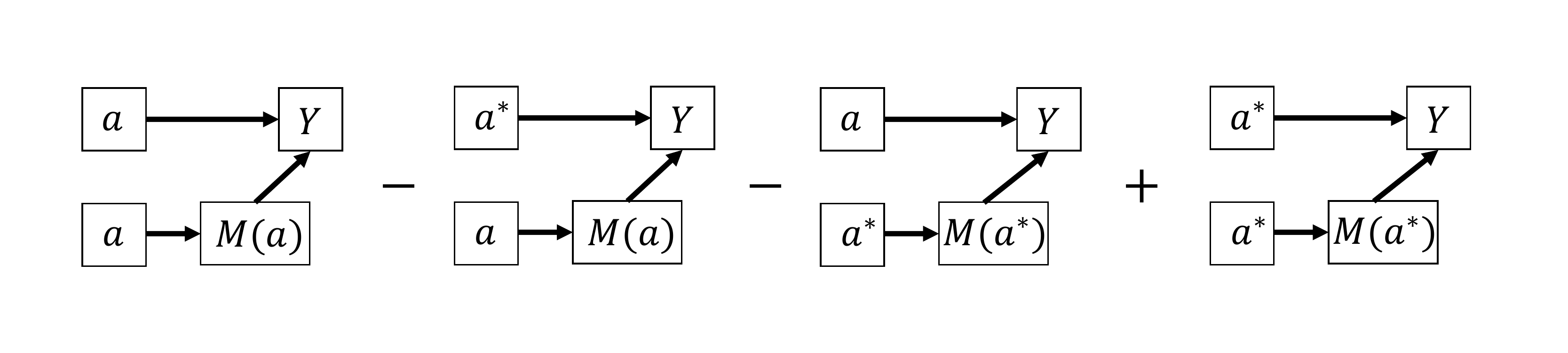}}
\caption{The nature of natural counterfactual interaction effect or mediated interaction effect
in a single-mediator scenario satisfies the classical definition of additive interaction effect with
corresponding counterfactual formulas of $Y$ with fixed treatment level of $A$ at $a$ and fixed
reference level of $A$ at $a^\ast$ as well as with natural treatment level of $M$ at $M(a)$ and natural reference level of $M$ at $M(a^\ast)$.}
\label{fig3}
\end{figure}

\clearpage
\begin{figure}[!h]
\centering
\scalebox{0.5}[0.5]{\includegraphics{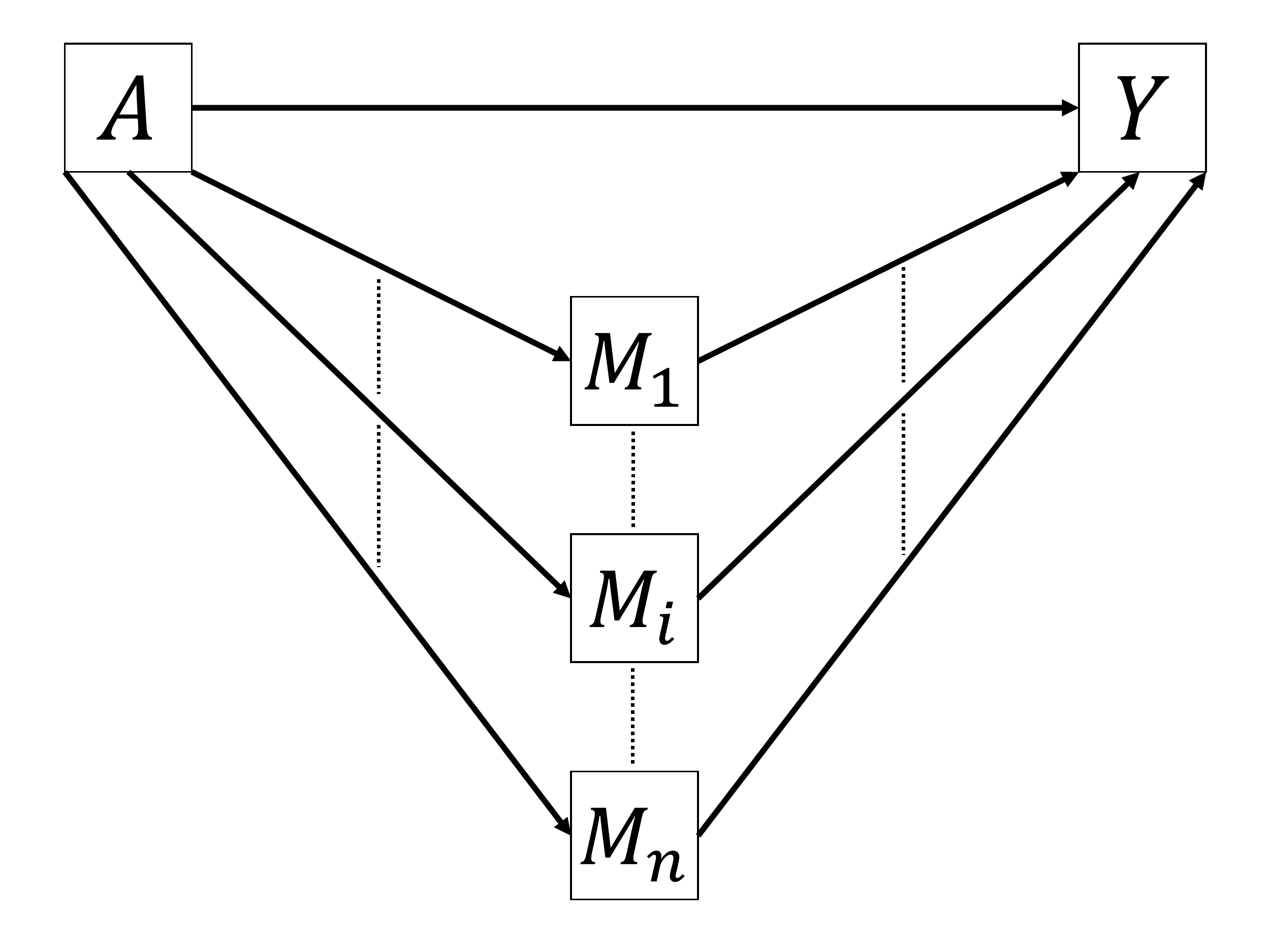}}
\caption{The directed acyclic graph of a non-sequential multiple-mediator scenario.}
\label{fig4}
\end{figure}

\clearpage
\begin{figure}[!h]
\centering
\scalebox{0.5}[0.5]{\includegraphics{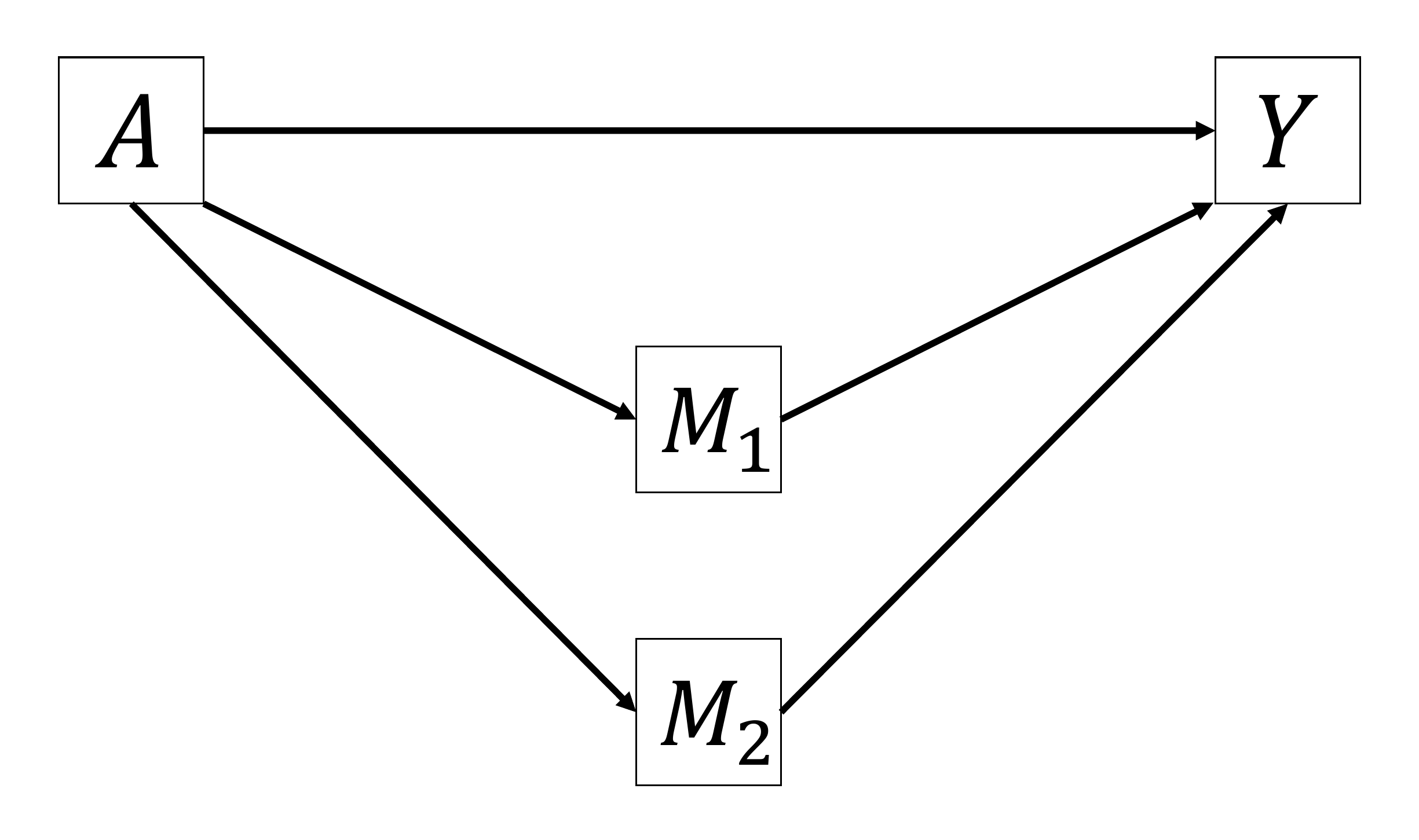}}
\caption{The directed acyclic graph with two mediators in no sequential order.}
\label{fig5}
\end{figure}

\clearpage
\begin{figure}[!h]
\centering
\scalebox{0.5}[0.5]{\includegraphics{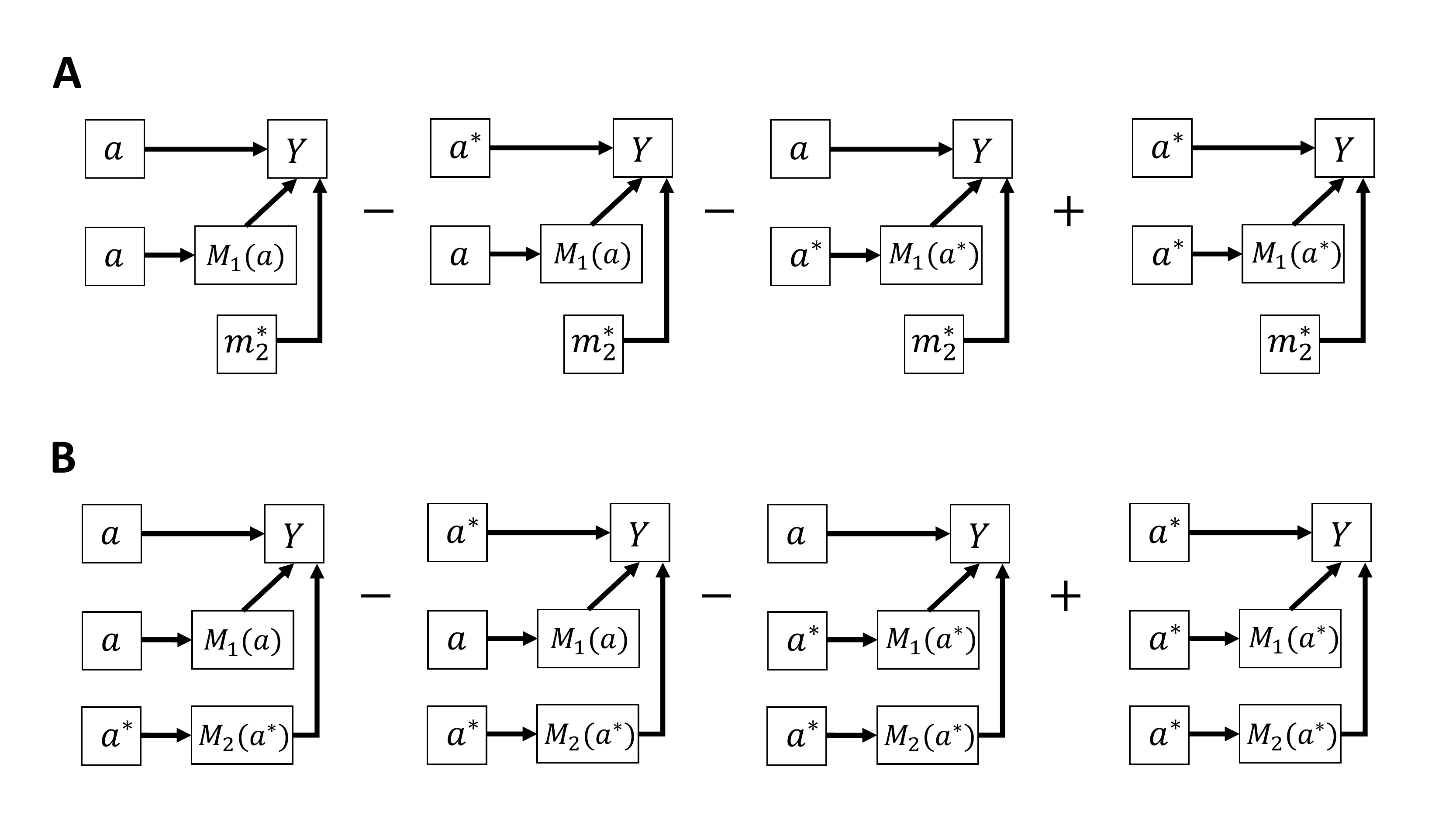}}
\caption{A comparison between mediated interaction effect and natural counterfactual interaction effect in a non-sequential two-mediator scenario. Figure 6A illustrates the mediated interaction effect between $A$ and $M_1$, where $M_2$ is assigned a fixed value at $m_2^\ast$ as well as $M_1$ naturally varies with exposure $A$. Figure 6B illustrates the natural counterfactual interaction effect between $A$ and $M_1$, where both $M_1$ and $M_2$ naturally vary with exposure $A$.}
\label{fig6}
\end{figure}

\clearpage
\begin{figure}[!h]
\centering
\scalebox{0.5}[0.5]{\includegraphics{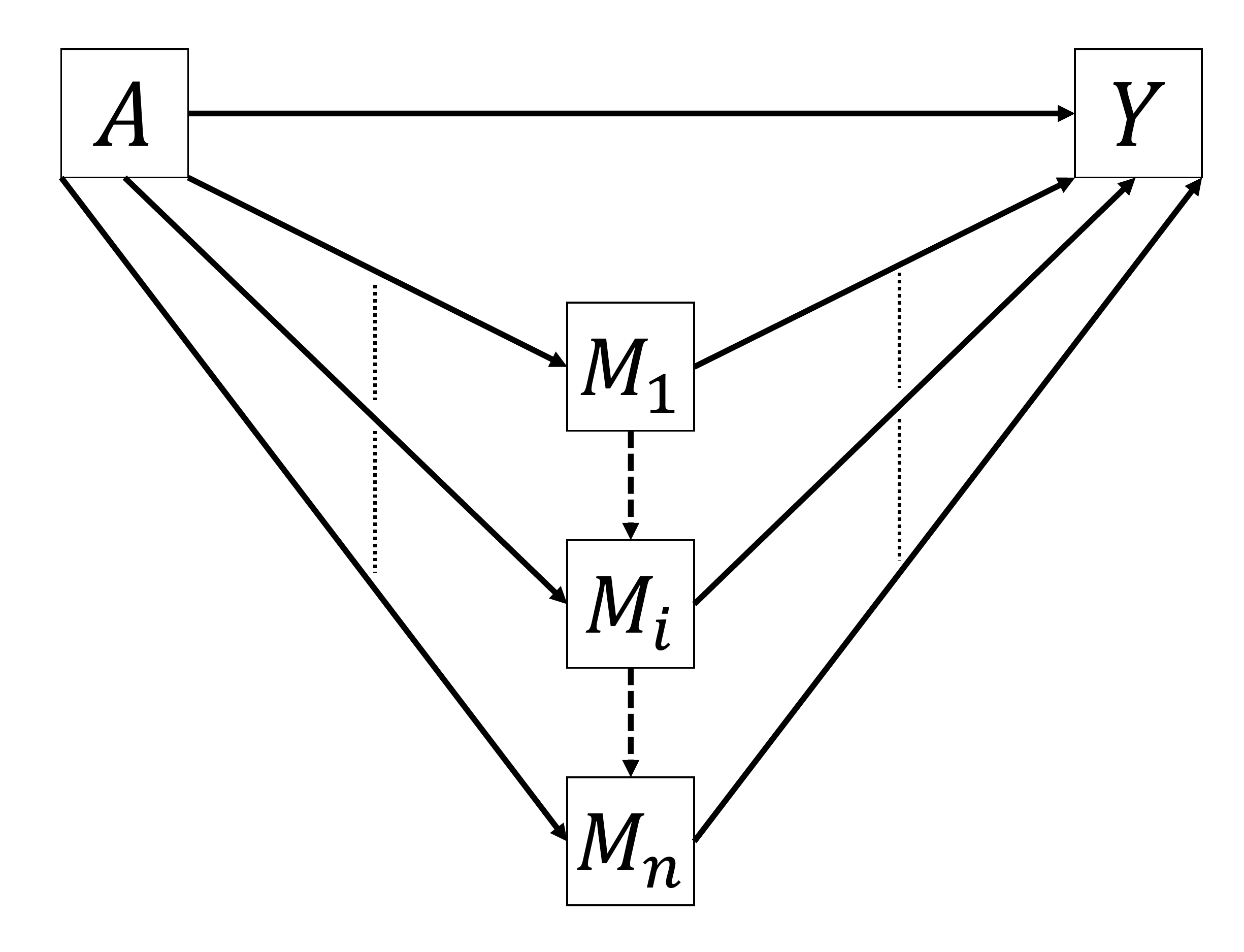}}
\caption{The directed acyclic graph of a one-path sequential multiple-mediator scenario.}
\label{fig7}
\end{figure}

\clearpage
\begin{figure}[!h]
\centering
\scalebox{0.5}[0.5]{\includegraphics{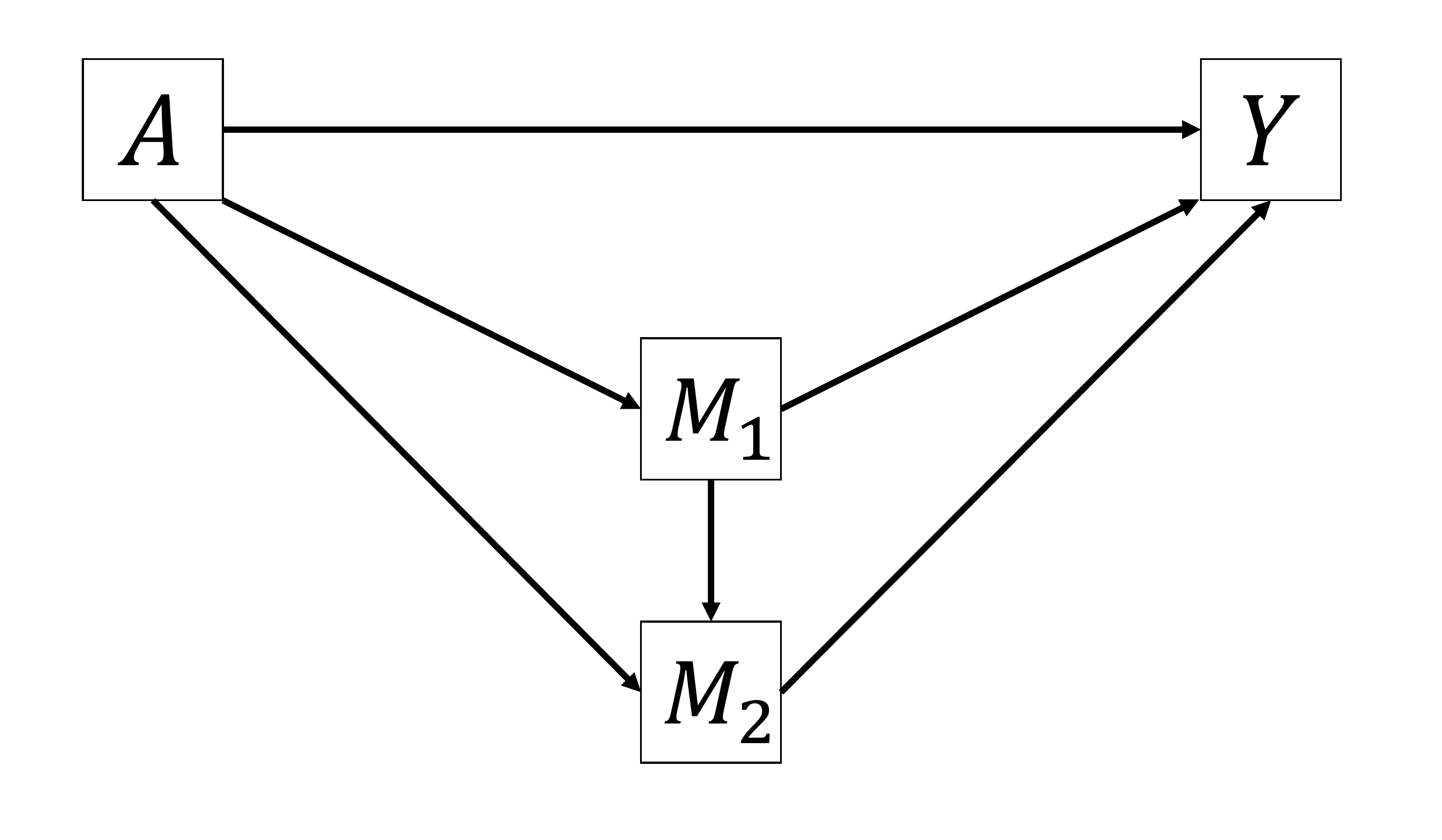}}
\caption{A directed acyclic graph of a sequential two-mediator scenario.}
\label{fig8}
\end{figure}

\clearpage
\begin{figure}[!h]
\centering
\scalebox{0.5}[0.5]{\includegraphics{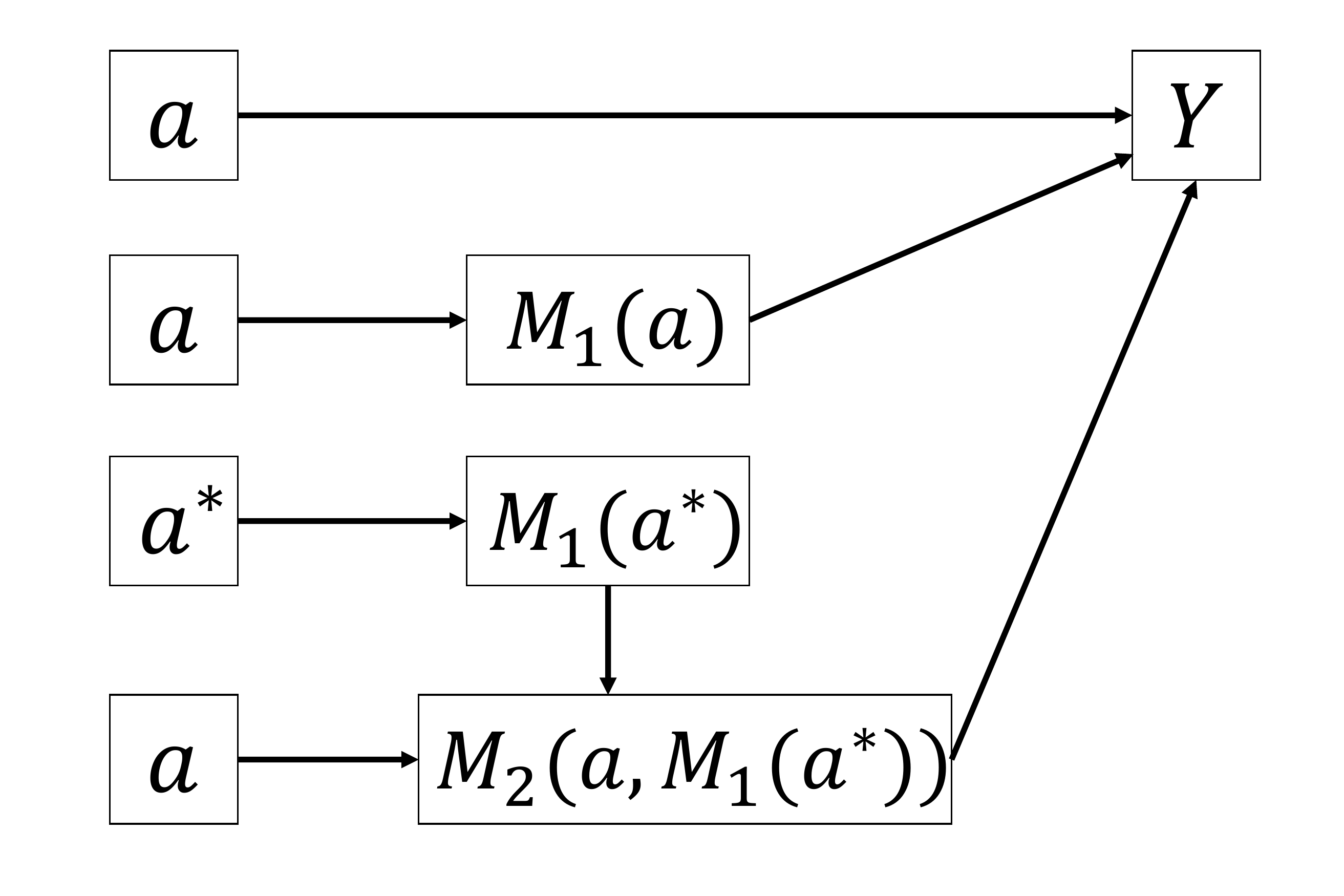}}
\caption{The graph illustrates the counterfactual formula $Y\left(a,M_1(a),M_2(a,M_1(a^\ast))\right)$. This type of counterfactual formula is non-identifiable since $M_1$ is being activated by $a$ and $a^\ast$ in the mean time, where $a \neq a^\ast$.}
\label{fig9}
\end{figure}

\clearpage
\begin{figure}[!h]
\centering
\scalebox{0.5}[0.5]{\includegraphics{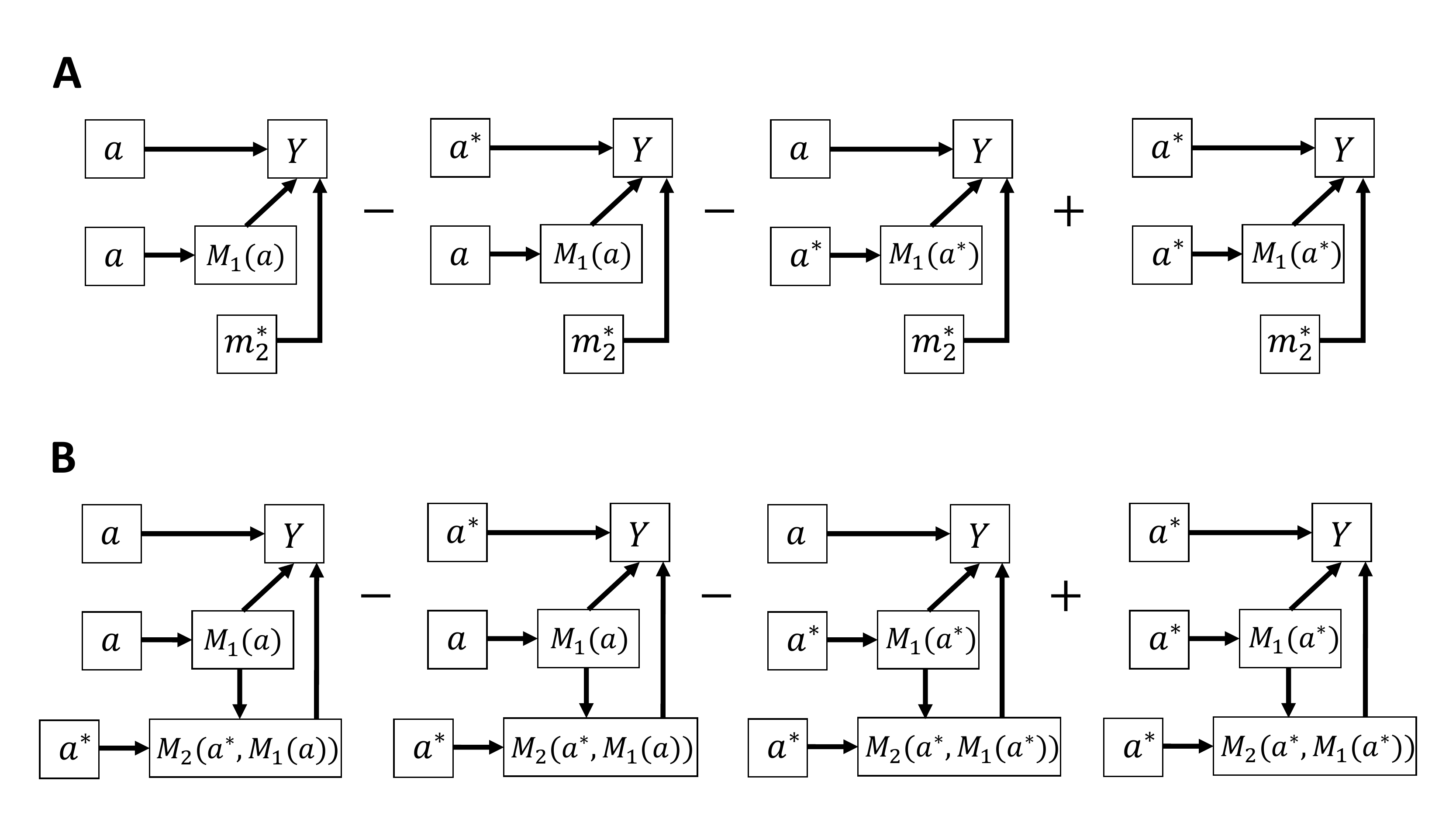}}
\caption{A comparison between mediated interaction effect and natural counterfactual interaction effect in a sequential two-mediator scenario. Figure 10A illustrates the mediated interaction effect between $A$ and $M_1$, which is tantamount to Figure 6A and therefore results in losing the important feature, dependence of $M_2$ on $M_1$. Figure 10B illustrates the natural counterfactual interaction effect between $A$ and $M_1$ which well accounts for the dependence.}
\label{fig10}
\end{figure}

\clearpage
\begin{figure}[!h]
\centering
\scalebox{0.5}[0.5]{\includegraphics{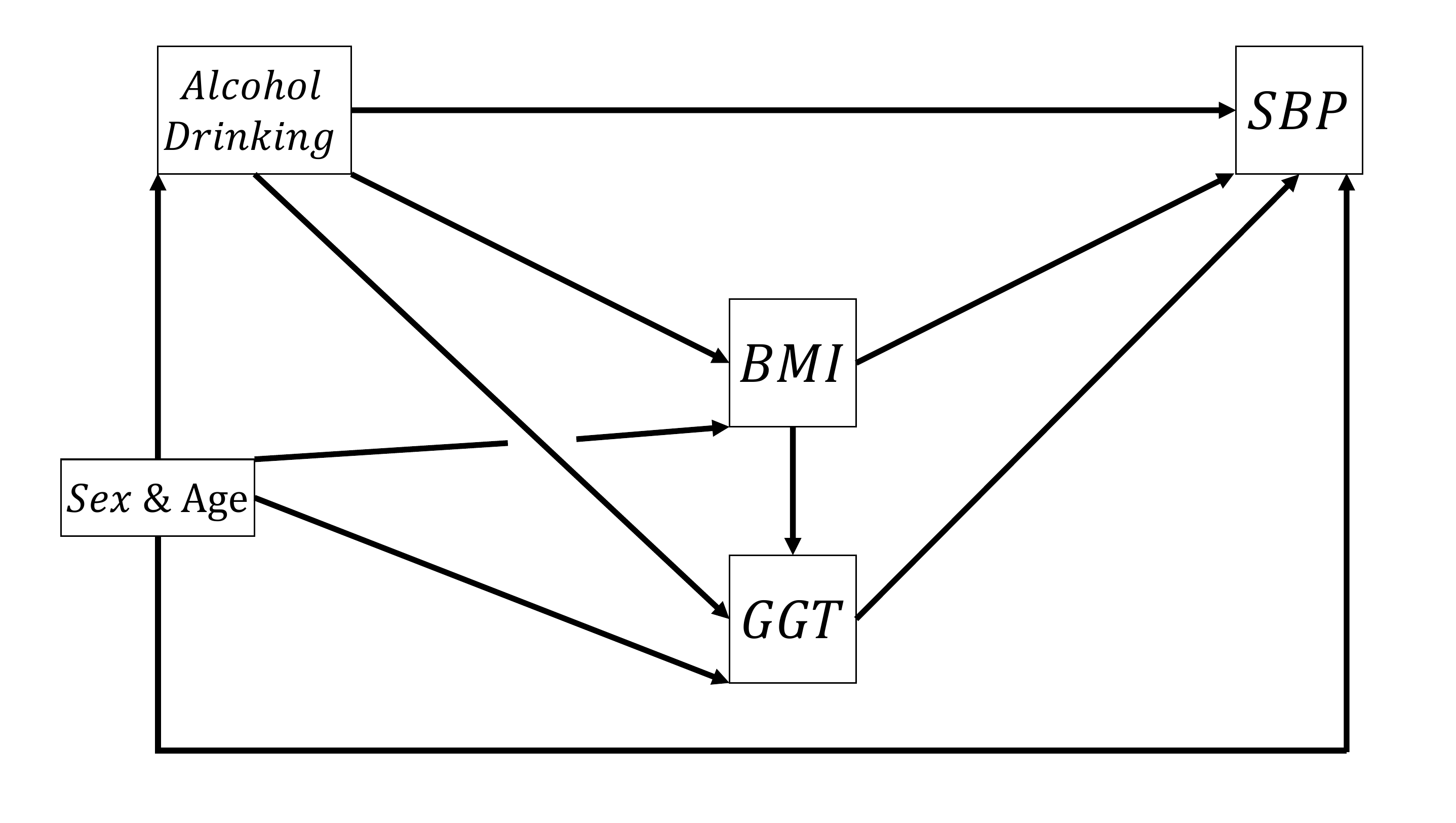}}
\caption{The directed acyclic graph for population based study, which focused on the hazard of drinking alcohol as a contribution to the abnormal pattern in mortality.}
\label{fig11}
\end{figure}

\clearpage
\begin{figure}[!h]
\centering
\scalebox{0.5}[0.5]{\includegraphics{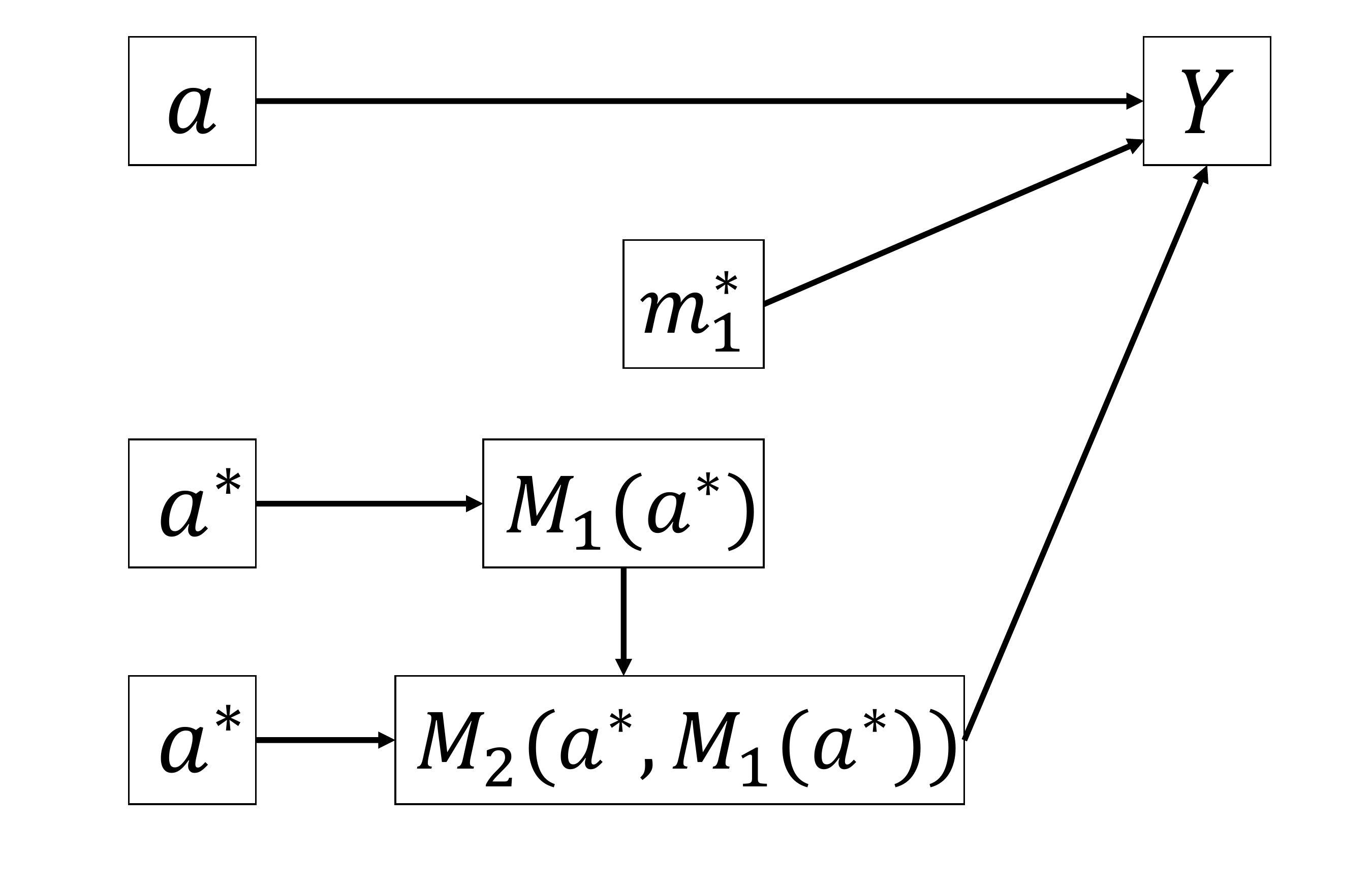}}
\caption{The graph illustrates the counterfactual formula $Y(a,m_1^\ast,M_2(a^\ast,M_1(a^\ast)))$. This type of counterfactual formula is non-identifiable. The proof is shown in Appendix D.}
\label{fig12}
\end{figure}
\clearpage
\section*{Acknowledgments}
This research was partially supported by UNM Comprehensive Cancer Center Support Grant NCI P30CA118100, the Biostatistics shared resource and UNM METALS Superfund Research Center (1P42ES025589).

\clearpage

\clearpage
\section*{Appendix A. Decomposition of total effect with the notion of natural counterfactual interaction effect in a non-sequential two-mediator scenario}

Suppose we have a directed acyclic graph as shown in Figure 5. We need to show that the total effect can be decomposed into the following 10 components at individual level:
\begin{eqnarray*}
  TE & = & CDE(m_1^\ast,m_2^\ast)+INT_{ref\mbox{-}AM_1}(m_1^\ast,m_2^\ast)+INT_{ref\mbox{-}AM_2}(m_1^\ast,m_2^\ast)\\
  & & +INT_{ref\mbox{-}AM_1M_2}(m_1^\ast,m_2^\ast)+ NatINT_{AM_1} + NatINT_{AM_2}+ NatINT_{AM_1M_2}\\
  & & + NatINT_{M_1M_2} + PIE_{M_1} + PIE_{M_2},
\end{eqnarray*}
where the natural counterfactual interaction effects satisfy Definition 2.\\

\noindent
\emph{Proof}: 

We first decompose the total effect into total direct effect ($TDE$) \cite{riden}, seminatural indirect effect through $M_1$ ($SIE_{M_1}$) \cite{p14} and pure indirect effect (path-specific effect) through $M_2$ ($PIE_{M_2}$) \cite{riden,p01}.
\begin{eqnarray*}
  TE & = & Y(a)-Y(a^\ast)\\
     \\
     & = & Y(a,M_1(a),M_2(a))-Y(a^\ast,M_1(a^\ast),M_2(a^\ast))\\
     \\
     & = & Y(a,M_1(a),M_2(a))-Y(a^\ast,M_1(a),M_2(a))\\
     & & +Y(a^\ast,M_1(a),M_2(a))-Y(a^\ast,M_1(a^\ast),M_2(a))\\
     & & +Y(a^\ast,M_1(a^\ast),M_2(a))-Y(a^\ast,M_1(a^\ast),M_2(a^\ast)),
\end{eqnarray*}
where the second equality follows by the composition axiom \cite{vbook,a} and the third equality follows by adding and subtracting the same counterfactual formulas. 

The formulas of $TDE$, $SIE_{M_1}$ and $PIE_{M_2}$ are presented as follows: 
\begin{eqnarray*}
  TDE & = & Y(a,M_1(a),M_2(a))-Y(a^\ast,M_1(a),M_2(a))\\
     \\
  SIE_{M_1} & = & Y(a^\ast,M_1(a),M_2(a))-Y(a^\ast,M_1(a^\ast),M_2(a))\\
     \\
  PIE_{M_2} & = & Y(a^\ast,M_1(a^\ast),M_2(a))-Y(a^\ast,M_1(a^\ast),M_2(a^\ast)),
\end{eqnarray*}
where $TE = TDE+SIE_{M_1}+PIE_{M_2}$.

We focus on $TDE$ for next step and try to decompose it into natural counterfactual interaction effects and pure direct effect ($PDE$) \cite{riden,p01} by subtracting $PDE$ from $TDE$, where $PDE$ satisfies the definition of path-specific effect \cite{p01} and equals the following contrast of two counterfactual formulas:
\begin{eqnarray*}
  PDE & = & Y(a,M_1(a^\ast),M_2(a^\ast))-Y(a^\ast,M_1(a^\ast),M_2(a^\ast)).
\end{eqnarray*}

We have the following results:
\begin{eqnarray*}
  TDE-PDE & = & Y(a,M_1(a),M_2(a))-Y(a^\ast,M_1(a),M_2(a))\\
  & & -Y(a,M_1(a^\ast),M_2(a^\ast))+Y(a^\ast,M_1(a^\ast),M_2(a^\ast))\\
  \\
  & = & Y(a,M_1(a),M_2(a))-Y(a^\ast,M_1(a),M_2(a))\\
  & & -Y(a,M_1(a^\ast),M_2(a^\ast))+Y(a^\ast,M_1(a^\ast),M_2(a^\ast))\\
  & & +Y(a^\ast,M_1(a^\ast),M_2(a^\ast))-Y(a^\ast,M_1(a^\ast),M_2(a^\ast))\\
  & & +Y(a^\ast,M_1(a^\ast),M_2(a))-Y(a^\ast,M_1(a^\ast),M_2(a))\\
  & & +Y(a^\ast,M_1(a),M_2(a^\ast))-Y(a^\ast,M_1(a),M_2(a^\ast))\\
  & & +Y(a,M_1(a^\ast),M_2(a^\ast))-Y(a,M_1(a^\ast),M_2(a^\ast))\\
  & & +Y(a,M_1(a^\ast),M_2(a))-Y(a,M_1(a^\ast),M_2(a))\\
  & & +Y(a,M_1(a),M_2(a^\ast))-Y(a,M_1(a),M_2(a^\ast))\\
  \\
  & = & Y(a,M_1(a),M_2(a^\ast))-Y(a^\ast,M_1(a),M_2(a^\ast))\\
  & & - Y(a,M_1(a^\ast),M_2(a^\ast))+Y(a^\ast,M_1(a^\ast),M_2(a^\ast))\\
  & & + Y(a,M_1(a^\ast),M_2(a))-Y(a,M_1(a^\ast),M_2(a^\ast))\\
  & & - Y(a^\ast,M_1(a^\ast),M_2(a))+Y(a^\ast,M_1(a^\ast),M_2(a^\ast))\\
  & & + Y(a,M_1(a),M_2(a))-Y(a,M_1(a),M_2(a^\ast))\\
  & & - Y(a,M_1(a^\ast),M_2(a))-Y(a^\ast,M_1(a),M_2(a))\\
  & & +Y(a^\ast,M_1(a^\ast),M_2(a))+Y(a^\ast,M_1(a),M_2(a^\ast))\\
  & & +Y(a,M_1(a^\ast),M_2(a^\ast))-Y(a^\ast,M_1(a^\ast),M_2(a^\ast)),
\end{eqnarray*}
where the second equality follows by adding and subtracting the same counterfactual formulas, and the third equality follows by rearranging all the terms to satisfy the definition of counterfactual interaction effects. 

Therefore, we have the following formulas satisfying Definition 2:
\begin{eqnarray*}
  NatINT_{AM_1} & = & Y(a,M_1(a),M_2(a^\ast))-Y(a^\ast,M_1(a),M_2(a^\ast))\\
  & & - Y(a,M_1(a^\ast),M_2(a^\ast))+Y(a^\ast,M_1(a^\ast),M_2(a^\ast))\\
  \\
  NatINT_{AM_2} & = & Y(a,M_1(a^\ast),M_2(a))-Y(a,M_1(a^\ast),M_2(a^\ast))\\
  & & - Y(a^\ast,M_1(a^\ast),M_2(a))+Y(a^\ast,M_1(a^\ast),M_2(a^\ast))\\
  \\
  NatINT_{AM_1M_2} & = & Y(a,M_1(a),M_2(a))-Y(a,M_1(a),M_2(a^\ast))\\
  & & - Y(a,M_1(a^\ast),M_2(a))-Y(a^\ast,M_1(a),M_2(a))\\
  & & +Y(a^\ast,M_1(a^\ast),M_2(a))+Y(a^\ast,M_1(a),M_2(a^\ast))\\
  & & +Y(a,M_1(a^\ast),M_2(a^\ast))-Y(a^\ast,M_1(a^\ast),M_2(a^\ast)).
\end{eqnarray*}

Accordingly, $TDE$ can be decomposed into the following components:
\begin{eqnarray*}
TDE &=& PDE + NatINT_{AM_1}+NatINT_{AM_2}+NatINT_{AM_1M_2}.
\end{eqnarray*}

We next focus on $PDE$ (path-specific effect) and try to decompose it into $CDE$ and reference interaction effects \cite{v4, b}:
\begin{eqnarray*}
PDE & = & Y(a,M_1(a^\ast),M_2(a^\ast))-Y(a^\ast,M_1(a^\ast),M_2(a^\ast))\\
\\
    & = & \sum_{m_2}\sum_{m_1}Y(a,m_1,m_2)\times I(M_1(a^\ast)=m_1)\times I(M_2(a^\ast)=m_2)\\
    & & - \sum_{m_2}\sum_{m_1}Y(a^\ast,m_1,m_2)\times I(M_1(a^\ast)=m_1)\times I(M_2(a^\ast)=m_2)\\
    \\
    & = & \sum_{m_2}\sum_{m_1}[Y(a,m_1,m_2)-Y(a^\ast,m_1,m_2)]\times I(M_1(a^\ast)=m_1)\times I(M_2(a^\ast)=m_2)\\
    \\
    & = & \sum_{m_2}\sum_{m_1}[Y(a,m_1,m_2)-Y(a^\ast,m_1,m_2)-Y(a,m_1^\ast,m_2^\ast)+Y(a^\ast,m_1^\ast,m_2^\ast)]\\
    & & \times I(M_1(a^\ast)=m_1)\times I(M_2(a^\ast)=m_2)\\
    & & + \sum_{m_2}\sum_{m_1}[Y(a,m_1^\ast,m_2^\ast)-Y(a^\ast,m_1^\ast,m_2^\ast)]\times I(M_1(a^\ast)=m_1)\times I(M_2(a^\ast)=m_2)\\
    \\
    & = & \sum_{m_2}\sum_{m_1}[Y(a,m_1,m_2)-Y(a^\ast,m_1,m_2)-Y(a,m_1^\ast,m_2^\ast)+Y(a^\ast,m_1^\ast,m_2^\ast)]\\
    & & \times I(M_1(a^\ast)=m_1)\times I(M_2(a^\ast)=m_2)\\
    & & + Y(a,m_1^\ast,m_2^\ast)-Y(a^\ast,m_1^\ast,m_2^\ast)\\
    \\
    & = & \sum_{m_2}\sum_{m_1}[Y(a,m_1,m_2)-Y(a^\ast,m_1,m_2)-Y(a,m_1^\ast,m_2^\ast)+Y(a^\ast,m_1^\ast,m_2^\ast)\\
    & & +Y(a^\ast,m_1^\ast,m_2^\ast)-Y(a^\ast,m_1^\ast,m_2^\ast)+Y(a^\ast,m_1^\ast,m_2)-Y(a^\ast,m_1^\ast,m_2)\\
    & & +Y(a^\ast,m_1,m_2^\ast)-Y(a^\ast,m_1,m_2^\ast)+Y(a,m_1^\ast,m_2^\ast)-Y(a,m_1^\ast,m_2^\ast)\\
    & & +Y(a,m_1^\ast,m_2)-Y(a,m_1^\ast,m_2)+Y(a,m_1,m_2^\ast)-Y(a,m_1,m_2^\ast)]\\
    & & \times I(M_1(a^\ast)=m_1)\times I(M_2(a^\ast)=m_2)\\
    & & + Y(a,m_1^\ast,m_2^\ast)-Y(a^\ast,m_1^\ast,m_2^\ast)\\
    \\
    & = & \sum_{m_2}\sum_{m_1}[Y(a,m_1,m_2^\ast)-Y(a,m_1^\ast,m_2^\ast)-Y(a^\ast,m_1,m_2^\ast)+Y(a^\ast,m_1^\ast,m_2^\ast)]\\
    & & \times I(M_1(a^\ast)=m_1)\times I(M_2(a^\ast)=m_2)\\
    & & + \sum_{m_2}\sum_{m_1}[Y(a,m_1^\ast,m_2)-Y(a,m_1^\ast,m_2^\ast)-Y(a^\ast,m_1^\ast,m_2)+Y(a^\ast,m_1^\ast,m_2^\ast)]\\
    & & \times I(M_1(a^\ast)=m_1)\times I(M_2(a^\ast)=m_2)\\
    & & + \sum_{m_2}\sum_{m_1}[Y(a,m_1,m_2)-Y(a,m_1,m_2^\ast)-Y(a,m_1^\ast,m_2)-Y(a^\ast,m_1,m_2)\\
    & & +Y(a^\ast,m_1^\ast,m_2)+Y(a^\ast,m_1,m_2^\ast)+Y(a,m_1^\ast,m_2^\ast)-Y(a^\ast,m_1^\ast,m_2^\ast)]\\
    & & \times I(M_1(a^\ast)=m_1)\times I(M_2(a^\ast)=m_2)\\
    & & + Y(a,m_1^\ast,m_2^\ast)-Y(a^\ast,m_1^\ast,m_2^\ast)\\
    \\
    & = & \sum_{m_1}[Y(a,m_1,m_2^\ast)-Y(a,m_1^\ast,m_2^\ast)-Y(a^\ast,m_1,m_2^\ast)+Y(a^\ast,m_1^\ast,m_2^\ast)]\\
    & & \times I(M_1(a^\ast)=m_1)\\
    & & + \sum_{m_2}[Y(a,m_1^\ast,m_2)-Y(a,m_1^\ast,m_2^\ast)-Y(a^\ast,m_1^\ast,m_2)+Y(a^\ast,m_1^\ast,m_2^\ast)]\\
    & & \times I(M_2(a^\ast)=m_2)\\
    & & + \sum_{m_2}\sum_{m_1}[Y(a,m_1,m_2)-Y(a,m_1,m_2^\ast)-Y(a,m_1^\ast,m_2)-Y(a^\ast,m_1,m_2)\\
    & & +Y(a^\ast,m_1^\ast,m_2)+Y(a^\ast,m_1,m_2^\ast)+Y(a,m_1^\ast,m_2^\ast)-Y(a^\ast,m_1^\ast,m_2^\ast)]\\
    & & \times I(M_1(a^\ast)=m_1)\times I(M_2(a^\ast)=m_2)\\
    & & + Y(a,m_1^\ast,m_2^\ast)-Y(a^\ast,m_1^\ast,m_2^\ast).
\end{eqnarray*}

According to the derivation above, the following formulas can be obtained:
\begin{eqnarray*}
CDE(m_1^\ast,m_2^\ast) & = & Y(a,m_1^\ast,m_2^\ast)-Y(a^\ast,m_1^\ast,m_2^\ast)\\
\\
INT_{ref\mbox{-}AM_1}(m_1^\ast,m_2^\ast) & = & \sum_{m_1}[Y(a,m_1,m_2^\ast)-Y(a,m_1^\ast,m_2^\ast)-Y(a^\ast,m_1,m_2^\ast)+Y(a^\ast,m_1^\ast,m_2^\ast)]\\
& & \times I(M_1(a^\ast)=m_1)\\
\\
INT_{ref\mbox{-}AM_2}(m_1^\ast,m_2^\ast) & = & \sum_{m_2}[Y(a,m_1^\ast,m_2)-Y(a,m_1^\ast,m_2^\ast)-Y(a^\ast,m_1^\ast,m_2)+Y(a^\ast,m_1^\ast,m_2^\ast)]\\
& & \times I(M_2(a^\ast)=m_2)\\
\\
INT_{ref\mbox{-}AM_1M_2}(m_1^\ast,m_2^\ast) & = & \sum_{m_2}\sum_{m_1}[Y(a,m_1,m_2)-Y(a,m_1,m_2^\ast)-Y(a,m_1^\ast,m_2)-Y(a^\ast,m_1,m_2)\\
    & & +Y(a^\ast,m_1^\ast,m_2)+Y(a^\ast,m_1,m_2^\ast)+Y(a,m_1^\ast,m_2^\ast)-Y(a^\ast,m_1^\ast,m_2^\ast)]\\
    & & \times I(M_1(a^\ast)=m_1)\times I(M_2(a^\ast)=m_2).
\end{eqnarray*}

Therefore, $PDE$ can be decomposed into the following components:
\begin{eqnarray*}
PDE & = & CDE(m_1^\ast,m_2^\ast)+INT_{ref\mbox{-}AM_1}(m_1^\ast,m_2^\ast)+INT_{ref\mbox{-}AM_2}(m_1^\ast,m_2^\ast)+INT_{ref\mbox{-}AM_1M_2}(m_1^\ast,m_2^\ast).
\end{eqnarray*}

$TDE$ can be decomposed into the following components:
\begin{eqnarray*}
TDE & = & PDE+NatINT_{AM_1}+NatINT_{AM_2}+NatINT_{AM_1M_2}\\
\\
& = & CDE(m_1^\ast,m_2^\ast)+INT_{ref\mbox{-}AM_1}(m_1^\ast,m_2^\ast)+INT_{ref\mbox{-}AM_2}(m_1^\ast,m_2^\ast)+INT_{ref\mbox{-}AM_1M_2}(m_1^\ast,m_2^\ast)\\
& & +NatINT_{AM_1}+NatINT_{AM_2}+NatINT_{AM_1M_2}.
\end{eqnarray*}

We next focus on $SIE_{M_1}$ and try to decompose it into $PIE_{M_1}$ and $NatINT_{M_1M_2}$ by subtracting $PIE_{M_1}$ from $SIE_{M_1}$:
\begin{eqnarray*}
SIE_{M_1}-PIE_{M_1} & = & Y(a^\ast,M_1(a),M_2(a))-Y(a^\ast,M_1(a),M_2(a^\ast))\\
& & -Y(a^\ast,M_1(a^\ast),M_2(a))+Y(a^\ast,M_1(a^\ast),M_2(a^\ast))\\
\\
& = & NatINT_{M_1M_2},
\end{eqnarray*}
where $NatINT_{M_1M_2}$ satisfies Definition 2.

Therefore, $SIE_{M_1}$ can be decomposed into the following components:
\begin{eqnarray*}
SIE_{M_1} & = & PIE_{M_1}+ NatINT_{M_1M_2}.
\end{eqnarray*}

Combining all the derivations above, we have the decomposition of total effect as follows:
\begin{eqnarray*}
  TE & = & CDE(m_1^\ast,m_2^\ast)+INT_{ref\mbox{-}AM_1}(m_1^\ast,m_2^\ast)+INT_{ref\mbox{-}AM_2}(m_1^\ast,m_2^\ast)\\
  & & +INT_{ref\mbox{-}AM_1M_2}(m_1^\ast,m_2^\ast)+ NatINT_{AM_1} + NatINT_{AM_2}+ NatINT_{AM_1M_2}\\
  & & + NatINT_{M_1M_2} + PIE_{M_1} + PIE_{M_2}.\qquad\qquad\qquad\qquad\qquad\qquad\qquad\qquad\square
\end{eqnarray*}

\clearpage
\section*{Appendix B. The nature of mediated interaction effect between $A$ and $M_1$ in a non-sequential two-mediator scenario}

Suppose we have a directed acyclic graph as shown in Figure 5, we need to show that the mediated interaction effect between $A$ and $M_1$ proposed by Bellavia and Valeri \cite{b} is equivalent to assigning $M_2$ a fixed reference level at $m_2^\ast$ and allowing $M_1$ to naturally change with $A$ as illustrated in Figure 6A. \\

\noindent
\emph{Proof}: 

First we need to find the sum of of $INT_{med\mbox{-}AM_1}$, $INT_{med\mbox{-}AM_2}$ and $INT_{med\mbox{-}AM_1M_2}$. From Appendix A we know that:
\begin{eqnarray*}
  TDE-PDE & = & Y(a,M_1(a),M_2(a))-Y(a^\ast,M_1(a),M_2(a))\\
  & & -Y(a,M_1(a^\ast),M_2(a^\ast))+Y(a^\ast,M_1(a^\ast),M_2(a^\ast))\\
  \\
  & = & \sum_{m_2}\sum_{m_1}Y(a,m_1,m_2)\times I(M_1(a)=m_1)\times I(M_2(a)=m_2)\\
  & & - \sum_{m_2}\sum_{m_1}Y(a^\ast,m_1,m_2)\times I(M_1(a)=m_1)\times I(M_2(a)=m_2)\\
  & & - \sum_{m_2}\sum_{m_1}Y(a,m_1,m_2)\times I(M_1(a^\ast)=m_1)\times I(M_2(a^\ast)=m_2)\\
  & & + \sum_{m_2}\sum_{m_1}Y(a^\ast,m_1,m_2)\times I(M_1(a^\ast)=m_1)\times I(M_2(a^\ast)=m_2)\\
  \\
  & = & \sum_{m_2}\sum_{m_1}[Y(a,m_1,m_2)-Y(a^\ast,m_1,m_2)]\\
  & & \times [I(M_1(a)=m_1)\times I(M_2(a)=m_2)-I(M_1(a^\ast)=m_1)\times I(M_2(a^\ast)=m_2)]\\
  \\
  & = & \sum_{m_2}\sum_{m_1}[Y(a,m_1,m_2)-Y(a^\ast,m_1,m_2)-Y(a,m_1^\ast,m_2^\ast)+Y(a^\ast,m_1^\ast,m_2^\ast)]\\
  & & \times [I(M_1(a)=m_1)\times I(M_2(a)=m_2)-I(M_1(a^\ast)=m_1)\times I(M_2(a^\ast)=m_2)]\\
  \\
  & = & \sum_{m_2}\sum_{m_1}[Y(a,m_1,m_2)-Y(a^\ast,m_1,m_2)-Y(a,m_1^\ast,m_2^\ast)+Y(a^\ast,m_1^\ast,m_2^\ast)\\
  & & +Y(a^\ast,m_1^\ast,m_2^\ast)-Y(a^\ast,m_1^\ast,m_2^\ast)+Y(a^\ast,m_1^\ast,m_2)-Y(a^\ast,m_1^\ast,m_2)\\
  & & +Y(a^\ast,m_1,m_2^\ast)-Y(a^\ast,m_1,m_2^\ast)+Y(a,m_1^\ast,m_2^\ast)-Y(a,m_1^\ast,m_2^\ast)\\
  & & +Y(a,m_1^\ast,m_2)-Y(a,m_1^\ast,m_2)+Y(a,m_1,m_2^\ast)-Y(a,m_1,m_2^\ast)]\\
  & & \times [I(M_1(a)=m_1)\times I(M_2(a)=m_2)-I(M_1(a^\ast)=m_1)\times I(M_2(a^\ast)=m_2)]\\
  \\
& = & \sum_{m_2}\sum_{m_1}[Y(a,m_1,m_2^\ast)-Y(a,m_1^\ast,m_2^\ast)-Y(a^\ast,m_1,m_2^\ast)+Y(a^\ast,m_1^\ast,m_2^\ast)]\\
& & \times [I(M_1(a)=m_1)\times I(M_2(a)=m_2)-I(M_1(a^\ast)=m_1)\times I(M_2(a^\ast)=m_2)]\\
& & + \sum_{m_2}\sum_{m_1}[Y(a,m_1^\ast,m_2)-Y(a,m_1^\ast,m_2^\ast)-Y(a^\ast,m_1^\ast,m_2)+Y(a^\ast,m_1^\ast,m_2^\ast)]\\
& & \times [I(M_1(a)=m_1)\times I(M_2(a)=m_2)-I(M_1(a^\ast)=m_1)\times I(M_2(a^\ast)=m_2)]\\
& & + \sum_{m_2}\sum_{m_1}[Y(a,m_1,m_2)-Y(a,m_1,m_2^\ast)-Y(a,m_1^\ast,m_2)-Y(a^\ast,m_1,m_2)\\
& & + Y(a^\ast,m_1^\ast,m_2)+Y(a^\ast,m_1,m_2^\ast)+Y(a,m_1^\ast,m_2^\ast)-Y(a^\ast,m_1^\ast,m_2^\ast)]\\
& & \times [I(M_1(a)=m_1)\times I(M_2(a)=m_2)-I(M_1(a^\ast)=m_1)\times I(M_2(a^\ast)=m_2)],
\end{eqnarray*}
where the fourth equality follows by the fact that $Y(a,m_1^\ast,m_2^\ast)$ and $Y(a^\ast,m_1^\ast,m_2^\ast)$ are constants and can be canceled out through the double summation, and the fifth equality follows by adding and subtracting the same counterfactual formulas.

The mediated interaction effect between $A$ and $M_1$ can be obtained as:
\begin{eqnarray*}
 INT_{med\mbox{-}AM_1}(m_2^\ast) & = & \sum_{m_2}\sum_{m_1}[Y(a,m_1,m_2^\ast)-Y(a,m_1^\ast,m_2^\ast)-Y(a^\ast,m_1,m_2^\ast)+Y(a^\ast,m_1^\ast,m_2^\ast)]\\
& & \times [I(M_1(a)=m_1)\times I(M_2(a)=m_2)-I(M_1(a^\ast)=m_1)\times I(M_2(a^\ast)=m_2)]\\
\\
& = & \sum_{m_1}[Y(a,m_1,m_2^\ast)-Y(a,m_1^\ast,m_2^\ast)-Y(a^\ast,m_1,m_2^\ast)+Y(a^\ast,m_1^\ast,m_2^\ast)]\\
& & \times [I(M_1(a)=m_1)-I(M_1(a^\ast)=m_1)]\\
\\
& = & \sum_{m_1}[Y(a,m_1,m_2^\ast)-Y(a^\ast,m_1,m_2^\ast)]\times [I(M_1(a)=m_1)-I(M_1(a^\ast)=m_1)]\\
\\
& = & \sum_{m_1}Y(a,m_1,m_2^\ast)\times I(M_1(a)=m_1)-\sum_{m_1}Y(a,m_1,m_2^\ast)\times I(M_1(a^\ast)=m_1)\\
& & - \sum_{m_1}Y(a^\ast,m_1,m_2^\ast)\times I(M_1(a)=m_1)+\sum_{m_1}Y(a^\ast,m_1,m_2^\ast)\times I(M_1(a^\ast)=m_1)\\
\\
& = & Y(a,M_1(a),m_2^\ast)-Y(a,M_1(a^\ast),m_2^\ast)-Y(a^\ast,M_1(a),m_2^\ast)+Y(a^\ast,M_1(a^\ast),m_2^\ast)\\
\\
& = & Y(a,M_1(a),m_2^\ast)-Y(a^\ast,M_1(a),m_2^\ast)-Y(a,M_1(a^\ast),m_2^\ast)+Y(a^\ast,M_1(a^\ast),m_2^\ast),
\end{eqnarray*}
where the second equality follows by the fact that $m_2^\ast$ is a constant so the summation indexed by $m_2$ can be dropped, and the third equality follows by the fact that $Y(a,m_1^\ast,m_2^\ast)$ and $Y(a^\ast,m_1^\ast,m_2^\ast)$ are constants and can be canceled out through the summation. 

The last equality indicates that the mediated interaction effect between $A$ and $M_1$ is equivalent to assigning a fixed reference level $m_2^\ast$ to $M_2$ and allowing $M_1$ to naturally change with the exposure $A$ as shown in Figure 6A.$\qquad\qquad\qquad\qquad\square$

\clearpage
\section*{Appendix C. Decomposition of total effect with the notion of natural counterfactual interaction effect in a sequential two-mediator scenario}

Suppose we have a directed acyclic graph as shown in Figure 8. We need to show that the total effect can be decomposed into the following 9 components at individual level:
\begin{eqnarray*}
  TE & = & CDE(m_1^\ast,m_2^\ast)+INT_{ref\mbox{-}AM_1}(m_1^\ast,m_2^\ast)+INT_{ref\mbox{-}AM_2+AM_1M_2}(m_2^\ast)\\
  & & + NatINT_{AM_1} + NatINT_{AM_2}+ NatINT_{AM_1M_2}+ NatINT_{M_1M_2}\\
  & & + PIE_{M_1} + PIE_{M_2},
\end{eqnarray*}
where all the natural counterfactual interaction effects satisfy Definition 3. \\

\noindent
\emph{Proof}: 

The proof is similar to Appendix A. We first decompose the total effect into total direct effect ($TDE$) \cite{riden}, seminatural indirect effect through $M_1$ ($SIE_{M_1}$) \cite{p14} and pure indirect effect (path-specific effect) through $M_2$ ($PIE_{M_2}$) \cite{riden,p01}.
\begin{eqnarray*}
  TE & = & Y(a)-Y(a^\ast)\\
  \\
  & = & Y(a,M_1(a),M_2(a,M_1(a)))-Y(a^\ast,M_1(a^\ast),M_2(a^\ast,M_1(a^\ast)))\\
  \\
  & = & Y(a,M_1(a),M_2(a,M_1(a)))-Y(a^\ast,M_1(a),M_2(a,M_1(a)))\\
  & & + Y(a^\ast,M_1(a),M_2(a,M_1(a))) - Y(a^\ast,M_1(a^\ast),M_2(a,M_1(a^\ast)))\\
  & & + Y(a^\ast,M_1(a^\ast),M_2(a,M_1(a^\ast)))-Y(a^\ast,M_1(a^\ast),M_2(a^\ast,M_1(a^\ast))),
\end{eqnarray*}
where the second equality follows by the composition axiom \cite{vbook,a} and the third equality follows by adding and subtracting the same identifiable counterfactual formulas. 

The formulas of $TDE$, $SIE_{M_1}$ and $PIE_{M_2}$ are presented below:
\begin{eqnarray*}
  TDE & = & Y(a,M_1(a),M_2(a,M_1(a)))-Y(a^\ast,M_1(a),M_2(a,M_1(a)))\\
  \\
  SIE_{M_1} & = & Y(a^\ast,M_1(a),M_2(a,M_1(a))) - Y(a^\ast,M_1(a^\ast),M_2(a,M_1(a^\ast)))\\
  \\
  PIE_{M_2} & = & Y(a^\ast,M_1(a^\ast),M_2(a,M_1(a^\ast)))-Y(a^\ast,M_1(a^\ast),M_2(a^\ast,M_1(a^\ast))),
\end{eqnarray*}
where $TE = TDE + SIE_{M_1}+PIE_{M_2}$.

We next focus on $TDE$ and decompose it into natural counterfactual interaction effects and pure direct effect ($PDE$) \cite{riden,p01} by subtracting $PDE$ from $TDE$, where $PDE$ satisfies the definition of path-specific effect \cite{p01} and equals the following difference of two identifiable counterfactual formulas:
\begin{eqnarray*}
  PDE & = & Y(a,M_1(a^\ast),M_2(a^\ast,M_1(a^\ast)))-Y(a^\ast,M_1(a^\ast),M_2(a^\ast,M_1(a^\ast))).
\end{eqnarray*}

We have the following results:
\begin{eqnarray*}
  TDE-PDE & = & Y(a,M_1(a),M_2(a,M_1(a)))-Y(a^\ast,M_1(a),M_2(a,M_1(a)))\\
  & & -Y(a,M_1(a^\ast),M_2(a^\ast,M_1(a^\ast)))+Y(a^\ast,M_1(a^\ast),M_2(a^\ast,M_1(a^\ast)))\\
  \\
  & = & Y(a,M_1(a),M_2(a,M_1(a)))-Y(a^\ast,M_1(a),M_2(a,M_1(a)))\\
  & & -Y(a,M_1(a^\ast),M_2(a^\ast,M_1(a^\ast)))+Y(a^\ast,M_1(a^\ast),M_2(a^\ast,M_1(a^\ast)))\\
  & & +Y(a^\ast,M_1(a^\ast),M_2(a^\ast,M_1(a^\ast)))-Y(a^\ast,M_1(a^\ast),M_2(a^\ast,M_1(a^\ast)))\\
  & & +Y(a^\ast,M_1(a^\ast),M_2(a,M_1(a^\ast)))-Y(a^\ast,M_1(a^\ast),M_2(a,M_1(a^\ast)))\\
  & & +Y(a^\ast,M_1(a),M_2(a^\ast,M_1(a)))-Y(a^\ast,M_1(a),M_2(a^\ast,M_1(a)))\\
  & & +Y(a,M_1(a^\ast),M_2(a^\ast,M_1(a^\ast)))-Y(a,M_1(a^\ast),M_2(a^\ast,M_1(a^\ast)))\\
  & & +Y(a,M_1(a^\ast),M_2(a,M_1(a^\ast)))-Y(a,M_1(a^\ast),M_2(a,M_1(a^\ast)))\\
  & & +Y(a,M_1(a),M_2(a^\ast,M_1(a)))-Y(a,M_1(a),M_2(a^\ast,M_1(a)))\\
  \\
  & = & Y(a,M_1(a),M_2(a^\ast,M_1(a)))-Y(a^\ast,M_1(a),M_2(a^\ast,M_1(a)))\\
  & & -Y(a,M_1(a^\ast),M_2(a^\ast,M_1(a^\ast)))+Y(a^\ast,M_1(a^\ast),M_2(a^\ast,M_1(a^\ast)))\\
  & & +Y(a,M_1(a^\ast),M_2(a,M_1(a^\ast)))-Y(a,M_1(a^\ast),M_2(a^\ast,M_1(a^\ast)))\\
  & & -Y(a^\ast,M_1(a^\ast),M_2(a,M_1(a^\ast)))+Y(a^\ast,M_1(a^\ast),M_2(a^\ast,M_1(a^\ast)))\\
  & & +Y(a,M_1(a),M_2(a,M_1(a)))-Y(a,M_1(a),M_2(a^\ast,M_1(a)))\\
  & & -Y(a,M_1(a^\ast),M_2(a,M_1(a^\ast)))-Y(a^\ast,M_1(a),M_2(a,M_1(a)))\\
  & & +Y(a^\ast,M_1(a^\ast),M_2(a,M_1(a^\ast)))+Y(a^\ast,M_1(a),M_2(a^\ast,M_1(a)))\\
  & & +Y(a,M_1(a^\ast),M_2(a^\ast,M_1(a^\ast)))-Y(a^\ast,M_1(a^\ast),M_2(a^\ast,M_1(a^\ast)))
\end{eqnarray*}
where the second equality follows by adding and subtracting the same identifiable counterfactual formulas and the third equality follows by rearranging all the terms to satisfy the definition of natural counterfactual interaction effects. 

Therefore, we have the following formulas satisfying Definition 3:
\begin{eqnarray*}
  NatINT_{AM_1} & = & Y(a,M_1(a),M_2(a^\ast,M_1(a)))-Y(a^\ast,M_1(a),M_2(a^\ast,M_1(a)))\\
  & & -Y(a,M_1(a^\ast),M_2(a^\ast,M_1(a^\ast)))+Y(a^\ast,M_1(a^\ast),M_2(a^\ast,M_1(a^\ast)))\\
  \\
  NatINT_{AM_2} & = & Y(a,M_1(a^\ast),M_2(a,M_1(a^\ast)))-Y(a,M_1(a^\ast),M_2(a^\ast,M_1(a^\ast)))\\
  & & -Y(a^\ast,M_1(a^\ast),M_2(a,M_1(a^\ast)))+Y(a^\ast,M_1(a^\ast),M_2(a^\ast,M_1(a^\ast)))\\
  \\
  NatINT_{AM_1M_2} & = & Y(a,M_1(a),M_2(a,M_1(a)))-Y(a,M_1(a),M_2(a^\ast,M_1(a)))\\
  & & -Y(a,M_1(a^\ast),M_2(a,M_1(a^\ast)))-Y(a^\ast,M_1(a),M_2(a,M_1(a)))\\
  & & +Y(a^\ast,M_1(a^\ast),M_2(a,M_1(a^\ast)))+Y(a^\ast,M_1(a),M_2(a^\ast,M_1(a)))\\
  & & +Y(a,M_1(a^\ast),M_2(a^\ast,M_1(a^\ast)))-Y(a^\ast,M_1(a^\ast),M_2(a^\ast,M_1(a^\ast))).
\end{eqnarray*}

Accordingly, $TDE$ can be decomposed into the following components:
\begin{eqnarray*}
  TDE = PDE + NatINT_{AM_1} + NatINT_{AM_2} + NatINT_{AM_1M_2}.
\end{eqnarray*}

We next focus on $PDE$ (path-specific effect) and decompose it into $CDE$ and reference interaction effects \cite{b,v4}:

Accordingly, $TDE$ can be decomposed into the following components:
\begin{eqnarray*}
  PDE & = & Y(a,M_1(a^\ast),M_2(a^\ast,M_1(a^\ast)))-Y(a^\ast,M_1(a^\ast),M_2(a^\ast,M_1(a^\ast)))\\
  \\
  & = & \sum_{m_2}\sum_{m_1} Y(a,m_1,m_2)\times I(M_1(a^\ast)=m_1)\times I(M_2(a^\ast,m_1)=m_2)\\
  & & -\sum_{m_2}\sum_{m_1} Y(a^\ast,m_1,m_2)\times I(M_1(a^\ast)=m_1)\times I(M_2(a^\ast,m_1)=m_2)\\
  \\
  & = &\sum_{m_2}\sum_{m_1}[Y(a,m_1,m_2)-Y(a^\ast,m_1,m_2)]\times I(M_1(a^\ast)=m_1)\times I(M_2(a^\ast,m_1)=m_2)\\
  \\
  & = &\sum_{m_2}\sum_{m_1}[Y(a,m_1,m_2)-Y(a^\ast,m_1,m_2)-Y(a,m_1^\ast,m_2^\ast)+Y(a^\ast,m_1^\ast,m_2^\ast)]\\
  & & \times I(M_1(a^\ast)=m_1)\times I(M_2(a^\ast,m_1)=m_2)\\
  & & +\sum_{m_2}\sum_{m_1}[Y(a,m_1^\ast,m_2^\ast)-Y(a^\ast,m_1^\ast,m_2^\ast)]\times I(M_1(a^\ast)=m_1)\times I(M_2(a^\ast,m_1)=m_2)\\
  \\
  & = &\sum_{m_2}\sum_{m_1}[Y(a,m_1,m_2)-Y(a^\ast,m_1,m_2)-Y(a,m_1^\ast,m_2^\ast)+Y(a^\ast,m_1^\ast,m_2^\ast)]\\
  & & \times I(M_1(a^\ast)=m_1)\times I(M_2(a^\ast,m_1)=m_2)\\
  & & +Y(a,m_1^\ast,m_2^\ast)-Y(a^\ast,m_1^\ast,m_2^\ast)\\
  \\
  & = & \sum_{m_2}\sum_{m_1}[Y(a,m_1,m_2)-Y(a^\ast,m_1,m_2)-Y(a,m_1^\ast,m_2^\ast)+Y(a^\ast,m_1^\ast,m_2^\ast)\\
  & & +Y(a^\ast,m_1,m_2^\ast)-Y(a^\ast,m_1,m_2^\ast)+Y(a,m_1,m_2^\ast)-Y(a,m_1,m_2^\ast)]\\
  & & \times I(M_1(a^\ast)=m_1)\times I(M_2(a^\ast,m_1)=m_2)\\
  & & +Y(a,m_1^\ast,m_2^\ast)-Y(a^\ast,m_1^\ast,m_2^\ast)\\
  \\
  & = & \sum_{m_2}\sum_{m_1} [Y(a,m_1,m_2^\ast)-Y(a^\ast,m_1,m_2^\ast)-Y(a,m_1^\ast,m_2^\ast)+Y(a^\ast,m_1^\ast,m_2^\ast)]\\
  & & \times I(M_1(a^\ast)=m_1)\times I(M_2(a^\ast,m_1)=m_2)\\
  & & + \sum_{m_2}\sum_{m_1} [Y(a,m_1,m_2)-Y(a,m_1,m_2^\ast)-Y(a^\ast,m_1,m_2)+Y(a^\ast,m_1,m_2^\ast)]\\
  & & \times I(M_1(a^\ast)=m_1)\times I(M_2(a^\ast,m_1)=m_2)\\
  & & + Y(a,m_1^\ast,m_2^\ast)-Y(a^\ast,m_1^\ast,m_2^\ast)\\
  \\
  & = & \sum_{m_1} [Y(a,m_1,m_2^\ast)-Y(a^\ast,m_1,m_2^\ast)-Y(a,m_1^\ast,m_2^\ast)+Y(a^\ast,m_1^\ast,m_2^\ast)]\\
  & & \times I(M_1(a^\ast)=m_1)\\
  & & + \sum_{m_2}\sum_{m_1} [Y(a,m_1,m_2)-Y(a,m_1,m_2^\ast)-Y(a^\ast,m_1,m_2)+Y(a^\ast,m_1,m_2^\ast)]\\
  & & \times I(M_1(a^\ast)=m_1)\times I(M_2(a^\ast,m_1)=m_2)\\
  & & + Y(a,m_1^\ast,m_2^\ast)-Y(a^\ast,m_1^\ast,m_2^\ast).
\end{eqnarray*}

According to the derivation above, the following formulas can be obtained:
\begin{eqnarray*}
  CDE(m_1^\ast,m_2^\ast) & = & Y(a,m_1^\ast,m_2^\ast)-Y(a^\ast,m_1^\ast,m_2^\ast)\\
  \\
  INT_{ref\mbox{-}AM_1}(m_1^\ast,m_2^\ast) & = & \sum_{m_1} [Y(a,m_1,m_2^\ast)-Y(a^\ast,m_1,m_2^\ast)-Y(a,m_1^\ast,m_2^\ast)+Y(a^\ast,m_1^\ast,m_2^\ast)]\\
  & & \times I(M_1(a^\ast)=m_1)\\
  \\
  INT_{ref\mbox{-}AM_2+AM_1M_2}(m_2^\ast) & = & \sum_{m_2}\sum_{m_1} [Y(a,m_1,m_2)-Y(a,m_1,m_2^\ast)-Y(a^\ast,m_1,m_2)+Y(a^\ast,m_1,m_2^\ast)]\\
  & & \times I(M_1(a^\ast)=m_1)\times I(M_2(a^\ast,m_1)=m_2).
\end{eqnarray*}

It is worth noting that $INT_{ref\mbox{-}AM_2+AM_1M_2}(m_2^\ast)$ cannot be separated into $INT_{ref\mbox{-}AM_2}(m_1^\ast,m_2^\ast)$ and $INT_{ref\mbox{-}AM_1M_2}(m_1^\ast,m_2^\ast)$ since both of the two terms are non-identifiable, which will be discussed in details in Appendix D. 

Therefore, $PDE$ can be decomposed into the following components:
\begin{eqnarray*}
  PDE = CDE(m_1^\ast,m_2^\ast) + INT_{ref\mbox{-}AM_1}(m_1^\ast,m_2^\ast) + INT_{ref\mbox{-}AM_2+AM_1M_2}(m_2^\ast).
\end{eqnarray*}

$TDE$ can be decomposed into the following components:
\begin{eqnarray*}
  TDE & = & PDE + NatINT_{AM_1} + NatINT_{AM_2}+ NatINT_{AM_1M_2}\\
  \\
      & = & CDE(m_1^\ast,m_2^\ast) + INT_{ref\mbox{-}AM_1}(m_1^\ast,m_2^\ast) + INT_{ref\mbox{-}AM_2+AM_1M_2}(m_2^\ast)\\
      & & + NatINT_{AM_1} + NatINT_{AM_2}+ NatINT_{AM_1M_2}.
\end{eqnarray*}

We next focus on $SIE_{M_1}$ and decompose it into $PIE_{M_1}$ and $NatINT_{M_1M_2}$ by subtracting $PIE_{M_1}$ from $SIE_{M_1}$:
\begin{eqnarray*}
  SIE_{M_1}-PIE_{M_1} & = & Y(a^\ast,M_1(a),M_2(a,M_1(a)))-Y(a^\ast,M_1(a^\ast),M_2(a,M_1(a^\ast)))\\
  & & -Y(a^\ast,M_1(a),M_2(a^\ast,M_1(a)))+Y(a^\ast,M_1(a^\ast),M_2(a^\ast,M_1(a^\ast)))\\
  \\
  & = & Y(a^\ast,M_1(a),M_2(a,M_1(a)))-Y(a^\ast,M_1(a),M_2(a^\ast,M_1(a)))\\
  & & -Y(a^\ast,M_1(a^\ast),M_2(a,M_1(a^\ast)))+Y(a^\ast,M_1(a^\ast),M_2(a^\ast,M_1(a^\ast)))\\
  \\
  & = & NatINT_{M_1M_2},
\end{eqnarray*}
where $NatINT_{M_1M_2}$ satisfies Definition 3.

Therefore, $SIE_{M_1}$ can be decomposed into the following components:
\begin{eqnarray*}
  SIE_{M_1} = PIE_{M_1} + NatINT_{M_1M_2}.
\end{eqnarray*}

Combining all the derivations above, we have the decomposition of total effect as follows:
\begin{eqnarray*}
  TE & = & CDE(m_1^\ast,m_2^\ast)+INT_{ref\mbox{-}AM_1}(m_1^\ast,m_2^\ast)+INT_{ref\mbox{-}AM_2+AM_1M_2}(m_2^\ast)\\
  & & + NatINT_{AM_1} + NatINT_{AM_2}+ NatINT_{AM_1M_2}+ NatINT_{M_1M_2}\\
  & & + PIE_{M_1} + PIE_{M_2},\qquad\qquad\qquad\qquad\qquad\qquad\qquad\qquad\qquad\qquad\qquad\square
\end{eqnarray*}

\clearpage
\section*{Appendix D. The non-identifiability of $INT_{ref\mbox{-}AM_2}(m_1^\ast,m_2^\ast)$ and $INT_{ref\mbox{-}AM_1M_2}(m_1^\ast,m_2^\ast)$ in a sequential two-mediator scenario}

We need to show that the reference interaction effects, $INT_{ref\mbox{-}AM_2}(m_1^\ast,m_2^\ast)$ and $INT_{ref\mbox{-}AM_1M_2}(m_1^\ast,m_2^\ast)$, are non-identifiable in a sequential two-mediator scenario as shown in Figure 8. \\

\noindent
\emph{Proof}: 

We first need to decompose $INT_{ref\mbox{-}AM_2+AM_1M_2}(m_2^\ast)$ into $INT_{ref\mbox{-}AM_2}(m_1^\ast,m_2^\ast)$ and $INT_{ref\mbox{-}AM_1M_2}(m_1^\ast,m_2^\ast)$: 
\begin{eqnarray*}
INT_{ref\mbox{-}AM_2+AM_1M_2}(m_2^\ast) & = & \sum_{m_2}\sum_{m_1} [Y(a,m_1,m_2)-Y(a,m_1,m_2^\ast)-Y(a^\ast,m_1,m_2)+Y(a^\ast,m_1,m_2^\ast)]\\
  & & \times I(M_1(a^\ast)=m_1)\times I(M_2(a^\ast,m_1)=m_2)\\
  \\
  & = & \sum_{m_2}\sum_{m_1} [Y(a,m_1,m_2)-Y(a,m_1,m_2^\ast)-Y(a^\ast,m_1,m_2)+Y(a^\ast,m_1,m_2^\ast)\\
  & & + Y(a^\ast,m_1^\ast,m_2^\ast)-Y(a^\ast,m_1^\ast,m_2^\ast)+Y(a,m_1^\ast,m_2^\ast)-Y(a,m_1^\ast,m_2^\ast)\\
  & & +Y(a^\ast,m_1^\ast,m_2)-Y(a^\ast,m_1^\ast,m_2)+Y(a,m_1^\ast,m_2)-Y(a,m_1^\ast,m_2)]\\
  & & \times I(M_1(a^\ast)=m_1)\times I(M_2(a^\ast,m_1)=m_2)\\
  \\
  & = & \sum_{m_2}\sum_{m_1}[Y(a,m_1^\ast,m_2)-Y(a,m_1^\ast,m_2^\ast)-Y(a^\ast,m_1^\ast,m_2)+Y(a^\ast,m_1^\ast,m_2^\ast)]\\
  & & \times I(M_1(a^\ast)=m_1)\times I(M_2(a^\ast,m_1)=m_2)\\
  & & +\sum_{m_2}\sum_{m_1}[Y(a,m_1,m_2)-Y(a,m_1,m_2^\ast)-Y(a,m_1^\ast,m_2)-Y(a^\ast,m_1,m_2)\\
  & & +Y(a^\ast,m_1^\ast,m_2)+Y(a^\ast,m_1,m_2^\ast)+Y(a,m_1^\ast,m_2^\ast)-Y(a^\ast,m_1^\ast,m_2^\ast)]\\
  & & \times I(M_1(a^\ast)=m_1)\times I(M_2(a^\ast,m_1)=m_2).
\end{eqnarray*}

Therefore, we have the following formulas:
\begin{eqnarray*}
  INT_{ref\mbox{-}AM_2}(m_1^\ast,m_2^\ast) & = & \sum_{m_2}\sum_{m_1}[Y(a,m_1^\ast,m_2)-Y(a,m_1^\ast,m_2^\ast)-Y(a^\ast,m_1^\ast,m_2)+Y(a^\ast,m_1^\ast,m_2^\ast)]\\
  & & \times I(M_1(a^\ast)=m_1)\times I(M_2(a^\ast,m_1)=m_2)\\
  \\
  INT_{ref\mbox{-}AM_1M_2}(m_1^\ast,m_2^\ast) & = & \sum_{m_2}\sum_{m_1}[Y(a,m_1,m_2)-Y(a,m_1,m_2^\ast)-Y(a,m_1^\ast,m_2)-Y(a^\ast,m_1,m_2)\\
  & & +Y(a^\ast,m_1^\ast,m_2)+Y(a^\ast,m_1,m_2^\ast)+Y(a,m_1^\ast,m_2^\ast)-Y(a^\ast,m_1^\ast,m_2^\ast)]\\
  & & \times I(M_1(a^\ast)=m_1)\times I(M_2(a^\ast,m_1)=m_2).
\end{eqnarray*}

It can be seen that both formulas include the following term:
\begin{eqnarray*}
  \sum_{m_2}\sum_{m_1}Y(a,m_1^\ast,m_2)\times I(M_1(a^\ast)=m_1)\times I(M_2(a^\ast,m_1)=m_2),
\end{eqnarray*}
which can be rewritten as the counterfactual formula $Y(a,m_1^\ast,M_2(a^\ast,M_1(a^\ast)))$ and graphically illustrated in Figure 12. 

Since $m_1^\ast$ is an arbitrary reference level of $M_2$, let us consider an instance that there exists $a^{\ast\ast}\neq a^\ast$ such that $M_1(a^{\ast\ast})=m_1^\ast$. In this case, the counterfactual formula can be rewritten as $Y(a,M_1(a^{\ast\ast}),M_2(a^\ast,M_1(a^\ast)))$, where $M_1$ is being activated by two different values of exposure $A$ in the kite graph \cite{a} formed up by the path $A\rightarrow M_1\rightarrow Y$ and the path $A\rightarrow M_1\rightarrow M_2 \rightarrow Y$ in Figure 8. Avin et al. showed that such counterfactual formulas are non-identifiable and referred to as problematic counterfactual formulas \cite{a}. Because the instance cannot be ruled out in any certain population, $Y(a,m_1^\ast,M_2(a^\ast,M_1(a^\ast)))$ is non-identifiable. Therefore, $INT_{ref\mbox{-}AM_2}(m_1^\ast,m_2^\ast)$ and $INT_{ref\mbox{-}AM_1M_2}(m_1^\ast,m_2^\ast)$ are non-identifiable. $\qquad\qquad\quad\square$ 

\clearpage
\section*{Appendix E. The nature of $INT_{med\mbox{-}AM_1}(m_2^\ast)$ and the non-identifiability of $INT_{med\mbox{-}AM_2}(m_1^\ast)$ and $INT_{med\mbox{-}AM_1M_2}(m_1^\ast,m_2^\ast)$ in a sequential two-mediator scenario}

Suppose we have a directed acyclic graph as shown in Figure 8, we need to show that the nature of mediated interaction effect $INT_{med\mbox{-}AM_1}(m_2^\ast)$ is identical to the illustration as shown in Figure 6A and Figure 10A. Second, we need to show that the mediated interaction effects, $INT_{med\mbox{-}AM_2}(m_1^\ast)$ and $INT_{med\mbox{-}AM_1M_2}(m_1^\ast,m_2^\ast)$, are non-identifiable.\\

\noindent
\emph{Proof}: 

We try to find $INT_{med\mbox{-}AM_1}(m_2^\ast)$, $INT_{med\mbox{-}AM_2}(m_1^\ast)$ and $INT_{med\mbox{-}AM_1M_2}(m_1^\ast,m_2^\ast)$ from the contrast $TDE-PDE$ \cite{v4,b}:
\begin{eqnarray*}
TDE-PDE & = & Y(a,M_1(a),M_2(a,M_1(a)))-Y(a^\ast,M_1(a),M_2(a,M_1(a)))\\
& & -Y(a,M_1(a^\ast),M_2(a^\ast,M_1(a^\ast)))+Y(a^\ast,M_1(a^\ast),M_2(a^\ast,M_1(a^\ast)))\\
\\
& = & \sum_{m_2}\sum_{m_1}Y(a,m_1,m_2)\times I(M_1(a)=m_1)\times I(M_2(a,m_1)=m_2)\\
& & - \sum_{m_2}\sum_{m_1}Y(a^\ast,m_1,m_2)\times I(M_1(a)=m_1)\times I(M_2(a,m_1)=m_2)\\
& & - \sum_{m_2}\sum_{m_1}Y(a,m_1,m_2)\times I(M_1(a^\ast)=m_1)\times I(M_2(a^\ast,m_1)=m_2)\\
& & + \sum_{m_2}\sum_{m_1}Y(a^\ast,m_1,m_2)\times I(M_1(a^\ast)=m_1)\times I(M_2(a^\ast,m_1)=m_2)\\
\\
& = & \sum_{m_2}\sum_{m_1}[Y(a,m_1,m_2)-Y(a^\ast,m_1,m_2)]\\
& & \times [I(M_1(a)=m_1)\times I(M_2(a,m_1)=m_2)\\
& & -I(M_1(a^\ast)=m_1)\times I(M_2(a^\ast,m_1)=m_2)]\\
\\
& = & \sum_{m_2}\sum_{m_1}[Y(a,m_1,m_2)-Y(a^\ast,m_1,m_2)-Y(a,m_1^\ast,m_2^\ast)+Y(a^\ast,m_1^\ast,m_2^\ast)]\\
& & \times [I(M_1(a)=m_1)\times I(M_2(a,m_1)=m_2)\\
& & -I(M_1(a^\ast)=m_1)\times I(M_2(a^\ast,m_1)=m_2)]\\
\\
& = & \sum_{m_2}\sum_{m_1}[Y(a,m_1,m_2)-Y(a^\ast,m_1,m_2)-Y(a,m_1^\ast,m_2^\ast)+Y(a^\ast,m_1^\ast,m_2^\ast)\\
& & +Y(a^\ast,m_1^\ast,m_2^\ast)-Y(a^\ast,m_1^\ast,m_2^\ast)+Y(a^\ast,m_1^\ast,m_2)-Y(a^\ast,m_1^\ast,m_2)\\
& & +Y(a^\ast,m_1,m_2^\ast)-Y(a^\ast,m_1,m_2^\ast)+Y(a,m_1^\ast,m_2^\ast)-Y(a,m_1^\ast,m_2^\ast)\\
& & +Y(a,m_1^\ast,m_2)-Y(a,m_1^\ast,m_2)+Y(a,m_1,m_2^\ast)-Y(a,m_1,m_2^\ast)]\\
& & \times [I(M_1(a)=m_1)\times I(M_2(a,m_1)=m_2)\\
& & -I(M_1(a^\ast)=m_1)\times I(M_2(a^\ast,m_1)=m_2)]\\
\\
& = & \sum_{m_2}\sum_{m_1}[Y(a,m_1,m_2^\ast)-Y(a,m_1^\ast,m_2^\ast)-Y(a^\ast,m_1,m_2^\ast)+Y(a^\ast,m_1^\ast,m_2^\ast)]\\
& & \times [I(M_1(a)=m_1)\times I(M_2(a,m_1)=m_2)\\
& & -I(M_1(a^\ast)=m_1)\times I(M_2(a^\ast,m_1)=m_2)]\\
& & +\sum_{m_2}\sum_{m_1}[Y(a,m_1^\ast,m_2)-Y(a,m_1^\ast,m_2^\ast)-Y(a^\ast,m_1^\ast,m_2)+Y(a^\ast,m_1^\ast,m_2^\ast)]\\
& & \times [I(M_1(a)=m_1)\times I(M_2(a,m_1)=m_2)\\
& & -I(M_1(a^\ast)=m_1)\times I(M_2(a^\ast,m_1)=m_2)]\\
& & +\sum_{m_2}\sum_{m_1}[Y(a,m_1,m_2)-Y(a,m_1,m_2^\ast)-Y(a,m_1^\ast,m_2)-Y(a^\ast,m_1,m_2)\\
& & +Y(a^\ast,m_1^\ast,m_2)+Y(a^\ast,m_1,m_2^\ast)+Y(a,m_1^\ast,m_2^\ast)-Y(a^\ast,m_1^\ast,m_2^\ast)]\\
& & \times [I(M_1(a)=m_1)\times I(M_2(a,m_1)=m_2)\\
& & -I(M_1(a^\ast)=m_1)\times I(M_2(a^\ast,m_1)=m_2)].
\end{eqnarray*}

Therefore, we have the following formulas:
\begin{eqnarray*}
INT_{med\mbox{-}AM_1}(m_2^\ast) & = & \sum_{m_2}\sum_{m_1}[Y(a,m_1,m_2^\ast)-Y(a,m_1^\ast,m_2^\ast)-Y(a^\ast,m_1,m_2^\ast)+Y(a^\ast,m_1^\ast,m_2^\ast)]\\
& & \times [I(M_1(a)=m_1)\times I(M_2(a,m_1)=m_2)\\
& & -I(M_1(a^\ast)=m_1)\times I(M_2(a^\ast,m_1)=m_2)]\\
\\
INT_{med\mbox{-}AM_2}(m_1^\ast) & = & \sum_{m_2}\sum_{m_1}[Y(a,m_1^\ast,m_2)-Y(a,m_1^\ast,m_2^\ast)-Y(a^\ast,m_1^\ast,m_2)+Y(a^\ast,m_1^\ast,m_2^\ast)]\\
& & \times [I(M_1(a)=m_1)\times I(M_2(a,m_1)=m_2)\\
& & -I(M_1(a^\ast)=m_1)\times I(M_2(a^\ast,m_1)=m_2)]\\
\\
INT_{med\mbox{-}AM_1M_2}(m_1^\ast,m_2^\ast) & = & \sum_{m_2}\sum_{m_1}[Y(a,m_1,m_2)-Y(a,m_1,m_2^\ast)-Y(a,m_1^\ast,m_2)-Y(a^\ast,m_1,m_2)\\
& & +Y(a^\ast,m_1^\ast,m_2)+Y(a^\ast,m_1,m_2^\ast)+Y(a,m_1^\ast,m_2^\ast)-Y(a^\ast,m_1^\ast,m_2^\ast)]\\
& & \times [I(M_1(a)=m_1)\times I(M_2(a,m_1)=m_2)\\
& & -I(M_1(a^\ast)=m_1)\times I(M_2(a^\ast,m_1)=m_2)].
\end{eqnarray*}

The mediated interaction effect between $A$ and $M_1$, $INT_{med\mbox{-}AM_1}(m_2^\ast)$, can be rewritten as follows:
\begin{eqnarray*}
INT_{med\mbox{-}AM_1}(m_2^\ast) & = & \sum_{m_2}\sum_{m_1}[Y(a,m_1,m_2^\ast)-Y(a,m_1^\ast,m_2^\ast)-Y(a^\ast,m_1,m_2^\ast)+Y(a^\ast,m_1^\ast,m_2^\ast)]\\
& & \times [I(M_1(a)=m_1)\times I(M_2(a,m_1)=m_2)\\
& & -I(M_1(a^\ast)=m_1)\times I(M_2(a^\ast,m_1)=m_2)]\\
\\
& = & \sum_{m_1}[Y(a,m_1,m_2^\ast)-Y(a,m_1^\ast,m_2^\ast)-Y(a^\ast,m_1,m_2^\ast)+Y(a^\ast,m_1^\ast,m_2^\ast)]\\
& & \times [I(M_1(a)=m_1)-I(M_1(a^\ast)=m_1)]\\
\\
& = & \sum_{m_1}[Y(a,m_1,m_2^\ast)-Y(a^\ast,m_1,m_2^\ast)]\times [I(M_1(a)=m_1)-I(M_1(a^\ast)=m_1)]\\
\\
& = & \sum_{m_1}Y(a,m_1,m_2^\ast)I(M_1(a)=m_1)-\sum_{m_1}Y(a,m_1,m_2^\ast)I(M_1(a^\ast)=m_1)\\
& & -\sum_{m_1}Y(a^\ast,m_1,m_2^\ast)I(M_1(a)=m_1)+\sum_{m_1}Y(a^\ast,m_1,m_2^\ast)I(M_1(a^\ast)=m_1)\\
\\
& = & Y(a,M_1(a),m_2^\ast)-Y(a,M_1(a^\ast),m_2^\ast)\\
& & -Y(a^\ast,M_1(a),m_2^\ast)+Y(a^\ast,M_1(a^\ast),m_2^\ast)\\
\\
& = & Y(a,M_1(a),m_2^\ast)-Y(a^\ast,M_1(a),m_2^\ast)\\
& & -Y(a,M_1(a^\ast),m_2^\ast)+Y(a^\ast,M_1(a^\ast),m_2^\ast),
\end{eqnarray*}
where the second equality follows by the fact that $m_2^\ast$ is a constant so the summation indexed by $m_2$ can be dropped, and the third equality follows by the fact that $Y(a,m_1^\ast,m_2^\ast)$ and $Y(a^\ast,m_1^\ast,m_2^\ast)$ are constants and can be canceled out through the summation.

Therefore, the last equality indicates that the mediated interaction effect between $A$ and $M_1$ is tantamount to assigning a fixed reference level $m_2^\ast$ to $M_2$ and allowing $M_1$ to naturally vary with exposure $A$ as shown in Figure 6A and Figure 10A. 

Furthermore, it can be seen that both of the formulas of $INT_{med\mbox{-}AM_2}(m_1^\ast)$ and $INT_{med\mbox{-}AM_1M_2}(m_1^\ast,m_2^\ast)$ contain the term:
\begin{eqnarray*}
\sum_{m_2}\sum_{m_1}Y(a,m_1^\ast,m_2)\times I(M_1(a^\ast)=m_1)\times I(M_2(a^\ast,m_1)=m_2).
\end{eqnarray*}

According to Appendix D, both $INT_{med\mbox{-}AM_2}(m_1^\ast)$ and $INT_{med\mbox{-}AM_1M_2}(m_1^\ast,m_2^\ast)$ are non-identifiable. $\qquad\qquad\qquad\qquad\qquad\qquad\qquad\qquad\qquad\qquad\qquad\qquad\square$

\clearpage
\section*{Appendix F. Linear regression models with continuous outcome and continuous mediators in a sequential two-mediator scenario}

Suppose we have a directed acyclic graph as shown in Figure 8 and the following linear models for $Y$, $M_2$ and $M_1$:
\begin{eqnarray*}
E[Y|A,M_1,M_2,C] & = & \theta_0 + \theta_1A + \theta_2M_1 + \theta_3M_2 + \theta_4AM_1 + \theta_5AM_2 + \theta_6M_1M_2\\ 
& & + \theta_7AM_1M_2 + \theta_8^\prime C\\
\\
E[M_2|A,M_1,C] & = & \beta_0 + \beta_1A + \beta_2M_1 + \beta_3AM_1 + \beta_4^\prime C\\
\\
E[M_1|A,C] & = & \gamma_0 + \gamma_1A + \gamma_2^\prime C, 
\end{eqnarray*}
where $C$ is a sufficient confounding set that satisfies the identification assumptions $(A1)$-$(A6)$.

$\epsilon_Y$, $\epsilon_{M_2}$ and $\epsilon_{M_1}$ denote independent random error terms for $Y$, $M_2$ and $M_1$ and follow $N(0,\sigma_{Y}^2)$, $N(0,\sigma_{M_2}^2)$ and $N(0,\sigma_{M_1}^2)$, respectively. According to Appendix C, the total effect can be decomposed into the following components:
\begin{eqnarray*}
  TE & = & CDE(m_1^\ast,m_2^\ast)+INT_{ref\mbox{-}AM_1}(m_1^\ast,m_2^\ast)+INT_{ref\mbox{-}AM_2+AM_1M_2}(m_2^\ast)\\
  & & + NatINT_{AM_1} + NatINT_{AM_2}+ NatINT_{AM_1M_2}+ NatINT_{M_1M_2}\\
  & & + PIE_{M_1} + PIE_{M_2}.
\end{eqnarray*}

We need to find the expected value of each component conditional on the sufficient confounding set.

\subsection*{Controlled direct effect}
\begin{eqnarray*}
CDE(m_1^\ast,m_2^\ast) & = & Y(a,m_1^\ast,m_2^\ast)-Y(a^\ast,m_1^\ast,m_2^\ast)
\end{eqnarray*}
\begin{eqnarray*}
&\Rightarrow & E[CDE(m_1^\ast,m_2^\ast)|c]\\
\\
 & = & E[Y(a,m_1^\ast,m_2^\ast)-Y(a^\ast,m_1^\ast,m_2^\ast)|c]\\
 \\
 & = & E[Y(a,m_1^\ast,m_2^\ast)|c]-E[Y(a^\ast,m_1^\ast,m_2^\ast)|c]\\
 \\
 & = & E[Y(a,m_1^\ast,m_2^\ast)|a,c]-E[Y(a^\ast,m_1^\ast,m_2^\ast)|a^\ast,c]\quad by\; A1\\
 \\
 & = & E[Y(a,m_1^\ast,m_2^\ast)|a,m_1^\ast,m_2^\ast,c]-E[Y(a^\ast,m_1^\ast,m_2^\ast)|a^\ast,m_1^\ast,m_2^\ast,c]\quad by\; A2\\
 \\
 & = & E[Y|a,m_1^\ast,m_2^\ast,c]-E[Y|a^\ast,m_1^\ast,m_2^\ast,c]\quad by\; consistency\\
 \\
 &= & (\theta_0+\theta_1a+\theta_2m_1^\ast+\theta_3m_2^\ast+\theta_4am_1^\ast+\theta_5am_2^\ast+\theta_6m_1^\ast m_2^\ast+\theta_7am_1^\ast m_2^\ast+\theta_8^\prime c)\\
 & & -(\theta_0+\theta_1a^\ast+\theta_2m_1^\ast+\theta_3m_2^\ast+\theta_4a^\ast m_1^\ast+\theta_5a^\ast m_2^\ast+\theta_6m_1^\ast m_2^\ast+\theta_7a^\ast m_1^\ast m_2^\ast+\theta_8^\prime c)\\
 \\
 & = & (\theta_1a+\theta_4am_1^\ast+\theta_5am_2^\ast+\theta_7am_1^\ast m_2^\ast)-(\theta_1a^\ast+\theta_4a^\ast m_1^\ast+\theta_5a^\ast m_2^\ast+\theta_7a^\ast m_1^\ast m_2^\ast)\\
 \\
 & = & \theta_1\left(a-a^\ast\right)+\theta_4m_1^\ast\left(a-a^\ast\right)+\theta_5m_2^\ast\left(a-a^\ast\right)+\theta_7m_1^\ast m_2^\ast\left(a-a^\ast\right)\\
 \\
 & = & \left(\theta_1+\theta_4m_1^\ast+\theta_5m_2^\ast+\theta_7m_1^\ast m_2^\ast\right)\left(a-a^\ast\right).
\end{eqnarray*}
\\
\subsection*{Reference interaction effect between $A$ and $M_1$}
We first consider $M_1$ a categorical random variable.
\begin{eqnarray*}
INT_{ref\mbox{-}AM_1}(m_1^\ast,m_2^\ast) & = & \sum_{m_1} [Y(a,m_1,m_2^\ast)-Y(a^\ast,m_1,m_2^\ast)-Y(a,m_1^\ast,m_2^\ast)+Y(a^\ast,m_1^\ast,m_2^\ast)]\\
  & & \times I(M_1(a^\ast)=m_1)\\
\end{eqnarray*}
\begin{eqnarray*}
&\Rightarrow & E[INT_{ref\mbox{-}AM_1}(m_1^\ast,m_2^\ast)|c]\\
\\
& = & E\left[\sum_{m_1} [Y(a,m_1,m_2^\ast)-Y(a^\ast,m_1,m_2^\ast)-Y(a,m_1^\ast,m_2^\ast)+Y(a^\ast,m_1^\ast,m_2^\ast)]\times I(M_1(a^\ast)=m_1)\bigg| c\right]\\
  \\
  & = & \sum_{m_1}E\left[[Y(a,m_1,m_2^\ast)-Y(a^\ast,m_1,m_2^\ast)-Y(a,m_1^\ast,m_2^\ast)+Y(a^\ast,m_1^\ast,m_2^\ast)]\times I(M_1(a^\ast)=m_1)|c\right]\\
  \\
  & = & \sum_{m_1}E\left[Y(a,m_1,m_2^\ast)-Y(a^\ast,m_1,m_2^\ast)-Y(a,m_1^\ast,m_2^\ast)+Y(a^\ast,m_1^\ast,m_2^\ast)|c\right]\\
  & &\times E\left[I(M_1(a^\ast)=m_1)|c\right]\quad by\; A4\\
  \\
  & = & \sum_{m_1}E\left[Y(a,m_1,m_2^\ast)-Y(a^\ast,m_1,m_2^\ast)-Y(a,m_1^\ast,m_2^\ast)+Y(a^\ast,m_1^\ast,m_2^\ast)|c\right]\\
  & &\times \Pr(M_1(a^\ast)=m_1|c)\\
  \\
  & = & \sum_{m_1}E\left[Y(a,m_1,m_2^\ast)-Y(a^\ast,m_1,m_2^\ast)-Y(a,m_1^\ast,m_2^\ast)+Y(a^\ast,m_1^\ast,m_2^\ast)|c\right]\\
  & &\times \Pr(M_1(a^\ast)=m_1|a^\ast,c)\quad by\; A3\\
  \\
  & = & \sum_{m_1}E\left[Y(a,m_1,m_2^\ast)-Y(a^\ast,m_1,m_2^\ast)-Y(a,m_1^\ast,m_2^\ast)+Y(a^\ast,m_1^\ast,m_2^\ast)|c\right]\\
  & &\times \Pr(M_1=m_1|a^\ast,c)\quad by\; consistency\\
  \\
  & = & \sum_{m_1}{E\left[Y\left(a,m_1,m_2^\ast\right)\middle| c\right]}\Pr{\left(M_1=m_1\middle| a^\ast,c\right)}-\sum_{m_1}{E\left[Y\left(a^\ast,m_1,m_2^\ast\right)\middle| c\right]}\Pr{\left(M_1=m_1\middle| a^\ast,c\right)}\\
  & & -\sum_{m_1}{E\left[Y\left(a,m_1^\ast,m_2^\ast\right)\middle| c\right]}\Pr{\left(M_1=m_1\middle| a^\ast,c\right)}+\sum_{m_1}{E\left[Y\left(a^\ast,m_1^\ast,m_2^\ast\right)\middle| c\right]}\Pr{\left(M_1=m_1\middle| a^\ast,c\right)}\\
  \\
  & = & \sum_{m_1}{E\left[Y\left(a,m_1,m_2^\ast\right)\middle|a,m_1,m_2^\ast,c\right]}\Pr{\left(M_1=m_1\middle| a^\ast,c\right)}\\
  & & -\sum_{m_1}{E\left[Y\left(a^\ast,m_1,m_2^\ast\right)\middle|a^\ast,m_1,m_2^\ast, c\right]}\Pr{\left(M_1=m_1\middle| a^\ast,c\right)}\\
  & & -\sum_{m_1}{E\left[Y\left(a,m_1^\ast,m_2^\ast\right)\middle|a,m_1^\ast,m_2^\ast, c\right]}\Pr{\left(M_1=m_1\middle| a^\ast,c\right)}\\
  & & +\sum_{m_1}{E\left[Y\left(a^\ast,m_1^\ast,m_2^\ast\right)\middle|a^\ast,m_1^\ast,m_2^\ast, c\right]}\Pr{\left(M_1=m_1\middle| a^\ast,c\right)}\quad by\; A1\;A2\\
  \\
  & = & \sum_{m_1}{E\left[Y|a,m_1,m_2^\ast,c\right]}\Pr{\left(M_1=m_1\middle| a^\ast,c\right)}-\sum_{m_1}{E\left[Y|a^\ast,m_1,m_2^\ast, c\right]}\Pr{\left(M_1=m_1\middle| a^\ast,c\right)}\\
  & & -\sum_{m_1}{E\left[Y|a,m_1^\ast,m_2^\ast\, c\right]}\Pr{\left(M_1=m_1\middle| a^\ast,c\right)}+\sum_{m_1}{E\left[Y|a^\ast,m_1^\ast,m_2^\ast, c\right]}\Pr{\left(M_1=m_1\middle| a^\ast,c\right)}\; by\; consistency
\end{eqnarray*}

We next extend the formula to a continuous $M_1$. 
\begin{eqnarray*}
& & E[INT_{ref\mbox{-}AM_1}(m_1^\ast,m_2^\ast)|c] \\
\\
& = & \int_{m_1}{E\left[Y\middle| a,m_1,m_2^\ast,c\right]d\Pr{\left(M_1=m_1\middle| a^\ast,c\right)}}\\
& & -\int_{m_1}{E\left[Y\middle| a^\ast,m_1,m_2^\ast,c\right]d\Pr{\left(M_1=m_1\middle| a^\ast,c\right)}}\\
& & -\int_{m_1}{E\left[Y\middle| a,m_1^\ast,m_2^\ast,c\right]d\Pr{\left(M_1=m_1\middle| a^\ast,c\right)}}\\
& & +\int_{m_1}{E\left[Y\middle| a^\ast,m_1^\ast,m_2^\ast,c\right]d\Pr{\left(M_1=m_1\middle| a^\ast,c\right)}}\\
\\
& = & \int_{m_1}(\theta_0+\theta_1a+\theta_2m_1+\theta_3m_2^\ast+\theta_4am_1+\theta_5am_2^\ast\\
& & +\theta_6m_1m_2^\ast+\theta_7am_1m_2^\ast+\theta_8^\prime c)d\Pr{\left(M_1=m_1\middle| a^\ast,c\right)})\\
& & -\int_{m_1}(\theta_0+\theta_1a^\ast+\theta_2m_1+\theta_3m_2^\ast+\theta_4a^\ast m_1+\theta_5a^\ast m_2^\ast\\
& & +\theta_6m_1m_2^\ast+\theta_7a^\ast m_1m_2^\ast+\theta_8^\prime c)d\Pr{\left(M_1=m_1\middle| a^\ast,c\right)}\\
& & -\int_{m_1}(\theta_0+\theta_1a+\theta_2m_1^\ast+\theta_3m_2^\ast+\theta_4am_1^\ast+\theta_5am_2^\ast\\
& & +\theta_6m_1^\ast m_2^\ast+\theta_7am_1^\ast m_2^\ast+\theta_8^\prime c)d\Pr{\left(M_1=m_1\middle| a^\ast,c\right)}\\
& & + \int_{m_1}(\theta_0+\theta_1a^\ast+\theta_2m_1^\ast+\theta_3m_2^\ast+\theta_4a^\ast m_1^\ast+\theta_5a^\ast m_2^\ast\\
& & +\theta_6m_1^\ast m_2^\ast+\theta_7a^\ast m_1^\ast m_2^\ast+\theta_8^\prime c)d\Pr{\left(M_1=m_1\middle| a^\ast,c\right)}\\
\\
& = & \left(\theta_0+\theta_1a+\theta_3m_2^\ast+\theta_5am_2^\ast+\theta_8^\prime c\right)+\left(\theta_2+\theta_4a+\theta_6m_2^\ast+\theta_7am_2^\ast\right)\\
& & \times\left(\gamma_0+\gamma_1a^\ast+\gamma_2^\prime c\right)\\
& & -\left(\theta_0+\theta_1a^\ast+\theta_3m_2^\ast+\theta_5a^\ast m_2^\ast+\theta_8^\prime c\right)+\left(\theta_2+\theta_4a^\ast+\theta_6m_2^\ast+\theta_7a^\ast m_2^\ast\right)\\
& & \times\left(\gamma_0+\gamma_1a^\ast+\gamma_2^\prime c\right)\\
& & -\left(\theta_0+\theta_1a+\theta_2m_1^\ast+\theta_3m_2^\ast+\theta_4am_1^\ast+\theta_5am_2^\ast+\theta_6m_1^\ast m_2^\ast+\theta_7am_1^\ast m_2^\ast+\theta_8^\prime c\right)\\
& & +\left(\theta_0+\theta_1a^\ast+\theta_2m_1^\ast+\theta_3m_2^\ast+\theta_4a^\ast m_1^\ast+\theta_5a^\ast m_2^\ast+\theta_6m_1^\ast m_2^\ast+\theta_7a^\ast m_1^\ast m_2^\ast+\theta_8^\prime c\right)\\
\\
& = & \left(\gamma_0+\gamma_1a^\ast+\gamma_2^\prime c-m_1^\ast\right)\times\left(\theta_4+\theta_7m_2^\ast\right)\times\left(a-a^\ast\right).
\end{eqnarray*}
\\
\subsection*{The sum of two reference interaction effects: $INT_{ref\mbox{-}AM_2+AM_1M_2}(m_2^\ast)$}
\begin{eqnarray*}
INT_{ref\mbox{-}AM_2+AM_1M_2}(m_2^\ast) & = & \sum_{m_2}\sum_{m_1} [Y(a,m_1,m_2)-Y(a,m_1,m_2^\ast)-Y(a^\ast,m_1,m_2)+Y(a^\ast,m_1,m_2^\ast)]\\
  & & \times I(M_1(a^\ast)=m_1)\times I(M_2(a^\ast,m_1)=m_2)
\end{eqnarray*}
\begin{eqnarray*}
&\Rightarrow & E[INT_{ref\mbox{-}AM_2+AM_1M_2}(m_2^\ast)|c]\\
\\
& = & E\left[\sum_{m_2}\sum_{m_1} [Y(a,m_1,m_2)-Y(a,m_1,m_2^\ast)-Y(a^\ast,m_1,m_2)+Y(a^\ast,m_1,m_2^\ast)]\right. \\
& & \left.\times I(M_1(a^\ast)=m_1)\times I(M_2(a^\ast,m_1)=m_2)\bigg|c\right]\\
\\
& = & \sum_{m_2}\sum_{m_1}E\left[[Y(a,m_1,m_2)-Y(a,m_1,m_2^\ast)-Y(a^\ast,m_1,m_2)+Y(a^\ast,m_1,m_2^\ast)]\right.\\
& &\left. \times I(M_1(a^\ast)=m_1)\times I(M_2(a^\ast,m_1)=m_2)|c\right]\\
\\
& = & \sum_{m_2}\sum_{m_1}E[Y(a,m_1,m_2)-Y(a,m_1,m_2^\ast)-Y(a^\ast,m_1,m_2)+Y(a^\ast,m_1,m_2^\ast)|c]\\
& & \times E[I(M_1(a^\ast)=m_1)\times I(M_2(a^\ast,m_1)=m_2)|c]\; by\; A4\;A6\\
\\
& = & \sum_{m_2}\sum_{m_1}E[Y(a,m_1,m_2)-Y(a,m_1,m_2^\ast)-Y(a^\ast,m_1,m_2)+Y(a^\ast,m_1,m_2^\ast)|c]\\
& & \times \Pr(M_1(a^\ast)=m_1|c)\times \Pr(M_2(a^\ast,m_1)=m_2)|c)\\
\\
& = & \sum_{m_2}\sum_{m_1}E[Y(a,m_1,m_2)-Y(a,m_1,m_2^\ast)-Y(a^\ast,m_1,m_2)+Y(a^\ast,m_1,m_2^\ast)|c]\\
& & \times \Pr(M_1(a^\ast)=m_1|a^\ast,c)\times \Pr(M_2(a^\ast,m_1)=m_2)|a^\ast,m_1,c)\; by\; A3\;A5\\
\\
& = & \sum_{m_2}\sum_{m_1}E[Y(a,m_1,m_2)-Y(a,m_1,m_2^\ast)-Y(a^\ast,m_1,m_2)+Y(a^\ast,m_1,m_2^\ast)|c]\\
& & \times \Pr(M_1=m_1|a^\ast,c)\times \Pr(M_2=m_2|a^\ast,m_1,c)\; by\;consistency\\
\\
& = & \sum_{m_2}\sum_{m_1}E[Y(a,m_1,m_2)|c]\times \Pr(M_1=m_1|a^\ast,c)\times \Pr(M_2=m_2|a^\ast,m_1,c)\\
& & - \sum_{m_2}\sum_{m_1}E[Y(a,m_1,m_2^\ast)|c]\times \Pr(M_1=m_1|a^\ast,c)\times \Pr(M_2=m_2|a^\ast,m_1,c)\\
& & - \sum_{m_2}\sum_{m_1}E[Y(a^\ast,m_1,m_2)|c]\times \Pr(M_1=m_1|a^\ast,c)\times \Pr(M_2=m_2|a^\ast,m_1,c)\\
& & + \sum_{m_2}\sum_{m_1}E[Y(a^\ast,m_1,m_2^\ast)|c]\times \Pr(M_1=m_1|a^\ast,c)\times \Pr(M_2=m_2|a^\ast,m_1,c)\\
\\
& = & \sum_{m_2}\sum_{m_1}E[Y(a,m_1,m_2)|a,m_1,m_2,c]\times \Pr(M_1=m_1|a^\ast,c)\times \Pr(M_2=m_2|a^\ast,m_1,c)\\
& & - \sum_{m_2}\sum_{m_1}E[Y(a,m_1,m_2^\ast)|a,m_1,m_2^\ast,c]\times \Pr(M_1=m_1|a^\ast,c)\times \Pr(M_2=m_2|a^\ast,m_1,c)\\
& & - \sum_{m_2}\sum_{m_1}E[Y(a^\ast,m_1,m_2)|a^\ast,m_1,m_2,c]\times \Pr(M_1=m_1|a^\ast,c)\times \Pr(M_2=m_2|a^\ast,m_1,c)\\
& & + \sum_{m_2}\sum_{m_1}E[Y(a^\ast,m_1,m_2^\ast)|a^\ast,m_1,m_2^\ast,c]\times \Pr(M_1=m_1|a^\ast,c)\times \Pr(M_2=m_2|a^\ast,m_1,c)\; by\;A1\;A2\\
\\
& = & \sum_{m_2}\sum_{m_1}E[Y|a,m_1,m_2,c]\times \Pr(M_1=m_1|a^\ast,c)\times \Pr(M_2=m_2|a^\ast,m_1,c)\\
& & - \sum_{m_2}\sum_{m_1}E[Y|a,m_1,m_2^\ast,c]\times \Pr(M_1=m_1|a^\ast,c)\times \Pr(M_2=m_2|a^\ast,m_1,c)\\
& & - \sum_{m_2}\sum_{m_1}E[Y|a^\ast,m_1,m_2,c]\times \Pr(M_1=m_1|a^\ast,c)\times \Pr(M_2=m_2|a^\ast,m_1,c)\\
& & + \sum_{m_2}\sum_{m_1}E[Y|a^\ast,m_1,m_2^\ast,c]\times \Pr(M_1=m_1|a^\ast,c)\times \Pr(M_2=m_2|a^\ast,m_1,c)\; by\;consistency\\
\\
& = & \int_{m_2}\int_{m_1}{E\left[Y\middle| a,m_1,m_2,c\right]}d\Pr{\left(M_1=m_1\middle| a^\ast,c\right)}d\Pr{\left(M_2=m_2\middle| a^\ast,m_1,c\right)}\\
& & - \int_{m_2}\int_{m_1}{E\left[Y\middle| a,m_1,m_2^\ast,c\right]}d\Pr{\left(M_1=m_1\middle| a^\ast,c\right)}d\Pr{\left(M_2=m_2\middle| a^\ast,m_1,c\right)}\\
& & - \int_{m_2}\int_{m_1}{E\left[Y\middle| a^\ast,m_1,m_2,c\right]}d\Pr{\left(M_1=m_1\middle| a^\ast,c\right)}d\Pr{\left(M_2=m_2\middle| a^\ast,m_1,c\right)}\\
& & + \int_{m_2}\int_{m_1}{E\left[Y\middle| a^\ast,m_1,m_2^\ast,c\right]}d\Pr{\left(M_1=m_1\middle| a^\ast,c\right)}d\Pr{\left(M_2=m_2\middle| a^\ast,m_1,c\right)}\\
\\
& = & \int_{m_1}\int_{m_2}{E\left[Y\middle| a,m_1,m_2,c\right]}d\Pr{\left(M_2=m_2\middle| a^\ast,m_1,c\right)}d\Pr{\left(M_1=m_1\middle| a^\ast,c\right)}\\
& & - \int_{m_1}\int_{m_2}{E\left[Y\middle| a,m_1,m_2^\ast,c\right]}d\Pr{\left(M_2=m_2\middle| a^\ast,m_1,c\right)}d\Pr{\left(M_1=m_1\middle| a^\ast,c\right)}\\
& & - \int_{m_1}\int_{m_2}{E\left[Y\middle| a^\ast,m_1,m_2,c\right]}d\Pr{\left(M_2=m_2\middle| a^\ast,m_1,c\right)}d\Pr{\left(M_1=m_1\middle| a^\ast,c\right)}\\
& & + \int_{m_1}\int_{m_2}{E\left[Y\middle| a^\ast,m_1,m_2^\ast,c\right]}d\Pr{\left(M_2=m_2\middle| a^\ast,m_1,c\right)}d\Pr{\left(M_1=m_1\middle| a^\ast,c\right)}\\
\\
& = & \int_{m_1}\int_{m_2}(\theta_0+\theta_1a+\theta_2m_1+\theta_3m_2+\theta_4am_1+\theta_5am_2\\
& & +\theta_6m_1m_2+\theta_7am_1m_2+\theta_8^\prime c) d\Pr{\left(M_2=m_2\middle| a^\ast,m_1,c\right)}d\Pr{\left(M_1=m_1\middle| a^\ast,c\right)}\\
& & - \int_{m_1}\int_{m_2}(\theta_0+\theta_1a+\theta_2m_1+\theta_3m_2^\ast+\theta_4am_1+\theta_5am_2^\ast\\
& & +\theta_6m_1m_2^\ast+\theta_7am_1m_2^\ast+\theta_8^\prime c)d\Pr{\left(M_2=m_2\middle| a^\ast,m_1,c\right)}d\Pr{\left(M_1=m_1\middle| a^\ast,c\right)}\\
& & - \int_{m_1}\int_{m_2}(\theta_0+\theta_1a^\ast+\theta_2m_1+\theta_3m_2+\theta_4a^\ast m_1+\theta_5a^\ast m_2\\
& & +\theta_6m_1m_2+\theta_7a^\ast m_1m_2+\theta_8^\prime c)d\Pr{\left(M_2=m_2\middle| a^\ast,m_1,c\right)}d\Pr{\left(M_1=m_1\middle| a^\ast,c\right)}\\
& & + \int_{m_1}\int_{m_2} (\theta_0+\theta_1a^\ast+\theta_2m_1+\theta_3m_2^\ast+\theta_4a^\ast m_1+\theta_5a^\ast m_2^\ast\\
& & +\theta_6m_1m_2^\ast+\theta_7a^\ast m_1m_2^\ast+\theta_8^\prime c)d\Pr{\left(M_2=m_2\middle| a^\ast,m_1,c\right)}d\Pr{\left(M_1=m_1\middle| a^\ast,c\right)}\\
\\
& = & \int_{m_1}\left[(\theta_0+\theta_1a+\theta_2m_1+\theta_4am_1+\theta_8^\prime c)\right.\\
& & +\left.\left(\theta_3+\theta_5a+\theta_6m_1+\theta_7am_1\right)\times\left(\beta_0+\beta_1a^\ast+\beta_2m_1+\beta_3a^\ast m_1+\beta_4^\prime c\right)\right]d\Pr{\left(M_1=m_1\middle| a^\ast,c\right)}\\
& & - \int_{m_1} (\theta_0+\theta_1a+\theta_2m_1+\theta_3m_2^\ast+\theta_4am_1+\theta_5am_2^\ast\\
& & +\theta_6m_1m_2^\ast+\theta_7am_1m_2^\ast+\theta_8^\prime c)d\Pr{\left(M_1=m_1\middle| a^\ast,c\right)}\\
& & - \int_{m_1} \left[(\theta_0+\theta_1a^\ast+\theta_2m_1+\theta_4a^\ast m_1+\theta_8^\prime c)\right.\\
& & +\left.\left(\theta_3+\theta_5a^\ast+\theta_6m_1+\theta_7a^\ast m_1\right)\times\left(\beta_0+\beta_1a^\ast+\beta_2m_1+\beta_3a^\ast m_1+\beta_4^\prime c\right)\right]d\Pr{\left(M_1=m_1\middle| a^\ast,c\right)}\\
& & + \int_{m_1} (\theta_0+\theta_1a^\ast+\theta_2m_1+\theta_3m_2^\ast+\theta_4a^\ast m_1+\theta_5a^\ast m_2^\ast\\
& & +\theta_6m_1m_2^\ast+\theta_7a^\ast m_1m_2^\ast+\theta_8^\prime c)d\Pr{\left(M_1=m_1\middle| a^\ast,c\right)}\\
\\
& = & \left(\theta_0+\theta_1a+\theta_8^\prime c\right)+\left(\theta_3+\theta_5a\right)\left(\beta_0+\beta_1a^\ast+\beta_4^\prime c\right)\\
& & +\left(\theta_2+\theta_4a\right)\left(\gamma_0+\gamma_1a^\ast+\gamma_2^\prime c\right)\\
& & +\left(\theta_6+\theta_7a\right)\left(\beta_0+\beta_1a^\ast+\beta_4^\prime c\right)\left(\gamma_0+\gamma_1a^\ast+\gamma_2^\prime c\right)\\
& & +\left(\theta_3+\theta_5a\right)\left(\beta_2+\beta_3a^\ast\right)\left(\gamma_0+\gamma_1a^\ast+\gamma_2^\prime c\right)\\
& & +\left(\theta_6+\theta_7a\right)\left(\beta_2+\beta_3a^\ast\right)\left[\sigma_{M_1}^2+\left(\gamma_0+\gamma_1a^\ast+\gamma_2^\prime c\right)^2\right]\\
& & -\left(\theta_0+\theta_1a+\theta_3m_2^\ast+\theta_5am_2^\ast+\theta_8^\prime c\right)\\
& & -\left(\theta_2+\theta_4a+\theta_6m_2^\ast+\theta_7am_2^\ast\right)\left(\gamma_0+\gamma_1a^\ast+\gamma_2^\prime c\right)\\
& & -\left(\theta_0+\theta_1a^\ast+\theta_8^\prime c\right)-\left(\theta_3+\theta_5a^\ast\right)\left(\beta_0+\beta_1a^\ast+\beta_4^\prime c\right)\\
& & -\left(\theta_2+\theta_4a^\ast\right)\left(\gamma_0+\gamma_1a^\ast+\gamma_2^\prime c\right)\\
& & -\left(\theta_6+\theta_7a^\ast\right)\left(\beta_0+\beta_1a^\ast+\beta_4^\prime c\right)\left(\gamma_0+\gamma_1a^\ast+\gamma_2^\prime c\right)\\
& & -\left(\theta_3+\theta_5a^\ast\right)\left(\beta_2+\beta_3a^\ast\right)\left(\gamma_0+\gamma_1a^\ast+\gamma_2^\prime c\right)\\
& & -\left(\theta_6+\theta_7a^\ast\right)\left(\beta_2+\beta_3a^\ast\right)\left[\sigma_{M_1}^2+\left(\gamma_0+\gamma_1a^\ast+\gamma_2^\prime c\right)^2\right]\\
& & +\left(\theta_0+\theta_1a^\ast+\theta_3m_2^\ast+\theta_5a^\ast m_2^\ast+\theta_8^\prime c\right)\\
& & +\left(\theta_2+\theta_4a^\ast+\theta_6m_2^\ast+\theta_7a^\ast m_2^\ast\right)\left(\gamma_0+\gamma_1a^\ast+\gamma_2^\prime c\right)\\
\\
& = & \theta_1\left(a-a^\ast\right)+\theta_5\left(\beta_0+\beta_1a^\ast+\beta_4^\prime c\right)\left(a-a^\ast\right)\\
& & +\theta_4\left(\gamma_0+\gamma_1a^\ast+\gamma_2^\prime c\right)\left(a-a^\ast\right)\\
& & +\theta_7\left(\beta_0+\beta_1a^\ast+\beta_4^\prime c\right)\left(\gamma_0+\gamma_1a^\ast+\gamma_2^\prime c\right)\left(a-a^\ast\right)\\
& & +\theta_5\left(\beta_2+\beta_3a^\ast\right)\left(\gamma_0+\gamma_1a^\ast+\gamma_2^\prime c\right)\left(a-a^\ast\right)\\
& & +\theta_7\left(\beta_2+\beta_3a^\ast\right)\left[\sigma_{M_1}^2+\left(\gamma_0+\gamma_1a^\ast+\gamma_2^\prime c\right)^2\right]\left(a-a^\ast\right)\\
& & -\left(\theta_1+\theta_5m_2^\ast\right)\left(a-a^\ast\right)-(\theta_4+\theta_7m_2^\ast)\left(\gamma_0+\gamma_1a^\ast+\gamma_2^\prime c\right)(a-a^\ast)\\
\\
& = & \left\{\theta_1+\theta_5\left(\beta_0+\beta_1a^\ast+\beta_4^\prime c\right)\right.\\
& & \left.+\theta_7\left(\beta_0+\beta_1a^\ast+\beta_4^\prime c\right)\left(\gamma_0+\gamma_1a^\ast+\gamma_2^\prime c\right)\right. \\
& & \left.+\theta_5\left(\beta_2+\beta_3a^\ast\right)\left(\gamma_0+\gamma_1a^\ast+\gamma_2^\prime c\right)\right.\\
& & \left.+\theta_7\left(\beta_2+\beta_3a^\ast\right)\left[\sigma_{M_1}^2+\left(\gamma_0+\gamma_1a^\ast+\gamma_2^\prime c\right)^2\right]\right.\\
& & \left. -\left(\theta_1+\theta_5m_2^\ast\right)-\theta_7m_2^\ast\left(\gamma_0+\gamma_1a^\ast+\gamma_2^\prime c\right) \right\}(a-a^\ast).
\end{eqnarray*}
\\
\subsection*{Natural counterfactual interaction effects}
We can derive the the expected value of each counterfactual formula and find the corresponding combinations for each interaction effect. 

\begin{eqnarray*}
Y(a,M_1(a),M_2(a,M_1(a))) = \sum_{m_2}\sum_{m_1}Y(a,m_1,m_2)\times I(M_1(a)=m_1)\times I(M_2(a,m_1)=m_2)
\end{eqnarray*}
\begin{eqnarray*}
&\Rightarrow& E[Y(a,M_1(a),M_2(a,M_1(a)))|c]\\
\\
& = & E\left[\sum_{m_2}\sum_{m_1}Y(a,m_1,m_2)\times I(M_1(a)=m_1)\times I(M_2(a,m_1)=m_2) \bigg|c\right]\\
\\
& = & \sum_{m_2}\sum_{m_1}{E\left[Y\left(a,m_1,m_2\right)\times I\left(M_1\left(a\right)=m_1\right)\times I\left(M_2\left(a,m_1\right)=m_2\right)\middle| c\right]}\\
\\
& = & \sum_{m_2}\sum_{m_1}{E\left[Y\left(a,m_1,m_2\right)\middle| c\right]E\left[I\left(M_1\left(a\right)=m_1\right)\middle| c\right]E\left[I\left(M_2\left(a,m_1\right)=m_2\right)\middle| c\right]}\; by\; A4\;A6\\
\\
& = & \sum_{m_2}\sum_{m_1}{E\left[Y\left(a,m_1,m_2\right)\middle| c\right]\Pr{\left(M_1\left(a\right)=m_1\middle| c\right)}\Pr{\left(M_2\left(a,m_1\right)=m_2\middle| c\right)}}\\
\\
& = & \sum_{m_2}\sum_{m_1}{E\left[Y\left(a,m_1,m_2\right)\middle| c\right]\Pr{\left(M_1\left(a\right)=m_1\middle| a,c\right)}\Pr{\left(M_2\left(a,m_1\right)=m_2\middle| a,m_1,c\right)}}\; by\; A3\;A5\\
\\
& = & \sum_{m_2}\sum_{m_1}{E\left[Y\left(a,m_1,m_2\right)\middle| c\right]\Pr{\left(M_1=m_1\middle| a,c\right)}\Pr{\left(M_2=m_2\middle| a,m_1,c\right)}}\; by\; consistency\\
\\
& = & \sum_{m_2}\sum_{m_1}{E\left[Y\left(a,m_1,m_2\right)\middle| a,m_1,m_2,c\right]\Pr{\left(M_1=m_1\middle| a,c\right)}\Pr{\left(M_2=m_2\middle| a,m_1,c\right)}}\; by\; A1\;A2\\
\\
& = & \sum_{m_2}\sum_{m_1}{E\left[Y\middle| a,m_1,m_2,c\right]\Pr{\left(M_1=m_1\middle| a,c\right)}\Pr{\left(M_2=m_2\middle| a,m_1,c\right)}}\; by\; consistency\\
\\
& = & \int_{m_2}\int_{m_1}{E\left[Y\middle| a,m_1,m_2,c\right]d}\Pr{\left(M_1=m_1\middle| a,c\right)}d\Pr{\left(M_2=m_2\middle| a,m_1,c\right)}\\
\\
& = & \int_{m_2}\int_{m_1}(\theta_0+\theta_1a+\theta_2m_1+\theta_3m_2+\theta_4am_1+\theta_5am_2\\
& & +\theta_6m_1m_2+\theta_7am_1m_2+\theta_8^\prime c)d\Pr{\left(M_1=m_1\middle| a,c\right)}d\Pr{\left(M_2=m_2\middle| a,m_1,c\right)}\\
\\
& = & \int_{m_1}\int_{m_2}(\theta_0+\theta_1a+\theta_2m_1+\theta_3m_2+\theta_4am_1+\theta_5am_2\\
& & +\theta_6m_1m_2+\theta_7am_1m_2+\theta_8^\prime c)d\Pr{\left(M_2=m_2\middle| a,m_1,c\right)}d\Pr{\left(M_1=m_1\middle| a,c\right)}\\
\\
& = & \int_{m_1}\int_{m_2} [(\theta_0+\theta_1a+\theta_2m_1+\theta_4am_1+\theta_8^\prime c)\\
& & +\left(\theta_3+\theta_5a+\theta_6m_1+\theta_7am_1\right)m_2]d\Pr{\left(M_2=m_2\middle| a,m_1,c\right)}d\Pr{\left(M_1=m_1\middle| a,c\right)}\\
\\
& = & \int_{m_1} \left[\left(\theta_0+\theta_1a+\theta_2m_1+\theta_4am_1+\theta_8^\prime c\right)\right.\\
& & +\left.\left(\theta_3+\theta_5a+\theta_6m_1+\theta_7am_1\right)\left(\beta_0+\beta_1a+\beta_2m_1+\beta_3am_1+\beta_4^\prime c\right)\right]d\Pr\left(M_1=m_1\middle| a,c\right)\\
\\
& = & \int_{m_1} \left[\left(\theta_0+\theta_1a+\theta_8^\prime c\right)+\left(\theta_2+\theta_4a\right)m_1\right.\\
& & +\left.\left(\theta_3+\theta_5a+\left(\theta_6+\theta_7a\right)m_1\right)\left(\beta_0+\beta_1a+\beta_4^\prime c+\left(\beta_2+\beta_3a\right)m_1\right)\right]d\Pr\left(M_1=m_1\middle| a,c\right)\\
\\
& = & \left(\theta_0+\theta_1a+\theta_8^\prime c\right)+\left(\theta_3+\theta_5a\right)\left(\beta_0+\beta_1a+\beta_4^\prime c\right)\\
& & +\left(\theta_2+\theta_4a\right)\left(\gamma_0+\gamma_1a+\gamma_2^\prime c\right)+\left(\theta_6+\theta_7a\right)\left(\beta_0+\beta_1a+\beta_4^\prime c\right)\left(\gamma_0+\gamma_1a+\gamma_2^\prime c\right)\\
& & +\left(\theta_3+\theta_5a\right)\left(\beta_2+\beta_3a\right)\left(\gamma_0+\gamma_1a+\gamma_2^\prime c\right)\\
& & +\left(\theta_6+\theta_7a\right)\left(\beta_2+\beta_3a\right)\left[\sigma_{M_1}^2+\left(\gamma_0+\gamma_1a+\gamma_2^\prime c\right)^2\right].\qquad\qquad(W1)
\end{eqnarray*}

Similarly, we can obtain the following expected values for the rest of the counterfactual formulas. 
\begin{eqnarray*}
& & E[Y(a,M_1(a),M_2(a^\ast,M_1(a)))|c]\\
\\
& = & \left(\theta_0+\theta_1a+\theta_8^\prime c\right)+\left(\theta_3+\theta_5a\right)\left(\beta_0+\beta_1a^\ast+\beta_4^\prime c\right)\\
& & +\left(\theta_2+\theta_4a\right)\left(\gamma_0+\gamma_1a+\gamma_2^\prime c\right)+\left(\theta_6+\theta_7a\right)\left(\beta_0+\beta_1a^\ast+\beta_4^\prime c\right)\left(\gamma_0+\gamma_1a+\gamma_2^\prime c\right)\\
& & +\left(\theta_3+\theta_5a\right)\left(\beta_2+\beta_3a^\ast\right)\left(\gamma_0+\gamma_1a+\gamma_2^\prime c\right)\\
& & +\left(\theta_6+\theta_7a\right)\left(\beta_2+\beta_3a^\ast\right)\left[\sigma_{M_1}^2+\left(\gamma_0+\gamma_1a+\gamma_2^\prime c\right)^2\right]\qquad\qquad(W2)
\end{eqnarray*}

\begin{eqnarray*}
& & E[Y(a,M_1(a^\ast),M_2(a,M_1(a^\ast)))|c]\\
\\
& = & \left(\theta_0+\theta_1a+\theta_8^\prime c\right)+\left(\theta_3+\theta_5a\right)\left(\beta_0+\beta_1a+\beta_4^\prime c\right)\\
& & +\left(\theta_2+\theta_4a\right)\left(\gamma_0+\gamma_1a^\ast+\gamma_2^\prime c\right)+\left(\theta_6+\theta_7a\right)\left(\beta_0+\beta_1a+\beta_4^\prime c\right)\left(\gamma_0+\gamma_1a^\ast+\gamma_2^\prime c\right)\\
& & +\left(\theta_3+\theta_5a\right)\left(\beta_2+\beta_3a\right)\left(\gamma_0+\gamma_1a^\ast+\gamma_2^\prime c\right)\\
& & +\left(\theta_6+\theta_7a\right)\left(\beta_2+\beta_3a\right)\left[\sigma_{M_1}^2+\left(\gamma_0+\gamma_1a^\ast+\gamma_2^\prime c\right)^2\right]\qquad\qquad(W3)
\end{eqnarray*}

\begin{eqnarray*}
& & E[Y(a^\ast,M_1(a),M_2(a,M_1(a)))|c]\\
\\
& = & \left(\theta_0+\theta_1a^\ast+\theta_8^\prime c\right)+\left(\theta_3+\theta_5a^\ast\right)\left(\beta_0+\beta_1a+\beta_4^\prime c\right)\\
& & +\left(\theta_2+\theta_4a^\ast\right)\left(\gamma_0+\gamma_1a+\gamma_2^\prime c\right)+\left(\theta_6+\theta_7a^\ast\right)\left(\beta_0+\beta_1a+\beta_4^\prime c\right)\left(\gamma_0+\gamma_1a+\gamma_2^\prime c\right)\\
& & +\left(\theta_3+\theta_5a^\ast\right)\left(\beta_2+\beta_3a\right)\left(\gamma_0+\gamma_1a+\gamma_2^\prime c\right)\\
& & +\left(\theta_6+\theta_7a^\ast\right)\left(\beta_2+\beta_3a\right)\left[\sigma_{M_1}^2+\left(\gamma_0+\gamma_1a+\gamma_2^\prime c\right)^2\right]\qquad\qquad(W4)
\end{eqnarray*}

\begin{eqnarray*}
& & E[Y(a^\ast,M_1(a^\ast),M_2(a,M_1(a^\ast)))|c]\\
\\
& = & \left(\theta_0+\theta_1a^\ast+\theta_8^\prime c\right)+\left(\theta_3+\theta_5a^\ast\right)\left(\beta_0+\beta_1a+\beta_4^\prime c\right)\\
& & +\left(\theta_2+\theta_4a^\ast\right)\left(\gamma_0+\gamma_1a^\ast+\gamma_2^\prime c\right)\\
& & +\left(\theta_6+\theta_7a^\ast\right)\left(\beta_0+\beta_1a+\beta_4^\prime c\right)\left(\gamma_0+\gamma_1a^\ast+\gamma_2^\prime c\right)\\
& & +\left(\theta_3+\theta_5a^\ast\right)\left(\beta_2+\beta_3a\right)\left(\gamma_0+\gamma_1a^\ast+\gamma_2^\prime c\right)\\
& & +\left(\theta_6+\theta_7a^\ast\right)\left(\beta_2+\beta_3a\right)\left[\sigma_{M_1}^2+\left(\gamma_0+\gamma_1a^\ast+\gamma_2^\prime c\right)^2\right]\qquad\qquad(W5)
\end{eqnarray*}

\begin{eqnarray*}
& & E[Y(a^\ast,M_1(a),M_2(a^\ast,M_1(a)))|c]\\
\\
& = & \left(\theta_0+\theta_1a^\ast+\theta_8^\prime c\right)+\left(\theta_3+\theta_5a^\ast\right)\left(\beta_0+\beta_1a^\ast+\beta_4^\prime c\right)\\
& & +\left(\theta_2+\theta_4a^\ast\right)\left(\gamma_0+\gamma_1a+\gamma_2^\prime c\right)\\
& & +\left(\theta_6+\theta_7a^\ast\right)\left(\beta_0+\beta_1a^\ast+\beta_4^\prime c\right)\left(\gamma_0+\gamma_1a+\gamma_2^\prime c\right)\\
& & +\left(\theta_3+\theta_5a^\ast\right)\left(\beta_2+\beta_3a^\ast\right)\left(\gamma_0+\gamma_1a+\gamma_2^\prime c\right)\\
& & +\left(\theta_6+\theta_7a^\ast\right)\left(\beta_2+\beta_3a^\ast\right)\left[\sigma_{M_1}^2+\left(\gamma_0+\gamma_1a+\gamma_2^\prime c\right)^2\right]\qquad\qquad(W6)
\end{eqnarray*}

\begin{eqnarray*}
& & E[Y(a,M_1(a^\ast),M_2(a^\ast,M_1(a^\ast)))|c]\\
\\
& = & \left(\theta_0+\theta_1a+\theta_8^\prime c\right)+\left(\theta_3+\theta_5a\right)\left(\beta_0+\beta_1a^\ast+\beta_4^\prime c\right)\\
& & +\left(\theta_2+\theta_4a\right)\left(\gamma_0+\gamma_1a^\ast+\gamma_2^\prime c\right)\\
& & +\left(\theta_6+\theta_7a\right)\left(\beta_0+\beta_1a^\ast+\beta_4^\prime c\right)\left(\gamma_0+\gamma_1a^\ast+\gamma_2^\prime c\right)\\
& & +\left(\theta_3+\theta_5a\right)\left(\beta_2+\beta_3a^\ast\right)\left(\gamma_0+\gamma_1a^\ast+\gamma_2^\prime c\right)\\
& & +\left(\theta_6+\theta_7a\right)\left(\beta_2+\beta_3a^\ast\right)\left[\sigma_{M_1}^2+\left(\gamma_0+\gamma_1a^\ast+\gamma_2^\prime c\right)^2\right]\qquad\qquad(W7)
\end{eqnarray*}

\begin{eqnarray*}
& & E[Y(a^\ast,M_1(a^\ast),M_2(a^\ast,M_1(a^\ast)))|c]\\
\\
& = & \left(\theta_0+\theta_1a^\ast+\theta_8^\prime c\right)+\left(\theta_3+\theta_5a^\ast\right)\left(\beta_0+\beta_1a^\ast+\beta_4^\prime c\right)\\
& & +\left(\theta_2+\theta_4a^\ast\right)\left(\gamma_0+\gamma_1a^\ast+\gamma_2^\prime c\right)\\
& & +\left(\theta_6+\theta_7a^\ast\right)\left(\beta_0+\beta_1a^\ast+\beta_4^\prime c\right)\left(\gamma_0+\gamma_1a^\ast+\gamma_2^\prime c\right)\\
& & +\left(\theta_3+\theta_5a^\ast\right)\left(\beta_2+\beta_3a^\ast\right)\left(\gamma_0+\gamma_1a^\ast+\gamma_2^\prime c\right)\\
& & +\left(\theta_6+\theta_7a^\ast\right)\left(\beta_2+\beta_3a^\ast\right)\left[\sigma_{M_1}^2+\left(\gamma_0+\gamma_1a^\ast+\gamma_2^\prime c\right)^2\right].\qquad\qquad(W8)
\end{eqnarray*}

The formulas of natural counterfactual interaction effects can be obtained as follows:
\begin{eqnarray*}
& & E[NatINT_{AM_1}|c]\\
\\
& = & (W2)-(W6)-(W7)+(W8)\\
\\
& = & \left[\theta_4\gamma_1+\theta_7\gamma_1\left(\beta_0+\beta_1a^\ast+\beta_4^\prime c\right)+\theta_5\gamma_1\left(\beta_2+\beta_3a^\ast\right)\right.\\
& & +2\theta_7\gamma_1\left(\beta_2+\beta_3a^\ast\right)\left(\gamma_0+\gamma_2^\prime c\right)\\
& & +\left.\theta_7\gamma_1^2\left(\beta_2+\beta_3a^\ast\right)\left(a+a^\ast\right)\right](a-a^\ast)^2
\end{eqnarray*}

\begin{eqnarray*}
& & E[NatINT_{AM_2}|c]\\
\\
& = & (W3)-(W5)-(W7)+(W8)\\
\\
& = & \left[\theta_5\beta_1+\theta_7\beta_1\left(\gamma_0+\gamma_1a^\ast+\gamma_2^\prime c\right)+\theta_5\beta_3\left(\gamma_0+\gamma_1a^\ast+\gamma_2^\prime c\right)\right.\\
& & \left.+\theta_7\beta_3\left[\sigma_{M_1}^2+\left(\gamma_0+\gamma_1a^\ast+\gamma_2^\prime c\right)^2\right] \right](a-a^\ast)^2
\end{eqnarray*}

\begin{eqnarray*}
& & E[NatINT_{AM_1M_2}|c]\\
\\
& = & (W1)-(W2)-(W3)-(W4)+(W5)+(W6)+(W7)-(W8)\\
\\
& = & \left[\theta_7\beta_1\gamma_1+\theta_5\beta_3\gamma_1+2\theta_7\beta_3\gamma_1\left(\gamma_0+\gamma_2^\prime c\right)+\theta_7\beta_3\gamma_1^2\left(a+a^\ast\right) \right](a-a^\ast)^3
\end{eqnarray*}

\begin{eqnarray*}
& & E[NatINT_{M_1M_2}|c]\\
\\
& = & (W4)-(W5)-(W6)+(W8)\\
\\
& = & \left[\beta_1\gamma_1\left(\theta_6+\theta_7a^\ast\right)+\beta_3\gamma_1\left(\theta_3+\theta_5a^\ast\right)\right.\\
& & +2\beta_3\gamma_1\left(\theta_6+\theta_7a^\ast\right)\left(\gamma_0+\gamma_2^\prime c\right)\\
& & \left. +\beta_3\gamma_1^2\left(\theta_6+\theta_7a^\ast\right)\left(a+a^\ast\right)\right](a-a^\ast)^2.
\end{eqnarray*}
\\
\subsection*{Pure indirect effects}
The pure indirect effect through $M_1$ can be obtained by the following derivation:
\begin{eqnarray*}
PIE_{M_1} = Y(a^\ast,M_1(a),M_2(a^\ast,M_1(a)))-Y(a^\ast,M_1(a^\ast),M_2(a^\ast,M_1(a^\ast)))
\end{eqnarray*}
\begin{eqnarray*}
&\Rightarrow& E[PIE_{M_1}|c]\\
\\
& = & E[Y(a^\ast,M_1(a),M_2(a^\ast,M_1(a)))-Y(a^\ast,M_1(a^\ast),M_2(a^\ast,M_1(a^\ast)))|c]\\
\\
& = & E[Y(a^\ast,M_1(a),M_2(a^\ast,M_1(a)))|c]-E[Y(a^\ast,M_1(a^\ast),M_2(a^\ast,M_1(a^\ast)))|c]\\
\\
& = & (W6)-(W8)
\\
& = & \left[\gamma_1\left(\theta_2+\theta_4a^\ast\right)+\gamma_1\left(\theta_6+\theta_7a^\ast\right)\left(\beta_0+\beta_1a^\ast+\beta_4^\prime c\right)\right.\\
& & +\gamma_1\left(\theta_3+\theta_5a^\ast\right)\left(\beta_2+\beta_3a^\ast\right)\\
& & +2\gamma_1\left(\theta_6+\theta_7a^\ast\right)\left(\beta_2+\beta_3a^\ast\right)\left(\gamma_0+\gamma_2^\prime c\right)\\
& & \left.+\gamma_1^2\left(\theta_6+\theta_7a^\ast\right)\left(\beta_2+\beta_3a^\ast\right)\left(a+a^\ast\right)\right](a-a^\ast).
\end{eqnarray*}

Similarly, the pure indirect effect through $M_2$ can be obtained by the following derivation:
\begin{eqnarray*}
PIE_{M_2} = Y(a^\ast,M_1(a^\ast),M_2(a,M_1(a^\ast)))-Y(a^\ast,M_1(a^\ast),M_2(a^\ast,M_1(a^\ast)))
\end{eqnarray*}
\begin{eqnarray*}
&\Rightarrow& E[PIE_{M_2}|c]\\
\\
& = & E[Y(a^\ast,M_1(a^\ast),M_2(a,M_1(a^\ast)))-Y(a^\ast,M_1(a^\ast),M_2(a^\ast,M_1(a^\ast)))|c]\\
\\
& = & E[Y(a^\ast,M_1(a^\ast),M_2(a,M_1(a^\ast)))|c]-E[Y(a^\ast,M_1(a^\ast),M_2(a^\ast,M_1(a^\ast)))|c]\\
\\
& = & (W5)-(W8)
\\
& = & \left[\beta_1\left(\theta_3+\theta_5a^\ast\right)+\beta_1\left(\theta_6+\theta_7a^\ast\right)\left(\gamma_0+\gamma_1a^\ast+\gamma_2^\prime c\right)\right.\\
& & +\beta_3\left(\theta_3+\theta_5a^\ast\right)\left(\gamma_0+\gamma_1a^\ast+\gamma_2^\prime c\right)\\
& & \left.+\beta_3\left(\theta_6+\theta_7a^\ast\right)\left[\sigma_{M_1}^2+\left(\gamma_0+\gamma_1a^\ast+\gamma_2^\prime c\right)^2\right]\right](a-a^\ast).
\end{eqnarray*}
\\
\subsection*{Total effect}
\begin{eqnarray*}
TE = Y(a,M_1(a),M_2(a,M_1(a)))-Y(a^\ast,M_1(a^\ast),M_2(a^\ast,M_1(a^\ast)))
\end{eqnarray*}
\begin{eqnarray*}
&\Rightarrow& E[TE|c]\\
\\
& = & E[Y(a,M_1(a),M_2(a,M_1(a)))-Y(a^\ast,M_1(a^\ast),M_2(a^\ast,M_1(a^\ast)))|c]\\
\\
& = & E[Y(a,M_1(a),M_2(a,M_1(a)))|c]-E[Y(a^\ast,M_1(a^\ast),M_2(a^\ast,M_1(a^\ast)))|c]\\
\\
& = & (W1)-(W8)
\\
& = & \left[\theta_1+\theta_5\left(\beta_0+\beta_4^\prime c\right)+\beta_1\theta_3+\theta_4\left(\gamma_0+\gamma_2^\prime c\right)+\gamma_1\theta_2\right.\\
& & +\theta_7\left(\beta_0+\beta_4^\prime c\right)\left(\gamma_0+\gamma_2^\prime c\right)+\beta_1\theta_6\left(\gamma_0+\gamma_2^\prime c\right)\\
& & +\gamma_1\theta_6\left(\beta_0+\beta_4^\prime c\right)+\theta_5\beta_2\left(\gamma_0+\gamma_2^\prime\right)+\theta_3\beta_3\left(\gamma_0+\gamma_2^\prime c\right)\\
& & +\theta_3\beta_2\gamma_1+\theta_7\beta_2\sigma_{M_1}^2+\theta_6\beta_3\sigma_{M_1}^2+\theta_7\beta_2\left(\gamma_0+\gamma_2^\prime c\right)^2\\
& & \left. +\theta_6\beta_3\left(\gamma_0+\gamma_2^\prime c\right)^2+2\gamma_1\theta_6\beta_2\left(\gamma_0+\gamma_2^\prime c\right)\right](a-a^\ast)
\\
& & +\left[\beta_1\theta_5+\gamma_1\theta_4+\beta_1\theta_7\left(\gamma_0+\gamma_2^\prime c\right)\right.\\
& & +\gamma_1\theta_7\left(\beta_0+\beta_4^\prime c\right)+\gamma_1\beta_1\theta_6+\theta_5\beta_3\left(\gamma_0+\gamma_2^\prime c\right)\\
& & +\theta_5\beta_2\gamma_1+\theta_3\beta_3\gamma_1+\theta_7\beta_3\sigma_{M_1}^2+\theta_7\beta_3\left(\gamma_0+\gamma_2^\prime c\right)^2\\
& & \left.+2\gamma_1\theta_7\beta_2\left(\gamma_0+\gamma_2^\prime c\right)+2\gamma_1\theta_6\beta_3\left(\gamma_0+\gamma_2^\prime c\right)+\theta_6\beta_2\gamma_1^2\right]\left(a^2-{a^\ast}^2\right)\\
\\
& & + \left[\gamma_1\beta_1\theta_7+\theta_5\beta_3\gamma_1+2\gamma_1\theta_7\beta_3\left(\gamma_0+\gamma_2^\prime c\right)+\theta_7\beta_2\gamma_1^2+\theta_6\beta_3\gamma_1^2\right]\left(a^3-{a^\ast}^3\right)\\
\\
& & + \theta_7\beta_3\gamma_1^2\left(a^4-{a^\ast}^4\right).
\end{eqnarray*}

\end{document}